\documentclass[10pt,onecolumn,aps,prd,preprintnumbers,showpacs,superscriptaddress,nofootinbib,amsmath,amssymb,floats,floatfix,showkeys,notitlepage,longbibliography]{revtex4-1}

\usepackage{orcidlink}
\usepackage{comment}
\usepackage{lipsum}
\usepackage{graphicx}
\usepackage{subfigure}
\usepackage{palatino}
\usepackage{sans}
\usepackage{hyperref}
\hypersetup{colorlinks=true,linkcolor=blue,urlcolor=blue,citecolor=blue}
\usepackage[toc,page]{appendix}
\usepackage[normalem]{ulem}
\usepackage{adjustbox}
\usepackage{latexsym}
\usepackage{amsmath}
\usepackage{amssymb}
\usepackage{amsfonts}
\usepackage{dcolumn}
\usepackage{bm}
\usepackage{tikz}
\usetikzlibrary{decorations.pathmorphing}
\usepackage{bigints}
\usepackage{array,tabularx,multirow,booktabs}
\usepackage[tracking=true]{microtype}
\usepackage{soul} %for highlighting
\SetTracking{}{500}
\SetTracking{encoding={*}, shape=sc}{40}
\UseRawInputEncoding %for inputenc error%
\allowdisplaybreaks

\usepackage[utf8]{inputenc}
\usepackage{algorithm}
\usepackage{algorithmicx}
\usepackage{algpseudocode}
\usepackage{xcolor} % For colored text if needed

\begin{document} \sloppy

\title{New Construction of Black Hole Solution in Non-Commutative Geometry and Their Thermodynamic Properties}
\author{Abdellah Touati \orcidlink{0000-0003-4478-2529}
} 
\email{touati.abph@gmail.com}
\affiliation{Department of Physics, Faculty of Exact Sciences, University of Bouira, Algeria.}

\begin{abstract}

	In this work, we present a new construction of black hole solutions in non-commutative gauge theory by applying the Seiberg-Witten map directly to interaction potentials before solving Einstein’s equations. This approach provides a dynamical effect of spacetime non-commutativity that preserves gauge covariance. We obtain both NC Schwarzschild and charged Reissner-Nordström-like black hole solutions, showing that the charged sector exhibits a novel branch dependence between attractive and repulsive electric interactions absent in the commutative limit. We analyze the geometrical properties, energy conditions, and thermodynamic properties of these spacetimes. Our results reveal that non-commutativity eliminates the temperature divergence at the final evaporation stage, inducing a second-order phase transition, or a Hawking-Page-like phase transition in the presence of pressure. Additionally, linear response analysis indicates high sensitivity to the NC parameter for small black holes. Finally, quantum tunneling investigations for both thermal and non-thermal radiation demonstrate that the NC deformation suppresses the particle-number density and weakens correlations between successive emissions, acting as a barrier to particle escape and supports the formation of a cold finite remnant. From a cosmological standpoint, since these stable remnant possess a fixed Planck-scale mass ($M^{\text{min}}\simeq2.73 M_{P}$), they provide a dynamically generated, purely gravitational cold dark matter candidate that aligns with dark universe phenomenology while simultaneously resolving the black hole information loss paradox.
	
\end{abstract}

%\pacs{04.70.Bw, 04.50.Kd, 04.25.-g, 95.30.Sf}

\maketitle
\tableofcontents
\section{Introduction}\label{intro}

Recently there has been considerable interest in unification theories that attempt to unify gravity, as described by general relativity, with the electromagnetic, weak and strong interactions described by the Standard Model. These two frameworks are among the most successful theories in modern physics: general relativity describes gravitational interaction at the macroscopic scale, while the Standard Model describes the fundamental interactions at the quantum scale. However, when one attempts to apply the gravitational interaction to quantum systems, its effect is negligible compared to other interactions that dominate at this scale; this is due to the relative weakness of Newton's constant compared to the coupling strengths of the strong, weak and electromagnetic interactions, and it is one reason why unification is difficult. In this context, several theories and models have emerged that aim to describe gravity at the quantum scale and thereby make its effects significant enough to allow unification with the other interactions. Among the most promising approaches that provide theoretical backgrounds for quantum gravity (QG) are string theory (ST), which predicts the graviton as the mediator of the gravitational force \cite{ST1,ST2,ST3}; loop quantum gravity (LQG), which presents a non-perturbative framework based on the quantization of geometry (area and volume) \cite{LQG1,LQG2} and can be applied to black hole physics \cite{QLG1,QLG2}; and supergravity (SG) \cite{SG1,SG2,SG3,SG4}. We also mention the semi-classical approach pioneered by Hawking \cite{hawking1}. Unfortunately, none of these approaches has been confirmed experimentally, despite their impressive applications and theoretical results.

The first bridge toward a QG picture that combined aspects of quantum field theory and gravity in a semi-classical setting was proposed by S. Hawking in 1975 \cite{hawking1}. By applying quantum field theory near the event horizon, Hawking discovered black hole radiation and evaporation \cite{hawking1,hawking3}, analogous to black-body radiation. This phenomenon of radiation and evaporation enabled the application of thermodynamic laws to black holes \cite{bardeen1}, treating these objects as thermodynamical systems. Hawking's results opened the way to the large research area of black hole phenomenology \cite{hawking3,harms,vaz,haranas2,hansen1,jawad1,chen1,cavity0,cavity1,cavity2,cavity3,cavity4,cavity5}. Subsequently, several other approaches were developed to derive black hole radiation in parallel with the original model \cite{hawking1}. These include anomaly mechanisms methods near the horizon \cite{anomalies1,anomalies2,anomalies3}, which reproduce Hawking radiation, and quantum tunneling methods in a semi-classical framework \cite{tunn1,tunn3,tunn2,H-J1}, which describe particle tunneling across the horizon. These techniques have proved invaluable for studying radiation from a variety of static, spherically symmetric black holes \cite{tunn10,tunn13,tunn16,tunn5,tunn6,tunn8,tunn7,tunn4}.

Despite the successes of Hawking's semi-classical approach \cite{hawking1} in predicting black hole evaporation, it faces problems at the final stage of evaporation and does not resolve the singularities inside black holes. The semi-classical treatment is applicable only in the exterior near-horizon region; in the deep interior, thermodynamic laws break down and a full QG description becomes necessary. In this context many phenomenological models have been proposed to capture QG effects (alongside the theories mentioned above, such as ST, LQG and SG). Examples include rainbow gravity \cite{feng1,feng5}, generalized uncertainty principle (GUP) and extended uncertainty principle (EUP) frameworks \cite{gup2,lutfugup1,lutfueup2,lutfueup4,evap1,evap3,evap4,nozari03,nozari4}, and NC geometry \cite{noncommutative1}. These models typically predict a minimal length that acts as a natural cutoff removing singular behavior and is expected to be of the order of the Planck scale, where gravitational effects become significant.

In this context NC geometry is one of the most promising approaches to describe QG effects. The main idea is to quantize spacetime itself, thereby introducing QG effects through the non-commutativity of coordinates. Such QG effects are expected to be relevant in strong gravitational fields and negligible in the low-energy limit. This approach is motivated in part by string theory \cite{seiberg1,berger}, which provides tools to regularize Hawking divergences via spacetime quantization. Recently, there has been much interest in the study of thermodynamic properties, thermal stability, and phase transitions of black holes in this geometry using different implementations. These include deformed mass models using smeared energy distributions with various profiles, e.g., Lorentzian distributions \cite{louranziandistrNC2,nctunn3} and Gaussian distributions \cite{nicolini2,gaussiandistrNC2}, or taking into account both distributions \cite{anand2025}, geometric corrections via Bopp's shift \cite{nozari02,nozari01}, formulations based on NC gauge theory of gravity built from the SW map \cite{seiberg1} and the star product \cite{abdellah2,Hassanabadi1,abdellah4,abdellahPhD}\footnote{These papers were written using the model developed in Ref. \cite{cham1}, which is based on a missing term recently reported in Ref.~\cite{Tajron1}.}. Here we aim to study thermodynamic properties of black holes in the presence of NC geometry using the NC gauge-theory formulation based on the SW map.

Our work is situated in the broader context of describing QG effects through NC geometry. However, it differs from the standard smeared mass/charge picture \cite{nicolini2,nozari2013ncq,Sunny1} and from approaches that modify only the final metric \cite{nozari01,Tajron1}, because we treat non-commutativity as a dynamical modification of the interaction itself. In a QG model, the gauge fields that mediate gravity and electromagnetism should be modified at the level of the interaction, rather than merely smearing the source distribution or deforming the final metric. Motivated by the quantum-scale origin of spacetime non-commutativity and by earlier deformed-source constructions, we introduce a new black hole model within NC gauge theory to preserve gauge symmetry, in which the SW map deforms the Newton and Coulomb potentials before the Einstein equations are solved. The resulting potentials encode both the NC structure and the interaction dynamics, providing a more physical description of NC effects at high energies, where spacetime quantization should appear as corrections to field propagation and self-interaction. This framework also makes it possible to distinguish between attractive and repulsive interactions, revealing an asymmetry that is absent in standard smeared-source models and in direct metric-deformation approaches.

Furthermore, this dynamical interpretation of non-commutativity allows us to connect the microscopic and macroscopic aspects of black hole evaporation within a unified physical framework. Semi-classical quantum tunneling describes the emission process at the microscopic level, while the NC deformation of the interaction potentials gradually generates an effective quantum-like barrier against particle escape as evaporation proceeds. This mechanism naturally explains the regularization of the temperature profile, the emergence of a critical point in the heat capacity, and the appearance of a second-order phase transition toward a cold finite remnant in the macroscopic thermodynamic description. These findings align with recent theoretical cosmological models suggesting that stable remnants emerging from primordial black holes serve as ideal candidates for CDM \cite{CDM,CDM1,CDM2,CDM3}. Consequently, the present model provides a natural mechanism for generating a purely gravitational dark matter candidate, thereby linking non-commutative gauge theory to current dark universe phenomenology.

In this work, we derive the effective energy-momentum tensor generated by the NC-deformed interaction potentials and solve the Einstein equations to construct the resulting black hole solutions. We then analyze their geometrical properties and energy conditions, followed by a detailed study of the thermodynamic behavior of both the uncharged and charged cases, including the ADM mass, Hawking temperature, entropy, heat capacity, free energy, and susceptibility with respect to the NC parameter. Finally, we investigate the quantum tunneling process for both thermal and non-thermal radiation, together with the particle-number density and the statistical correlation between successive emissions. In this way, the tunneling analysis is not treated as a separate formal calculation, but as the microscopic dynamical mechanism underlying the thermodynamic evolution toward the critical point and the remnant phase.

This paper is organized as follows. In Sec. \ref{Sec:NCSBH}, we present a new model to construct an NC black hole based on NC gauge theory by solving the Einstein equations in the presence of an NC Newton potential at first order in the NC parameter, and we generalize the construction to the $U(1)$ potential for charged black holes. We then investigate some geometrical properties and energy conditions for this model. In Sec. \ref{Sec:TP}, we study thermodynamic properties of the new black hole solution, where we examine the impact of this geometry on the ADM mass, temperature, and entropy as functions of the NC parameter. We then investigate phase transitions and thermal properties by analyzing the profiles of the heat capacity and both Helmholtz and Gibbs free energies in the presence of this geometry. We also investigate the NC susceptibility function of this black hole to describe its sensitivity to changes in the deformation parameter. In Sec. \ref{Sec:QTP}, we study black hole radiation as a quantum-tunneling process of massless particles from the NC black hole in the context of NC gauge theory, considering two scenarios: thermal and non-thermal radiation. We also investigate particle number densities and correlations between successive emissions in this approach. Finally, Sec. \ref{Sec:Conc}, contains our concluding remarks.

%--------------------------------------------------------------------------------------------
\section{Construction of the non-commutative black hole via gauge theory}\label{Sec:NCSBH}
%--------------------------------------------------------------------------------------------

In this section, we briefly review NC geometry and NC gauge theory via the SW map, and we present a new construction of black hole solutions for studying the effects of this geometry by imposing an NC structure on the interaction potential prior to solving the Einstein equations via the SW map \cite{seiberg1}. In this case, the Einstein equations read
\begin{equation}
	G_{\mu\nu}=R_{\mu\nu}-\tfrac{1}{2}g_{\mu\nu}R=\frac{8\pi G}{c^2}\,\hat{T}_{\mu\nu},\label{eq:Eeq1}
\end{equation}
where $\hat{T}_{\mu\nu}$ is the energy-momentum tensor that encodes both quantum-geometry effects and the interaction dynamics.

In this work, we consider static, spherically symmetric black hole solutions with a Schwarzschild-like metric,
\begin{equation}
	ds^2=-f(r)\,c^2dt^2+\frac{dr^2}{f(r)}+r^2d\theta^2+r^2\sin^2\theta\,d\phi^2, \label{eq:metric1}
\end{equation}
which we use below to solve the Einstein equations in the presence of the deformed potential and thereby determine the lapse (curvature) function $f(r)$ for each case.

The basic idea of our approach is that one may probe QG effects by quantizing spacetime itself; NC geometry provides a convenient effective framework for this purpose. The geometry of NC spacetime is characterized by the following commutation relation between the coordinate operators \cite{noncommutative1}:
\begin{equation}\label{eq:1}
	\left[\hat{x}^\mu,\hat{x}^\nu\right]=i\Theta^{\mu\nu},
\end{equation}
where $\Theta^{\mu\nu}$ is a real, antisymmetric matrix. In this framework, the ordinary product of two functions $f(x)$ and $g(x)$ is replaced by the $\star$-product (Moyal product); defined by:
\begin{equation}\label{eq:2}
	f(x)\star g(x)
	=\left.\exp\!\left(\tfrac{i}{2}\,\Theta^{\mu\nu}\,\partial_\mu\,\partial'_\nu\right)f(x)\,g(x')\right|_{x'=x}.
\end{equation}

The NC gauge theory was developed by N. Seiberg and E. Witten \cite{seiberg1}, who introduced an important gauge transformation that establishes a correspondence between ordinary gauge fields and NC gauge fields. This transformation is called the SW map and is given by
\begin{equation}\label{eq:SW1}
	\hat{A}(A,\Theta)+\hat{\delta}_{\hat{\lambda}}\hat{A}(A,\Theta)=\hat{A}(A+\delta_{\lambda}A,\Theta),
\end{equation}
where $\delta_{\lambda}$ and $\delta_{\hat{\lambda}}$ denote the infinitesimal variations under commutative and NC gauge transformations, respectively. By using this map, we express the deformed gauge potential $\hat{A}$ as a power series in $\Theta$ up to first order \cite{seiberg1}:
\begin{equation}
	\hat{A}_\mu=A_\mu-\frac{g}{2}\,\Theta^{\alpha\beta}A_\alpha\left(\partial_\beta A_\mu+F_{\beta\mu}\right)+\mathcal{O}(\Theta^2),,
\end{equation}
where $g$ is the coupling constant of the interaction. Moreover, this geometry leads to the emergence of self-interactions of the gauge field, with the NC parameter acting as an effective coupling constant. Such interactions are not present in the commutative $U(1)$ gauge potential. This can therefore be interpreted as a new type of quantum interaction generated by the NC structure of spacetime itself.

In general, the choice of the NC matrix is motivated by its analogy with a magnetic field in the case of space-space non-commutativity \cite{Josif,Jackiw1,nair1,Mebarek1} and with an electric field in the case of space-time non-commutativity \cite{seiberg2000,zaim2013,zaim2015}. Together with its antisymmetric nature, we aim to present this matrix in a more general form similar to the electromagnetic tensor $F_{\mu\nu}$. In this case, the NC matrix in general form is given by:
\begin{equation}
	\Theta^{\mu\nu}=\begin{pmatrix}
		0	& a_1 & a_2 & a_3 \\
		-a_1 & 0 & d_1 & d_2 \\
		-a_2 & -d_1 & 0 & d_3\\
		-a_3	& -d_2 & -d_3 & 0
	\end{pmatrix}, \qquad \mu,\nu=0,1,2,3. \label{eq:NCM}
\end{equation}
where the space-time non-commutativity, given by the constant NC parameters $a_i$, can generate an effect equivalent to the presence of an electric field, while the space-space non-commutativity, given by the constants $d_i$, can generate an effect equivalent to the presence of a magnetic field.
In this work, we use only the first component of the gauge potential $A_t$, which is static and radial in spherical symmetry. Therefore, the only non-zero NC correction comes from $a_1=\Theta$, and the remaining components can be set to zero.

\subsection{Non-commutative Schwarzschild black hole}

To obtain the NC correction to a black hole, we begin by modifying the Newtonian potential for gravity via the SW map. To apply this map to the gravitational potential, we use the gravito-electromagnetic potential notation \(\mathcal{A}_\mu(r)=\big(\Phi_N(r)/c,0,0,0\big)\), where the scalar potential \(\Phi_N(r)\) is given by the Newtonian potential:
\begin{equation}
	\Phi_N(r)=-\frac{GM}{r}\,
\end{equation}
To describe the NC correction to this potential, we employ NC gauge theory via the SW map, which yields 
\begin{equation}\label{eq:NCNP1}
	\hat{\Phi}_N(r)=-\frac{GM}{r}+\frac{\Theta}{l_p c^2}\,\frac{G^2M^2}{r^3}\,
\end{equation}
where the coupling charge in the gravitational case is taken to be \(g=m_p/(\hbar c)\).\footnote{We have used the standard Planck-unit definitions \(l_p=\sqrt{G\hbar/c^3}\) and \(m_p=\sqrt{\hbar c/G}\). After simple algebra, one obtains \(m_p/(\hbar c)=1/(l_p c^2)\).} From now on, we set \(\tilde{\Theta}=\Theta/l_p\) as a new NC parameter with dimensions of length.

Next, we obtain the contribution of NC geometry to the Newtonian mass density. The Poisson relation \(\nabla^2\hat{\Phi}_N(r)=4\pi G\hat{\rho}\) for the exterior region \(r>0\) yields
\begin{equation}
	\hat{\rho}=3\frac{\tilde{\Theta}G M^2}{2\pi c^2r^5}.\label{eq:NCrhom1}
\end{equation}
For a spherically symmetric matter distribution, the energy-momentum tensor can be written in anisotropic-fluid form with diagonal components \cite{GR1}:
\begin{equation}\label{eq:EMT1}
	[\hat{T}^\mu{}_\nu]=\operatorname{diag}(-\hat{\rho},\hat{p}_r,\hat{p}_t,\hat{p}_t).
\end{equation}
To preserve the Schwarzschild-like form of the metric, the radial pressure must satisfy \(p_r=-\rho\), which, together with the conservation equations, implies a tangential pressure
\(\hat{p}_t=-\hat{\rho}-\tfrac{r}{2}\partial_r\hat{\rho}\). The temporal and radial components of the energy-momentum tensor in our case read
\begin{equation}
	\hat{T}^0_0= \hat{T}^r_r=-\frac{3GM^2}{2\pi r^5}\tilde{\Theta}.\label{eq:emtdmass}
\end{equation}
For a Schwarzschild-like metric, the lapse (curvature) function has the form
\begin{equation}
	f(r)=1-\frac{2G\tilde{m}(r)}{c^2r},\label{eq:NCCF1}
\end{equation} 
where \(\tilde{m}(r)\) is a function of \(r\). Using metric \eqref{eq:metric1} and \eqref{eq:NCCF1}, the non-vanishing components of the Einstein tensor are
\begin{equation}
	G^0_0=G_r^r=-\frac{2G\tilde{m}'(r)}{c^2r^2},\qquad G^\theta_\theta=G^\phi_\phi=-\frac{G\tilde{m}''(r)}{c^2r^2}.
\end{equation} 
Using the Einstein equations, we obtain
\begin{align}
	\tilde{m}(r)&=M-\frac{1}{c^2}\int 4\pi r'^2 \hat{T}^0_0\,dr',\notag\\
	&=M-\frac{3GM^2}{c^2r^2}\tilde{\Theta}= M\bigg(1-\frac{3m}{r^2}\tilde{\Theta}\bigg)\label{eq:smass1}
\end{align} 
where \(m=\dfrac{GM}{c^2}\) is the Schwarzschild mass of the black hole. Note that this expression is written to first order in the NC parameter; at higher orders, one obtains a different compact form\footnote{This expression is valid only at first order in \(\tilde{\Theta}\); higher-order corrections change the compact form.}. In a regularized form, similar to a regular-like black hole and reducing to \eqref{eq:smass1} at \(\mathcal{O}(\tilde{\Theta})\), one may write
\begin{align}
	\tilde{m}(r)=\frac{Mr^2}{\big(r^2+3m\tilde{\Theta}\big)}. \label{eq:smass2}
\end{align} 
Substituting this result into \eqref{eq:NCCF1} gives the NC Schwarzschild metric
\begin{subequations}
	\begin{align}
		f(r)&= 1-\frac{2mr}{\big(r^2+3m\tilde{\Theta}\big)}\label{eq:NCRFL1},\\
		&\simeq 1-\frac{2m}{r}+\frac{6m^2}{r^3}\tilde{\Theta}\label{eq:NCApp1}.
	\end{align}
\end{subequations}
The second line is the first-order expansion, equivalent to the solution obtained in Eq.~\eqref{eq:smass1}, whereas the first line has a form analogous to regular black hole solutions. In our construction, this solution is obtained from the gauge-theory approach to non-commutativity via the SW map correction; the compact form above is valid only to first order in the SW map parameter. In the limit \(\tilde{\Theta}\to0\), the commutative Schwarzschild metric is recovered. It is important to note that our compact form \eqref{eq:NCRFL1} coincides with the curvature function reported in Ref. \cite{bonanno1} for the case \(\gamma=0\), where the author defined a running Newton constant using the renormalization group. Their result differs from ours by a term proportional to \(\gamma\); when \(\gamma\) is set to zero, their metric reduces to our metric above with the following parameter defined as \(\tilde{\omega}=3M\tilde{\Theta}/c^2\), which is a QG parameter. This parameter represents the running of the Newton constant due to quantum vacuum fluctuations arising from the renormalization group, and the fact that this parameter can be related to the NC parameter of the geometry in our model connects quantum fluctuations with the quantization of spacetime. In particular, we can observe some similarity with the holographic screen metric \cite{nicolini-holographic}, while the difference lies in the constant parameter appearing in the numerator of their model. The overall behavior remains the same as in the above-mentioned models, which shows the consistency of this model with other QG approaches.

\subsubsection{Geometrical properties}\label{Sub:GP1}

In the following, we present some geometric properties of the new black hole solution in NC spacetime.
\vspace{0.5cm}
\paragraph{Event horizon:}

We first consider the event horizons, which are obtained by solving \(f(r_{\pm})=0\). The solutions are
\begin{equation}
	r_\pm=m\pm\sqrt{m^2-3m\tilde{\Theta}}\label{eq:eh1}
\end{equation}
Thus, the black hole possesses two event horizons, an inner horizon and an outer horizon. In the commutative limit \(\tilde{\Theta}\to0\), the two horizons merge, and one recovers the single Schwarzschild horizon.

\begin{figure}[h]
	\begin{center}
		\includegraphics[width=0.45\textwidth]{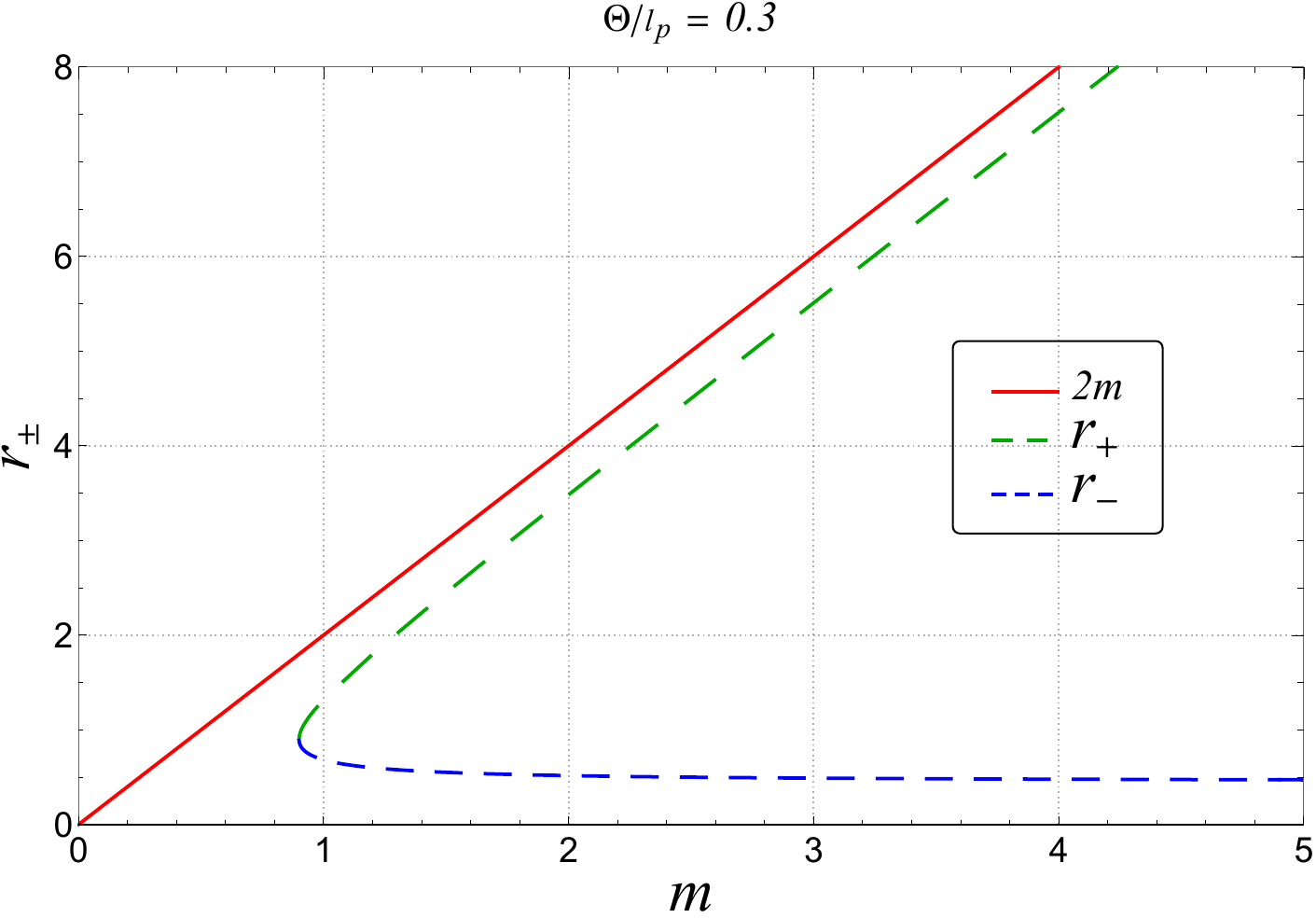}
		\includegraphics[width=0.45\textwidth]{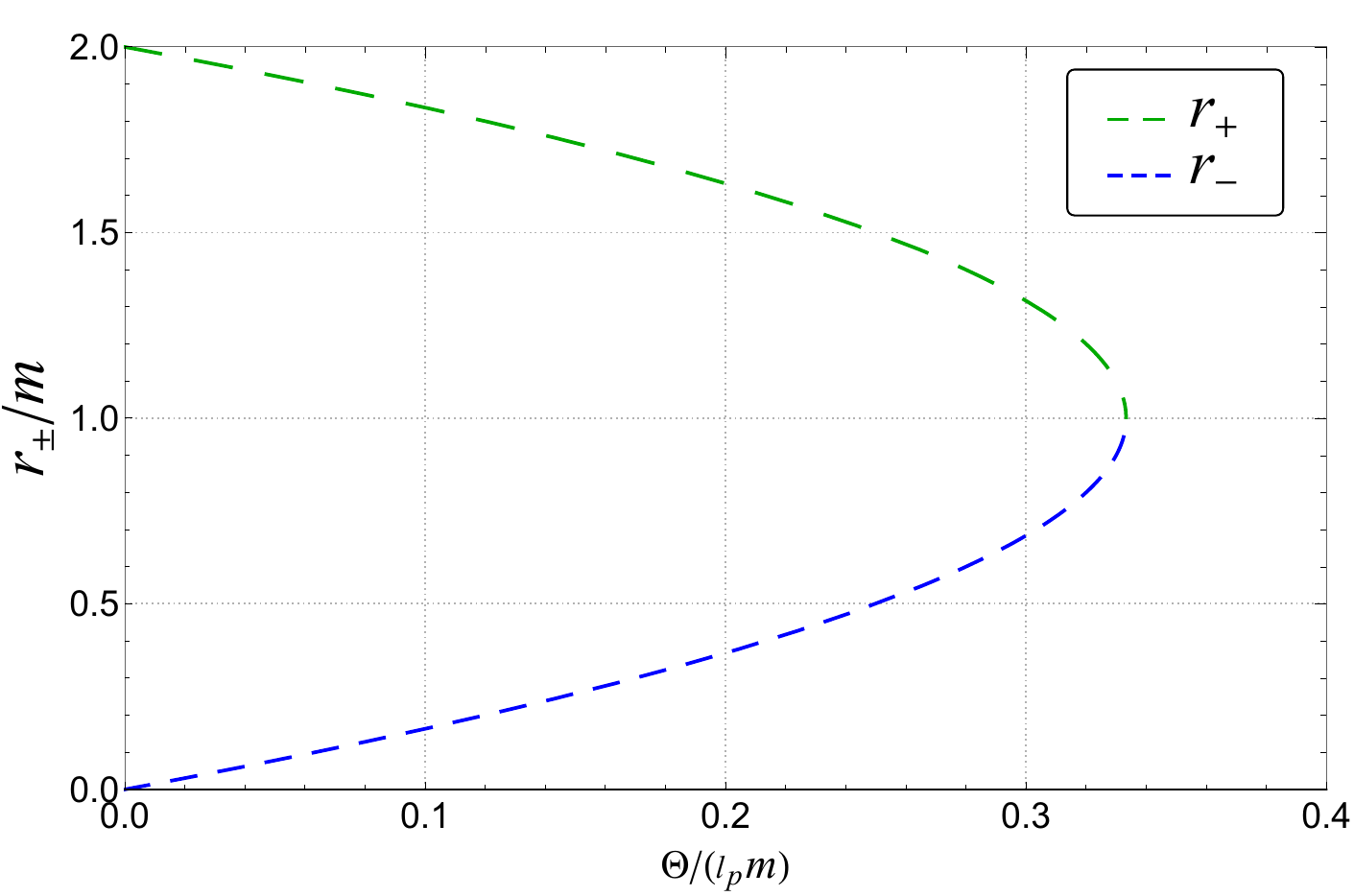}
		\caption{Behavior of the event horizons \(r_\pm\) as a function of the mass for \(\tilde{\Theta}=0.3\), compared with the commutative solution \(\tilde{\Theta}=0.0\) (solid red line) (left panel), and as a function of the NC parameter \(\tilde{\Theta}\) (right panel).} \label{fig:r1}
	\end{center}
\end{figure}
Fig.~\ref{fig:r1} shows the dependence of the outer and inner horizons \(r_\pm\) on the black hole mass \(m\), together with the commutative case. The qualitative behavior resembles that of the RN metric: the NC parameter plays a role analogous to the electric charge in producing a pair of horizons. In our solution, the NC event horizons are smaller than the commutative Schwarzschild radius for the chosen parameters; this trend agrees with previous NC gauge-theory results \cite{Tajron1}, although the magnitude of the effect depends on the adopted model. For other NC implementations, the effect can be reversed, with the horizon increasing with non-commutativity \cite{abdellah6}. From the right panel, we clearly see that increasing \(\tilde{\Theta}\) increases the inner horizon and decreases the outer horizon until they coincide. This coincidence corresponds to an extremal black hole, while larger values of \(\tilde{\Theta}\) (for fixed \(m\)) lead to no black hole solution.

\begin{figure}[h]
	\begin{center}
		\includegraphics[width=0.45\textwidth]{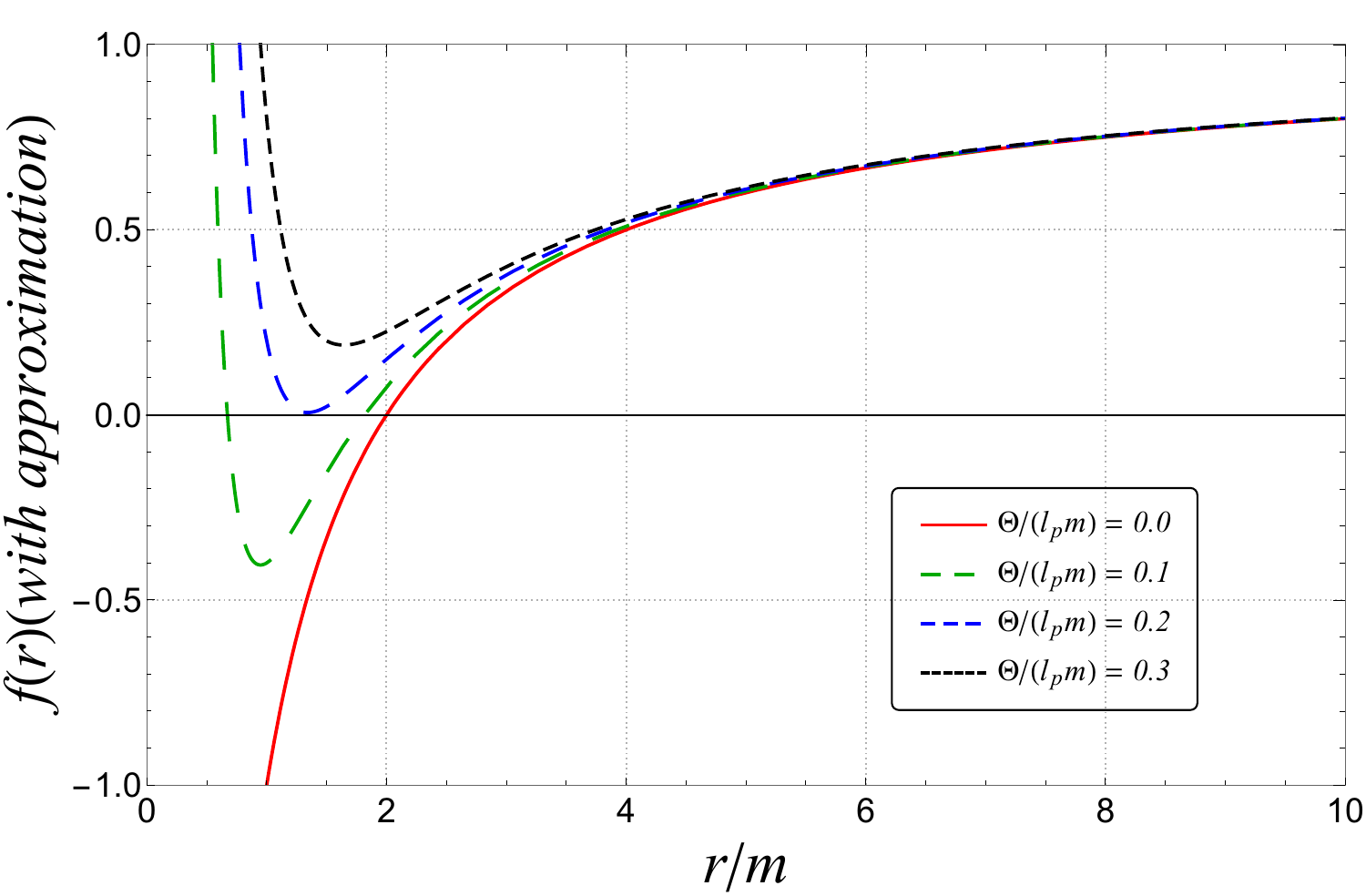}
		\includegraphics[width=0.45\textwidth]{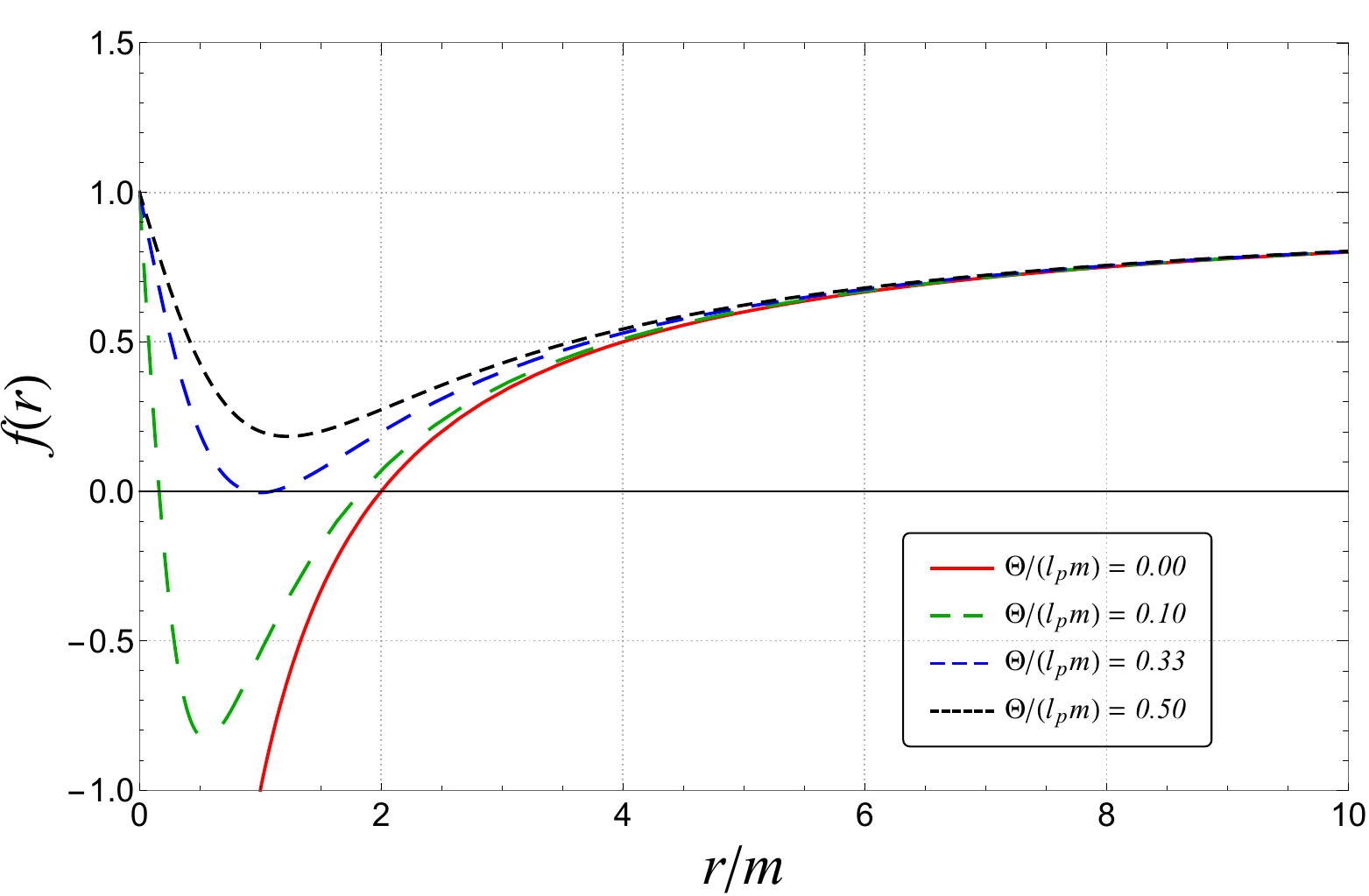}
		\caption{Behavior of the curvature function \(f(r)\) as a function of the radius \(r\) for different values of \(\tilde{\Theta}\). Left panel: leading order in \(\tilde{\Theta}\). Right panel: compact form.}\label{fig:1}
	\end{center} 
\end{figure}
Fig.~\ref{fig:1} displays \(f(r)\) in both the first-order expansion (left panel) and the compact form (right panel) as a function of the radius \(r/m\) for different values of \(\tilde{\Theta}\). In both representations, the solution admits two event horizons given by \eqref{eq:eh1}. The behavior is qualitatively similar to that of the RN family (see \cite{Chandr1,sharif1}): one finds three regimes determined by the sign of \(m-3\tilde{\Theta}\). If \(m>3\tilde{\Theta}\), the solution describes a black hole with two distinct horizons and a timelike curvature singularity (here \(f(r)>0\) at \(r=0\)); if \(m=3\tilde{\Theta}\), the solution is extremal with a double horizon; and if \(m<3\tilde{\Theta}\), no horizon exists and the solution corresponds to a naked singularity. The same qualitative regimes are apparent in the left panel. Our SW map construction therefore yields a horizon structure similar to that obtained from point-like NC matter distributions (Lorentzian or Gaussian) and also shows similarities with certain nonlinear electromagnetic black hole models \cite{non-linearEM1}.

\vspace{0.5cm}
\paragraph{Scalar invariants:} Now we consider two scalar invariants that contain information about the curvature of spacetime and its singularity. We begin with the Ricci scalar, which is obtained by contracting the Ricci tensor. For static and spherically symmetric spacetime, the Ricci scalar in terms of the curvature function \(f(r)\) is written as follows:
\begin{equation}
	R = -f''(r) - \frac{4}{r}f'(r) - \frac{2}{r^2}(f(r) - 1),\label{eq:RS1}
\end{equation}
By using the above regular-like form of \(f(r)\) in \eqref{eq:NCRFL1}, we find:
\begin{equation}
	R=\frac{12 a m^2 \left(-r^2+9 a m\right)}{r \left(3 a m+r^2\right)^3},\label{eq:NCRSM1}
\end{equation}
It is clear that, in NC geometry, the Ricci scalar is non-zero for the Schwarzschild black hole, contrary to the commutative case, which agrees with other models of NC-geometry corrections to the metric, where non-commutativity is introduced through the energy-momentum tensor as a source of the NC black hole. This scalar is not regular, even though the curvature function has a regular-like form. This approach softens the singularity and hides it inside the inner event horizon. This can be seen from the limit of the Ricci scalar, which gives:
\begin{equation}
	\lim_{r\to0}R=\frac{4}{\tilde{\Theta} r}\bigg|_{r=0},\label{eq:NCRSM2}
\end{equation}
which clearly has a divergent behavior at \(r=0\).

\begin{figure}[h]
	\begin{center}
		\includegraphics[width=0.45\textwidth]{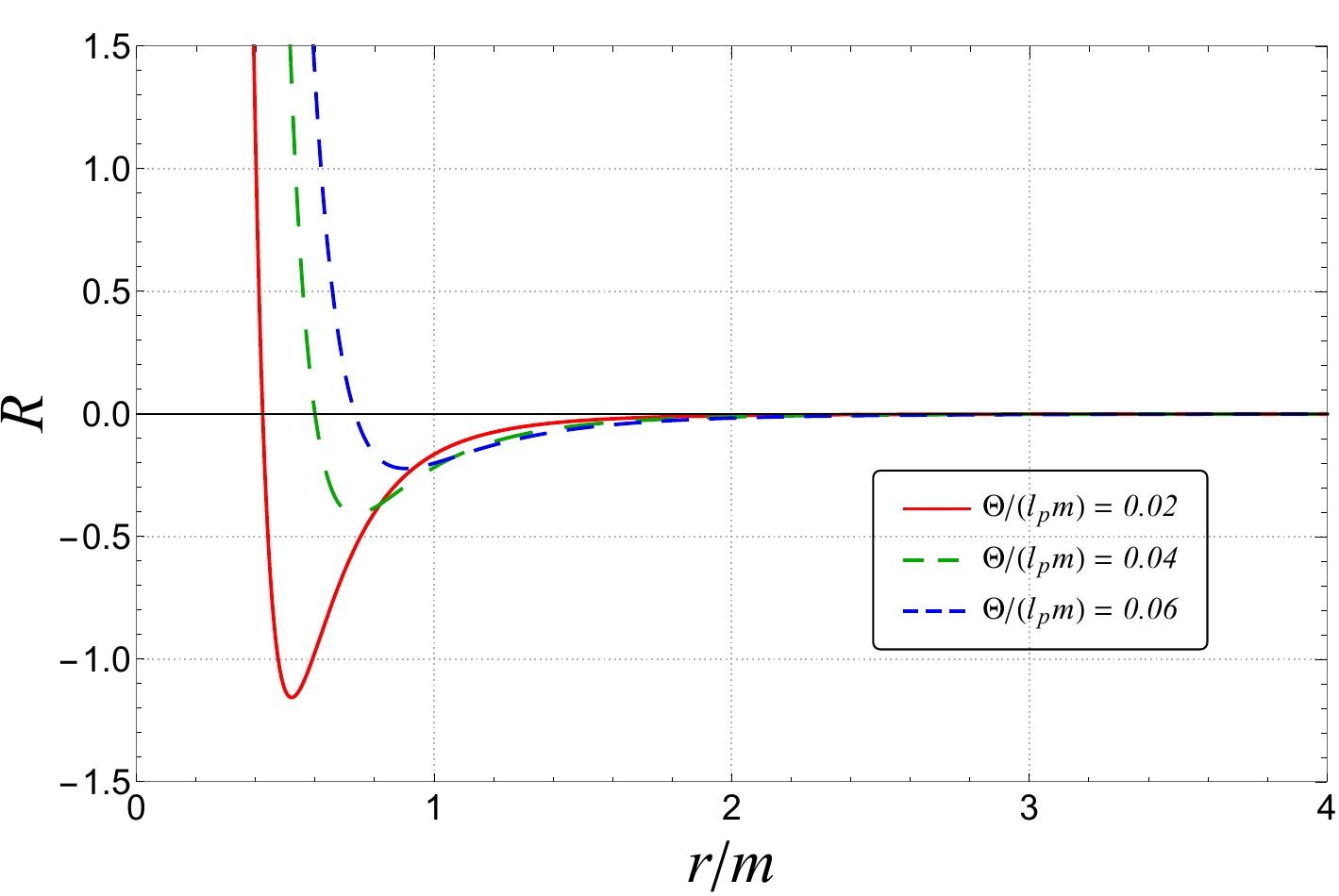}
		\caption{Behavior of the Ricci scalar \(R\) as a function of the radius \(r\) for different values of \(\tilde{\Theta}\).}\label{fig:R-1}
	\end{center} 
\end{figure}
In Fig.~\ref{fig:R-1}, we illustrate the behavior of the Ricci scalar \(R\) as a function of \(r\), for different values of the NC parameter. As shown in this figure, the Ricci scalar of the NC Schwarzschild black hole with the regular-like metric \eqref{eq:NCRFL1} shows a typical behavior of regular black holes or smeared-source models, in which a negative minimum appears in the Ricci scalar for small values of the NC parameter \(\tilde{\Theta}\). This is consistent with other singular black hole models, while the main difference lies at the origin, where our model shows a non-singular behavior at this radius. Since the Ricci scalar is related to the trace of the energy-momentum tensor, this minimum indicates the maximal curvature deformation of spacetime induced by the strongest NC correction at this radius. Unfortunately, even the strongest correction from this geometry is not enough to regularize the divergence at the origin.

The second scalar invariant is the Kretschmann scalar, which is defined in terms of the Riemann tensor as \(K=R_{\rho\mu\nu\sigma}R^{\rho\mu\nu\sigma}\), and for our metric case is given by:
\begin{equation}
	K=f''(r)^2+\frac{4}{r^2}f'(r)^2+\frac{4}{r^4}(1-f(r))^2,\label{eq:KS1}
\end{equation}
Using our above solution \eqref{eq:NCRFL1}, we find:
\begin{equation}
	K=\frac{48 m^2 \left(54 \tilde{\Theta}^4 m^4+36 \tilde{\Theta}^3 m^3 r^2+39 \tilde{\Theta}^2 m^2 r^4-2 \tilde{\Theta} m r^6+r^8\right)}{r^2 \left(3 \tilde{\Theta} m+r^2\right)^6},\label{eq:NCKSM1}
\end{equation}
For small radii, the dominant term is given by:
\begin{equation}
	\lim_{r\to0}K=\frac{32}{9\tilde{\Theta}^2 r^2}\bigg|_{r=0},\label{eq:NCKSM2}
\end{equation}
As we observe, the singularity still exists, but it is weakened by the NC correction compared with the commutative case \(K\propto1/r^6\), where it is reduced to \(K\propto1/r^2\). The regularization of the curvature function produced by NC geometry is not sufficient to remove the central singularity in the curvature of spacetime, but it succeeds in weakening this singularity. To address this problem, we should compute higher-order corrections and then compare them with the Taylor expansion to fix the compact form that satisfies the following two principal conditions:
\begin{itemize}
	\item It reduces to the SW map correction in the expansion series in powers of \(\tilde{\Theta}\).
	\item At small radii, the curvature function must show a de Sitter core: \(f(r)\sim 1-\beta r^q\), where \(q\geq2\).
\end{itemize}
Moreover, the generalization of our compact regular-like form metric \eqref{eq:NCRFL1} is given by:
\begin{equation}
	f(r)=1-\frac{2m r^s}{(r^2+ d m\tilde{\Theta})^p}, 
\end{equation}
where the parameters \(s\), \(d\), and \(p\) are fixed by the above two conditions. For greater accuracy, the NC corrections should be extended to higher orders rather than only the first order, since the first-order approximation only weakens the singularity in the Kretschmann scalar.

\subsection{Non-commutative charged black hole}

In the following, we consider a charged black hole whose electric potential is
\(A_\mu(r)=\big(\Phi_e(r)/c,0,0,0\big)\), with the Coulomb scalar potential
\begin{equation}
	\Phi_e(r)=\pm\frac{q}{4\pi\varepsilon_0 r}\,.
\end{equation}
where the sign $\pm$ describes the repulsive and attractive interactions, respectively. In analogy with the gravitational case, we apply the NC gauge-theory correction to the Coulomb potential via the SW map. To first order in \(\Theta\), the SW map for the gauge potential reads\footnote{With the coupling charge in the electromagnetic interaction taken as $g=q_p/\hbar$.}
\begin{equation}
	\hat{A}_\mu=A_\mu-\frac{q_p}{\hbar}\,\Theta^{\alpha\beta}A_\alpha\partial_\beta A_\mu+\mathcal{O}(\Theta^2),
\end{equation}
where \(q_p=\sqrt{4\pi\varepsilon_0\hbar c}\) is the Planck charge. After some algebra, we obtain the corrected time component
\begin{equation}
	\hat{A}_0=\pm\frac{q}{4\pi\varepsilon_0 c\, r}
	+\frac{\Theta}{l_p}\,\frac{\sqrt{G}\,q^2}{(4\pi\varepsilon_0)^{3/2}\, c^3\,r^3}\,.
\end{equation}
As before, we introduce \(\tilde{\Theta}=\Theta/l_p\) as the NC parameter with dimensions of length.

We now compute the energy-momentum tensor for the corrected gauge potential. In our conventions, the Maxwell energy-momentum tensor (SI units) may be written as
\begin{equation}
	(T^{\mathrm{Maxwell}})_{\mu\nu}
	=\varepsilon_0\Big(F_{\mu\sigma}F_{\nu}{}^{\sigma}-\tfrac{1}{4}g_{\mu\nu}F_{\rho\sigma}F^{\rho\sigma}\Big),
\end{equation}

The nonzero diagonal components of the corrected Maxwell energy-momentum tensor evaluated for our \(\hat{A}_0\) are
\begin{equation}
	(\hat{T}^{\mathrm{Maxwell}})_0{}^0=(\hat{T}^{\mathrm{Maxwell}})_r{}^r
	=-\frac{q^2}{32\pi^2\varepsilon_0\, r^4}
	\mp\tilde{\Theta}\,\frac{3\sqrt{G}\,q^3}{4\pi(4\pi\varepsilon_0)^{3/2} c^2\, r^6}.\label{eq:TNC_corrected}
\end{equation}
It is clear that the presence of the sign \(\pm\) in the NC contribution to the energy-momentum tensor distinguishes between attractive and repulsive interactions in NC geometry, which is not possible in commutative spacetime. In this geometry, these interactions can be distinguished, contrary to the commutative case. This result is observed only in this approach, whereas other NC models are based on the deformation of the final metric. 
Substituting \eqref{eq:TNC_corrected} into the Einstein equations for the Schwarzschild-like metric gives
\begin{align}
	\tilde{m}(r)&=M-\frac{1}{c^2}\int 4\pi r'^2 (\hat{T}^{\mathrm{Maxwell}})^0{}_0\,dr'\notag\\
	&=M-\frac{q^2}{8\pi\varepsilon_0 c^2\, r}
	\mp\frac{\tilde{\Theta}\sqrt{G}\,q^3}{(4\pi\varepsilon_0)^{3/2} c^4\, r^3}\notag\\
	&= M-\frac{q^2}{8\pi\varepsilon_0 c^2\, r}\Big(1\pm\frac{2\tilde{\Theta}Q}{r^2}\Big),\label{eq:smass3_corrected}
\end{align}
where we define the effective squared charge (with dimensions of area) as
\[
Q^2=\frac{G q^2}{4\pi\varepsilon_0 c^4}.
\]
A regular-like compact form that reproduces the first-order expression above is
\begin{equation}
	\tilde{m}(r)=M-\frac{q^2 r}{8\pi\varepsilon_0 c^2\big(r^2\mp2\tilde{\Theta}Q\big)},
\end{equation}
valid to first order in \(\tilde{\Theta}\).

Substituting this mass function into \(f(r)=1-2Gm(r)/(c^2 r)\) yields the NC RN-like metric function:
\begin{subequations}
	\begin{align}
		f^{(\mp)}_Q(r)&=1-\frac{2m}{r}+\frac{Q^2}{\big(r^2\mp2\tilde{\Theta}Q\big)}\label{eq:NCRN1}\\
		&\simeq 1-\frac{2m}{r}+\frac{Q^2}{r^2}\pm\frac{2\tilde{\Theta}Q^3}{r^4},\label{eq:NCRN2}
	\end{align}
\end{subequations}
where the second line is the first-order expansion in \(\tilde{\Theta}\). In the limit \(\tilde{\Theta}\to0\), one recovers the standard RN solution.

It is worth remarking that this SW-map-based deformation yields a comparatively simple corrected metric; other approaches that introduce NC gauge theory of gravity into the metric often lead to more complicated expressions that require additional approximations \cite{abdellahPhD}. 

One may also combine the deformed mass contribution (from the gravitational NC correction) with the deformed electromagnetic contribution by adding the corresponding energy-momentum components. Doing so gives the combined mass function
\begin{align}
	\tilde{m}(r)&=M-\frac{1}{c^2}\int 4\pi r'^2\big(\hat{T}^0{}_0+(\hat{T}^{\mathrm{Maxwell}})^0{}_0\big)\,dr'\notag\\
	&=\frac{M r^2}{\big(r^2+3m\tilde{\Theta}\big)}-\frac{q^2 r}{8\pi\varepsilon_0 c^2\big(r^2\mp2\tilde{\Theta}Q\big)}\notag\\
	&\simeq M\Big(1-\frac{3m}{r^2}\tilde{\Theta}\Big)-\frac{q^2}{8\pi\varepsilon_0 c^2\, r}\Big(1\pm\frac{2\tilde{\Theta}Q}{r^2}\Big).
\end{align}

Correspondingly, the curvature function in compact and first-order expansion forms becomes
\begin{subequations}
	\begin{align}
		f^{(\mp)}_Q(r)&=1-\frac{2mr}{\big(r^2+3m\tilde{\Theta}\big)}+\frac{Q^2}{(r^2\mp2\tilde{\Theta}Q)},\label{eq:NCRNM1}\\
		&\simeq 1-\frac{2m}{r}+\frac{Q^2}{r^2}+\Big(\frac{6m^2}{r^3}\pm\frac{2Q^3}{r^4}\Big)\tilde{\Theta}.\label{eq:NCRNM2}
	\end{align}
\end{subequations}
Here, \(f^{(-)}_Q(r)\) describes the NC RN-like black hole for the repulsive electric branch, while \(f^{(+)}_Q(r)\) corresponds to the attractive one. In the commutative limit \(\tilde{\Theta}\to 0\), both branches reduce to the standard RN geometry, which depends only on \(Q^2\) and is therefore insensitive to the sign or nature of the electric interaction. By contrast, in the NC case the geometry becomes explicitly branch-dependent, as reflected by the \(\pm\,2Q^3\tilde{\Theta}/r^4\) correction. This shows that non-commutativity induces an intrinsic asymmetry between attractive and repulsive electric interactions, leading to distinct short-distance behaviors for the two branches. To the best of our knowledge, this interaction-sensitive deformation of the RN metric provides a novel feature of the present NC model, since it distinguishes the attractive and repulsive electric branches already at first order in $\tilde{\Theta}$. Therefore, this branch-dependent deformation may have important implications for the innermost stable circular orbit (ISCO) and, consequently, for accretion-disk observables such as the inner-edge position, radiative efficiency, and near-horizon emission profile. In this way, the NC correction introduces a potentially observable distinction between the attractive and repulsive electric branches through its impact on the ISCO structure and the associated astrophysical signatures. 

\subsubsection{Geometrical properties}\label{Sub:GP2}
In what follows, we present some geometric properties of the new charged NC black hole, such as the event horizons, Ricci scalar, and Kretschmann scalar.

\vspace{0.5cm}
\paragraph{Event horizon:}

For the case without the deformed mass (Eq.~\eqref{eq:NCRN2}), one finds two solutions\footnote{Note that both solutions formally satisfy \(f(r_+)=0\); however, the inner-horizon solution may become inconsistent with numerical and graphical results due to the leading-order approximation in the NC parameter, and the same remark applies to the cases below. The perturbative expansion becomes unreliable near the extremal regime and for small radii, so the inner-horizon expression should be used with caution.} given by the following expression for the repulsive case:
\begin{equation}
	r_\pm^{(-)}=m\pm\sqrt{m^2-Q^2}\mp\bigg( \frac{Q^3}{\left(m\pm \sqrt{m^2-Q^2}\right) \left(m \left(\sqrt{m^2-Q^2}\pm m\right)+\mp Q^2\right)}\bigg)\tilde{\Theta}.
\end{equation}
and for the attractive case:
\begin{equation}
	r_\pm^{(+)}=m\pm\sqrt{m^2-Q^2}\pm\bigg( \frac{Q^3}{\left(m\pm \sqrt{m^2-Q^2}\right) \left(m \left(\sqrt{m^2-Q^2}\pm m\right)+\mp Q^2\right)}\bigg)\tilde{\Theta}.
\end{equation}
Notice that, in the commutative limit \(\tilde{\Theta}=0\), both the attractive and repulsive event horizons reduce to the commutative RN result \(r_\pm^{(+)}=r_\pm^{(-)}=m\pm\sqrt{m^2-Q^2}\). Moreover, for the uncharged black hole \(Q=0\), the above solution reduces to the commutative Schwarzschild event horizon \(r_\pm^{(+)}=r_\pm^{(-)}=2m\).

\begin{figure}[h]
	\begin{center}
		\includegraphics[width=0.45\textwidth]{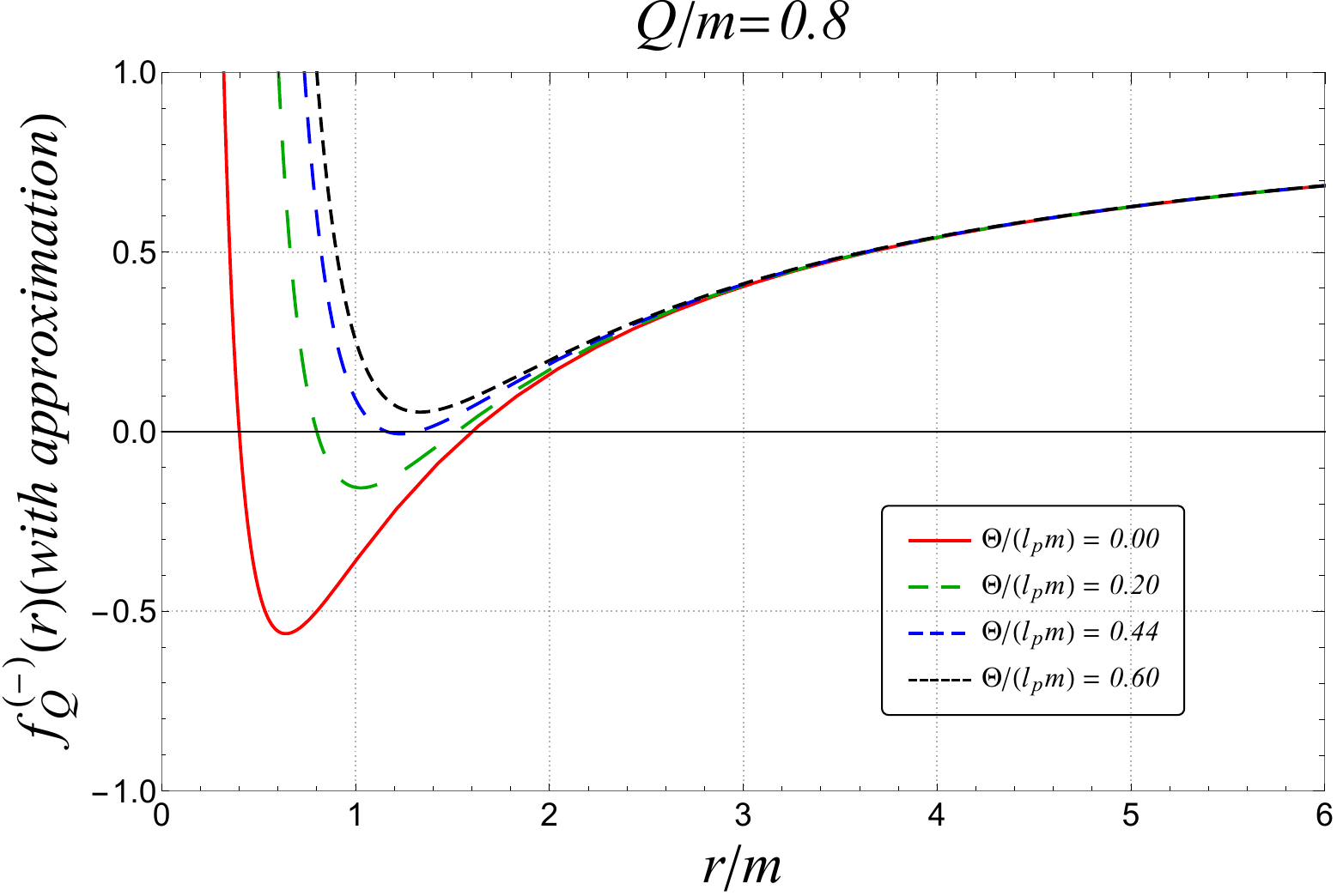}
		\includegraphics[width=0.45\textwidth]{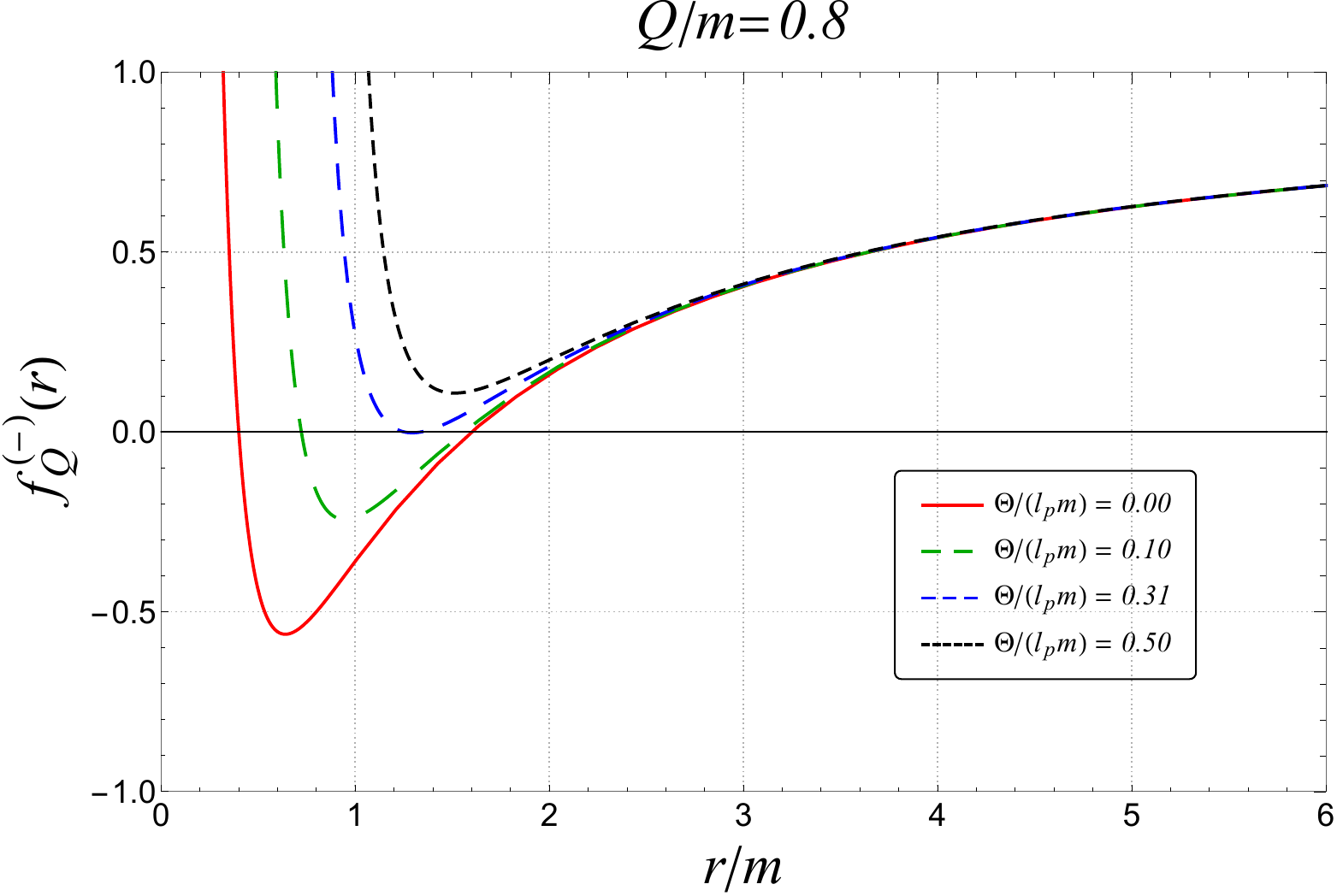}
		\includegraphics[width=0.45\textwidth]{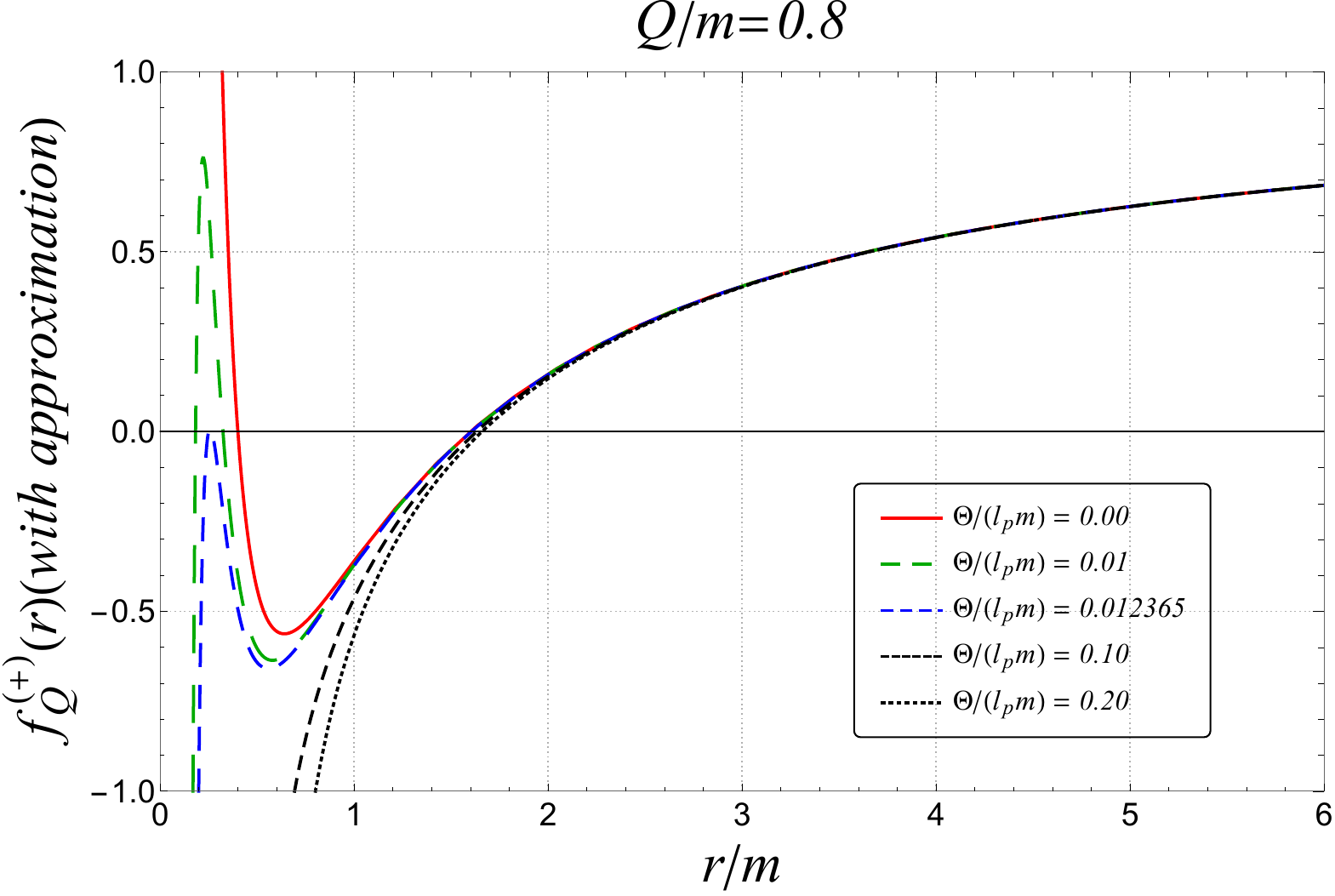}
		\includegraphics[width=0.45\textwidth]{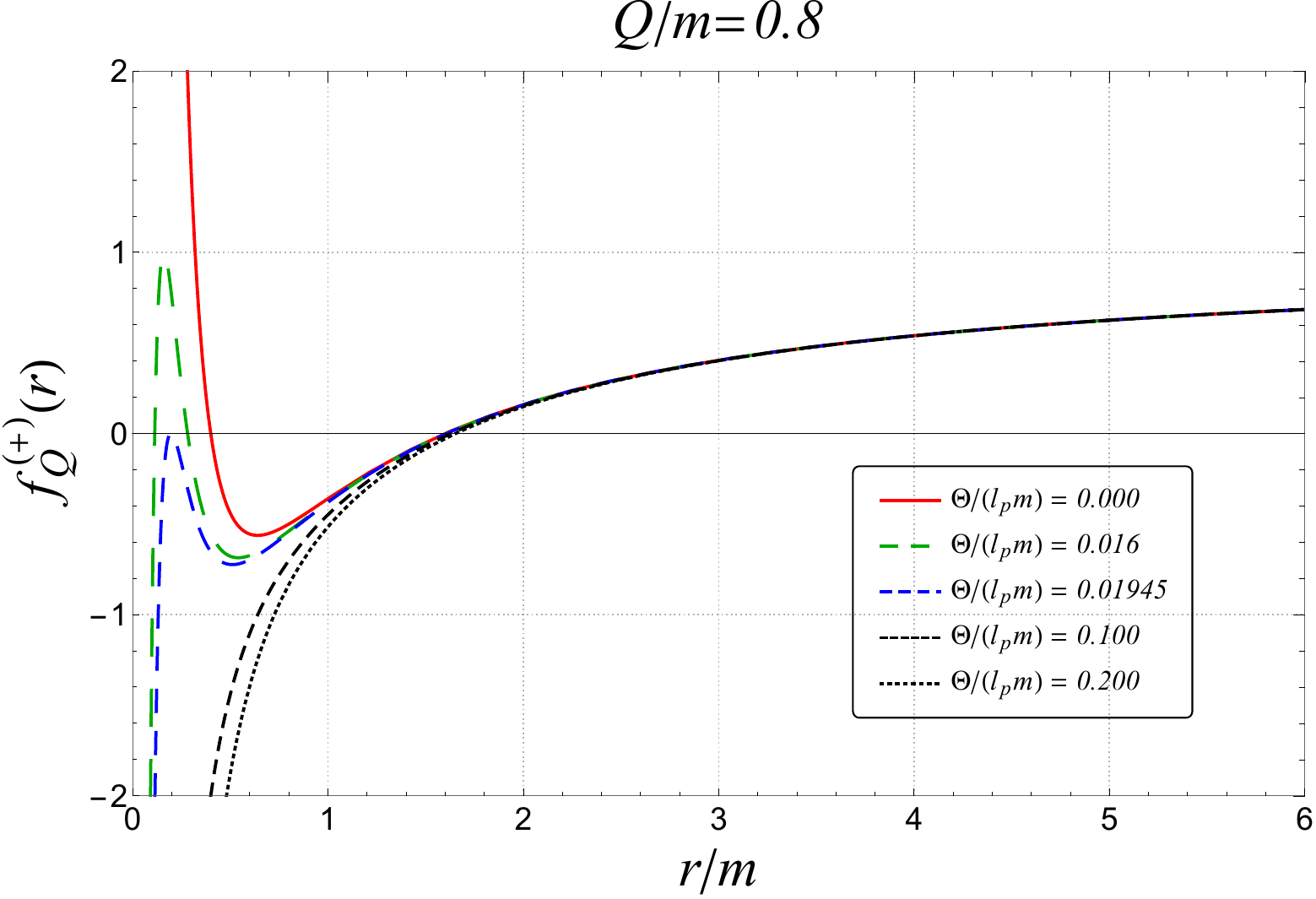}
		\caption{Behavior of the curvature function \(f_Q^\mp(r)\) without mass correction for the charged black hole (parameter \(Q/m\)) as a function of the radius \(r\) for different NC parameters \(\tilde{\Theta}\). Left panel: leading order in \(\tilde{\Theta}\). Right panel: compact form.}\label{fig:f2}
	\end{center} 
\end{figure}
Fig.~\ref{fig:f2} shows the variation of the curvature function \(f_Q^{\pm}(r)\) for the charged black hole as a function of \(r\) and for various values of the NC parameter \(\tilde{\Theta}\), without correction to the mass. From the left panel in the first row, we present the variation of the curvature function for the repulsive electric interaction \(f_Q^{(-)}(r)\), where we observe a qualitative similarity between the uncharged NC black hole (see the left panel of Fig.~\ref{fig:1}) and the charged case: non-commutativity plays a role analogous to that of the electric charge, an analogy noted in several works \cite{kim,abdellahPhD}. In the compact form (right panel), the profile is separated into two regions by a divergence at \(r=\sqrt{2\tilde{\Theta}Q}\). For the region \(r>\sqrt{2\tilde{\Theta}Q}\), the behavior mirrors the left panel: there are three regimes determined by the parameters, namely (i) a black hole solution with two horizons for \(\tilde{\Theta}<\tilde{\Theta}_c\), (ii) an extremal solution at \(\tilde{\Theta}=\tilde{\Theta}_c\) with a double horizon, and (iii) no black hole solution (naked singularity) for \(\tilde{\Theta}>\tilde{\Theta}_c\) when \(Q<m\). For \(Q\geqslant m\), one typically obtains no black hole solution\footnote{For the region \(r<\sqrt{2\tilde{\Theta}Q}\), the curvature function becomes negative for any value of the NC parameter \(\tilde{\Theta}\) and electric charge \(Q\), which is not allowed in the approximation.}. In the second row, we present the variation of the same curvature function for the attractive electric interaction \(f_Q^{(+)}(r)\). We observe new behavior: for some parameter values, there are three event horizons, one outer and two inner horizons, while beyond these values there is only one event horizon, which means there is always a black hole solution for this metric (\(\tilde{\Theta}\neq0\)). This is consistent with the results observed in the NC gauge-theory of gravity model \cite{abdellahPhD}. 

Now we investigate the case with the deformed mass (Eq.~\eqref{eq:NCRNM1}). One finds two solutions given by the following expression for the repulsive case:
\begin{equation}
	r_\pm^{(-)}=m\pm\sqrt{m^2-Q^2}+\bigg(\frac{\mp 3 m^3 \pm Q^3+2 m Q \sqrt{m^2-Q^2}+m^2 \left(\mp 2Q+3 \sqrt{m^2-Q^2}\right)}{Q^2 \sqrt{m^2-Q^2}}\bigg)\tilde{\Theta}.\label{eq:ehdmdq1}
\end{equation}
and for the attractive case we have:
\begin{equation}
	r^{(+)}_\pm=m\pm\sqrt{m^2-Q^2}+\bigg(\frac{\mp 3 m^3 \mp Q^3-2 m Q \sqrt{m^2-Q^2}+m^2 \left(\pm 2Q+3 \sqrt{m^2-Q^2}\right)}{Q^2 \sqrt{m^2-Q^2}}\bigg)\tilde{\Theta}.\label{eq:ehdmdq2}
\end{equation}
It is clear that the commutative solution is recovered when we set \(\tilde{\Theta}=0\). It is worth noting that, for sufficiently small electric charge \(\mathcal{O}(Q)\), the above relations reduce to
\begin{align}
	r^{(+)}_+=r^{(-)}_+&=2m-\frac{3}{2}\tilde{\Theta}+\mathcal{O}(Q,m\tilde{\Theta}),\notag\\
	r^{(+)}_- = r^{(-)}_-&=\frac{3}{2}\tilde{\Theta}+\mathcal{O}(Q,m\tilde{\Theta}),
\end{align}
which correspond to the leading-order event horizons of the NC Schwarzschild black hole given by Eq.~\eqref{eq:eh1}.

\begin{figure}[h]
	\begin{center}
		\includegraphics[width=0.45\textwidth]{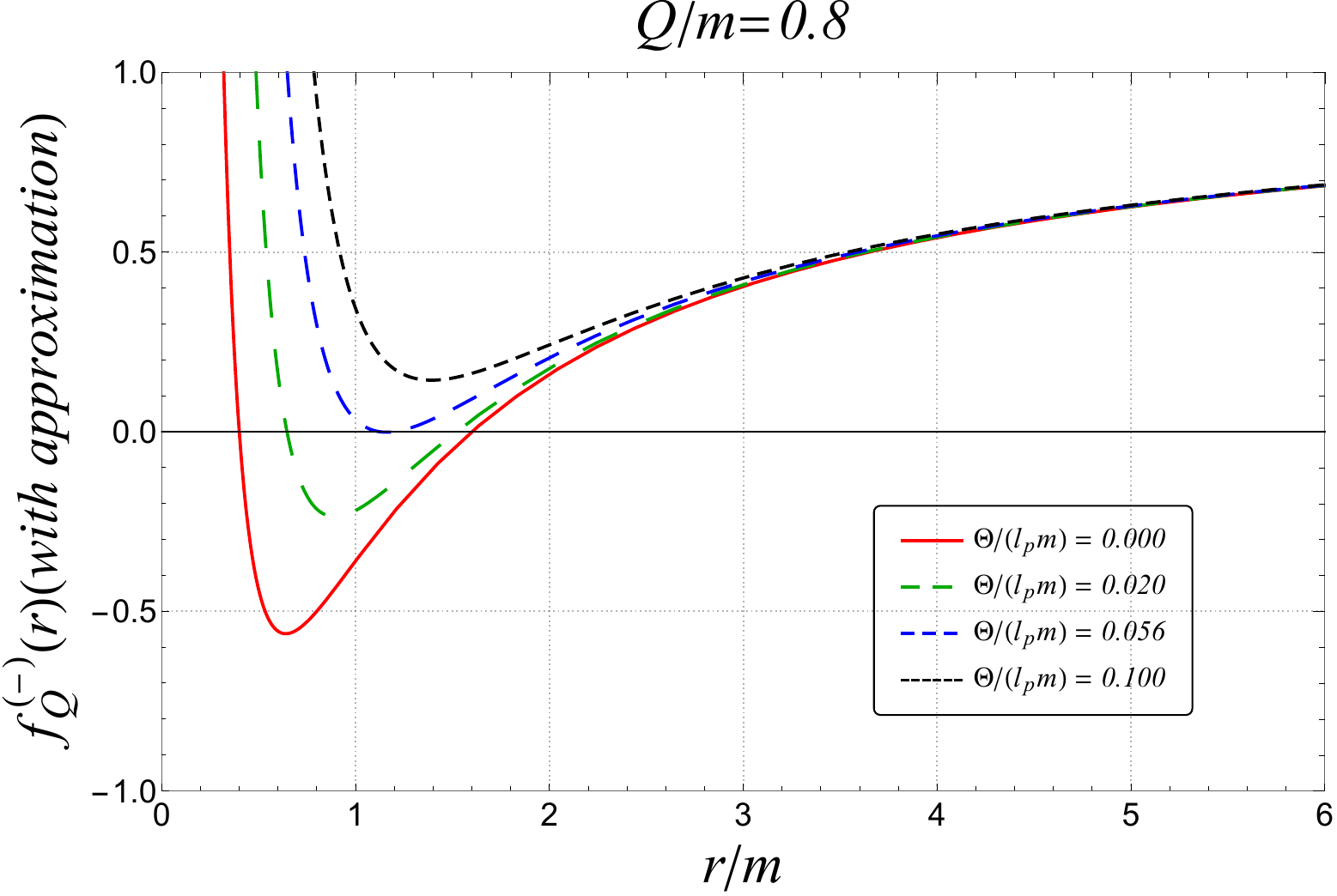}
		\includegraphics[width=0.45\textwidth]{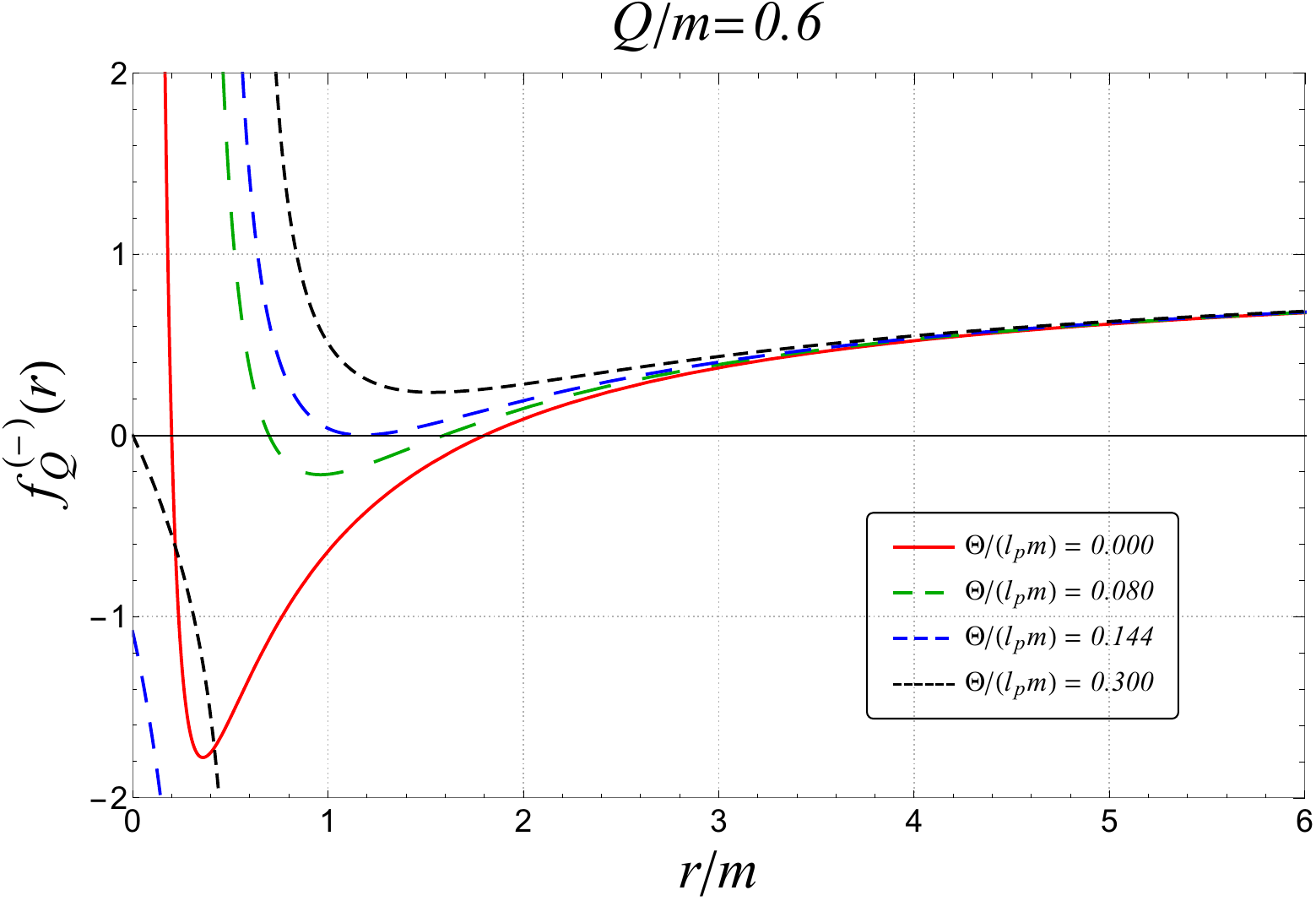}
		\includegraphics[width=0.45\textwidth]{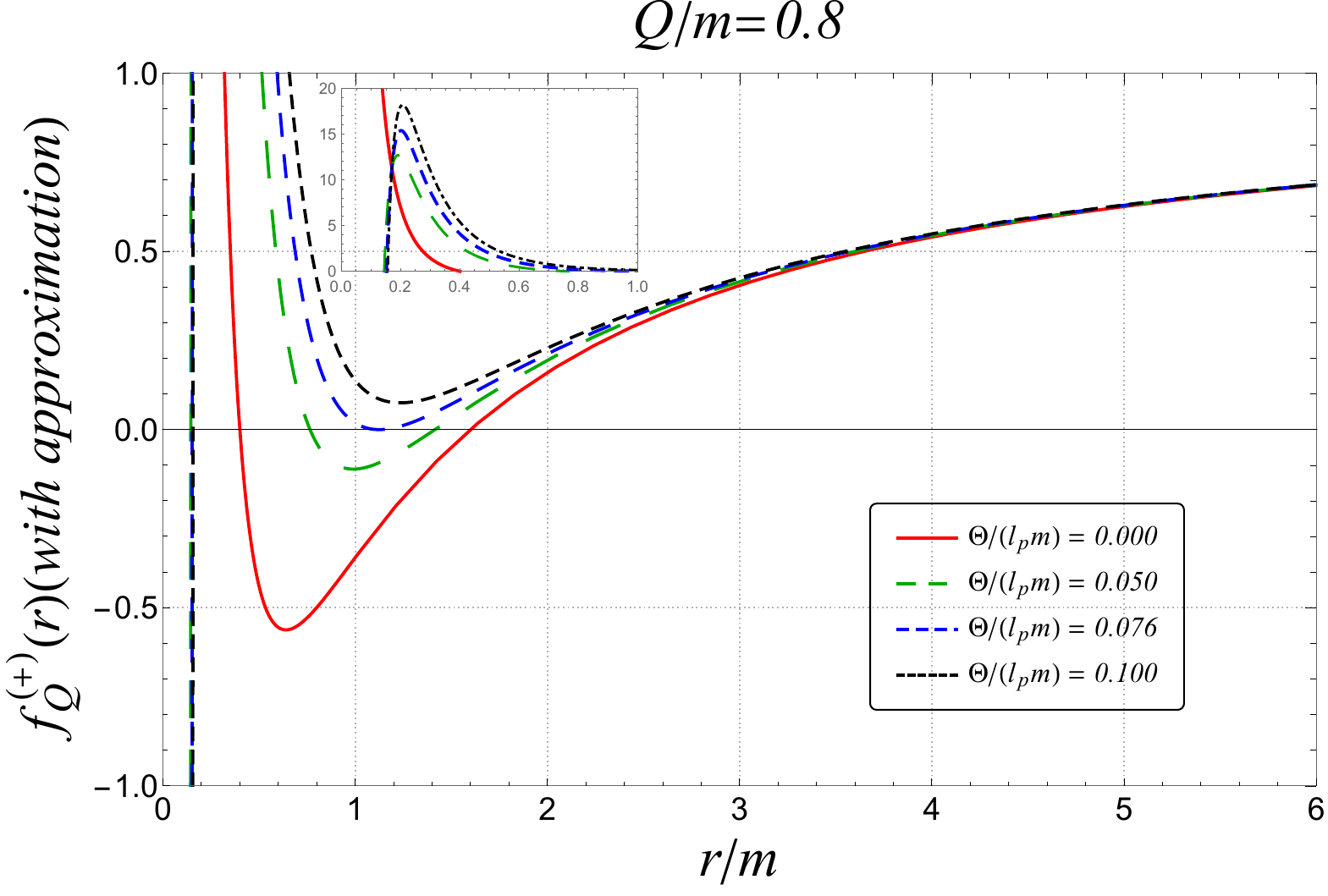}
		\includegraphics[width=0.45\textwidth]{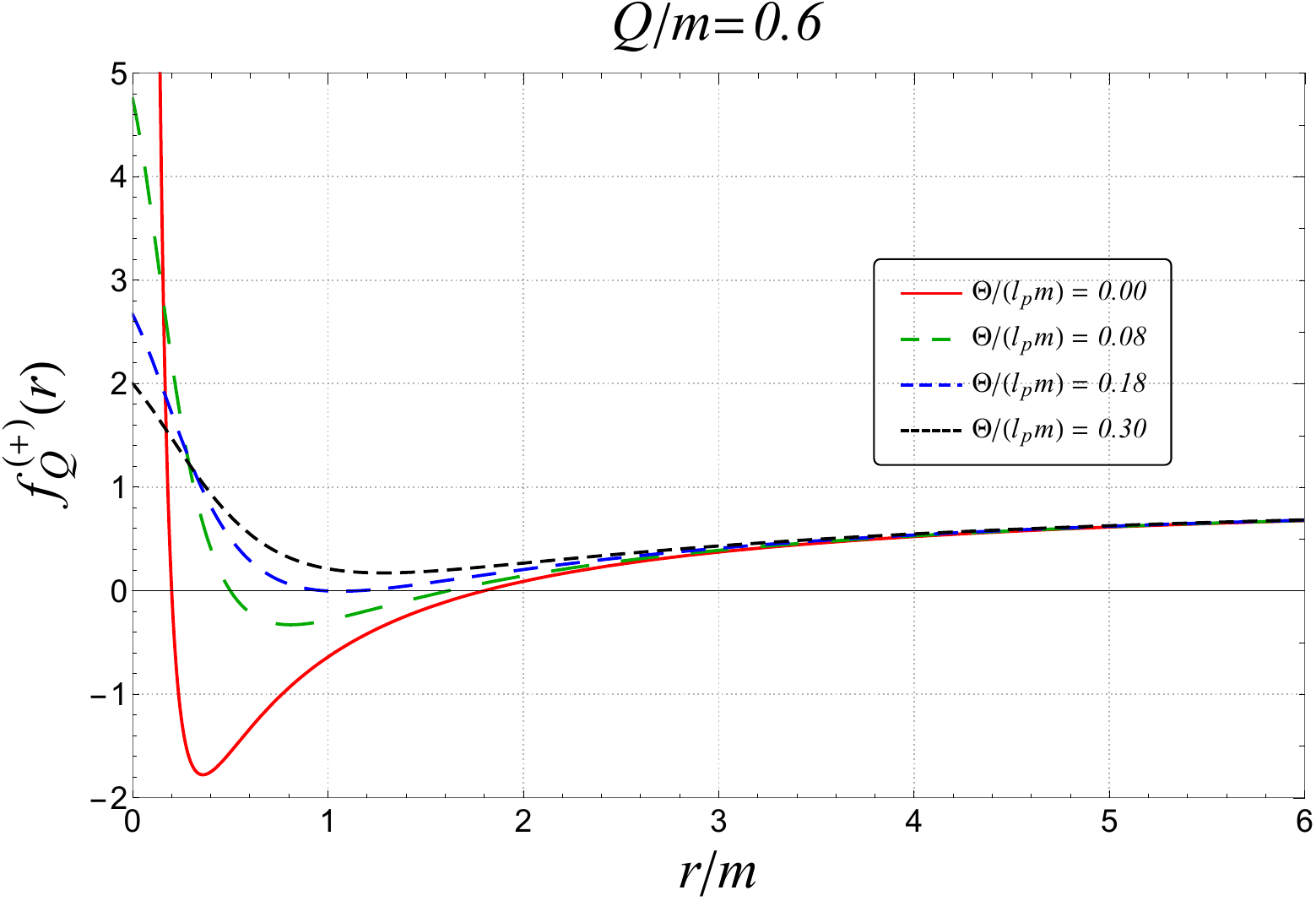}
		\caption{Behavior of the curvature function \(f_Q^\mp(r)\) for the charged black hole (parameter \(Q/m\)) with mass correction as a function of the radius \(r\) for different NC parameters \(\tilde{\Theta}\). Left panel: leading order in \(\tilde{\Theta}\). Right panel: compact form.}\label{fig:f3}
	\end{center} 
\end{figure}
In Fig.~\ref{fig:f3}, we illustrate the variation of the curvature function \(f_Q^\mp(r)\) for the RN-like black hole as a function of \(r\) and for different NC parameters with mass correction. In the first row, we represent the repulsive case, and this type of interaction shows behavior similar to the previous case, in which the left and right panels present qualitatively the same behavior, with a slight shift in the event horizons. The difference can be seen in the compact form in the right panel of the same row. This difference concerns the region beyond the divergence at the point \(r<\sqrt{2\tilde{\Theta}Q}\), where the function remains finite and the mass correction dominates over the NC correction to the electric contribution. For the attractive case, shown in the second row, both the leading-order approximation (left panel) and the compact form (right panel) of the curvature function exhibit new behavior in the NC RN-like black hole. The left panel shows three event horizons for some parameter values, and beyond those values the function presents one event horizon, regardless of the chosen parameters. Therefore, we can conclude that there is always a black hole solution, as in the repulsive case. Here, in the leading-order approximation, the NC term of the attractive electric interaction changes the divergence of the function from positive to negative at \(r=0\), which leads to a new maximum for the curvature function. For the attractive case in the compact form, we obtain a regular-like behavior similar to the profile of the NC correction to the Schwarzschild black hole (see Fig.~\ref{fig:1}); thus, there are three possible solutions depending on the parameter values of \(Q\) and \(\tilde{\Theta}\).

\vspace{0.5cm}
\paragraph{Scalar invariant:} The Ricci scalar for the new RN-like black hole is obtained using the same expression defined above in \eqref{eq:RS1}, together with the curvature function \(f_Q^\pm(r)\) given by Eq.~\eqref{eq:NCRNM1} for both the attractive and repulsive electric interactions. Then we get:
\begin{equation}
	R^\pm_Q=\frac{12 m^2 \tilde{\Theta} \left(9 \tilde{\Theta} m-r^2\right)}{r\left(3 \tilde{\Theta} m+r^2\right)^3}+\frac{4\tilde{\Theta}Q^3 \left(3 r^2\mp2 \tilde{\Theta} Q\right)}{r^2\left(2 \tilde{\Theta} Q\pm r^2\right)^3}
\end{equation}
where \(R^+_Q\) and \(R^-_Q\) are the Ricci scalars for the attractive and repulsive electric interactions, respectively. The first term corresponds to the NC correction to the matter contribution \eqref{eq:NCRSM2}, and in the case of the NC correction to the electric interaction, this expression reduces to the one obtained using the curvature function given by \eqref{eq:NCRN2}. This expression still contains a singularity at the origin \(r=0\), and, in addition, in the case of repulsive interaction, the Ricci scalar has an additional singularity at \(r=\sqrt{2\tilde{\Theta}Q}\).

For this case, the dominant term at the origin is given by:
\begin{equation}
	\lim_{r\to0}R^{(\pm)}_Q=\mp\frac{Q}{\tilde{\Theta} r^2}\bigg|_{r=0},\label{eq:NCRNRSM3}
\end{equation}
which clearly has divergent behavior at \(r=0\). It is worth noting that this limit depends on the nature of the electric interaction, which is a consequence of the present model.

\begin{figure}[h]
	\begin{center}
		\includegraphics[width=0.45\textwidth]{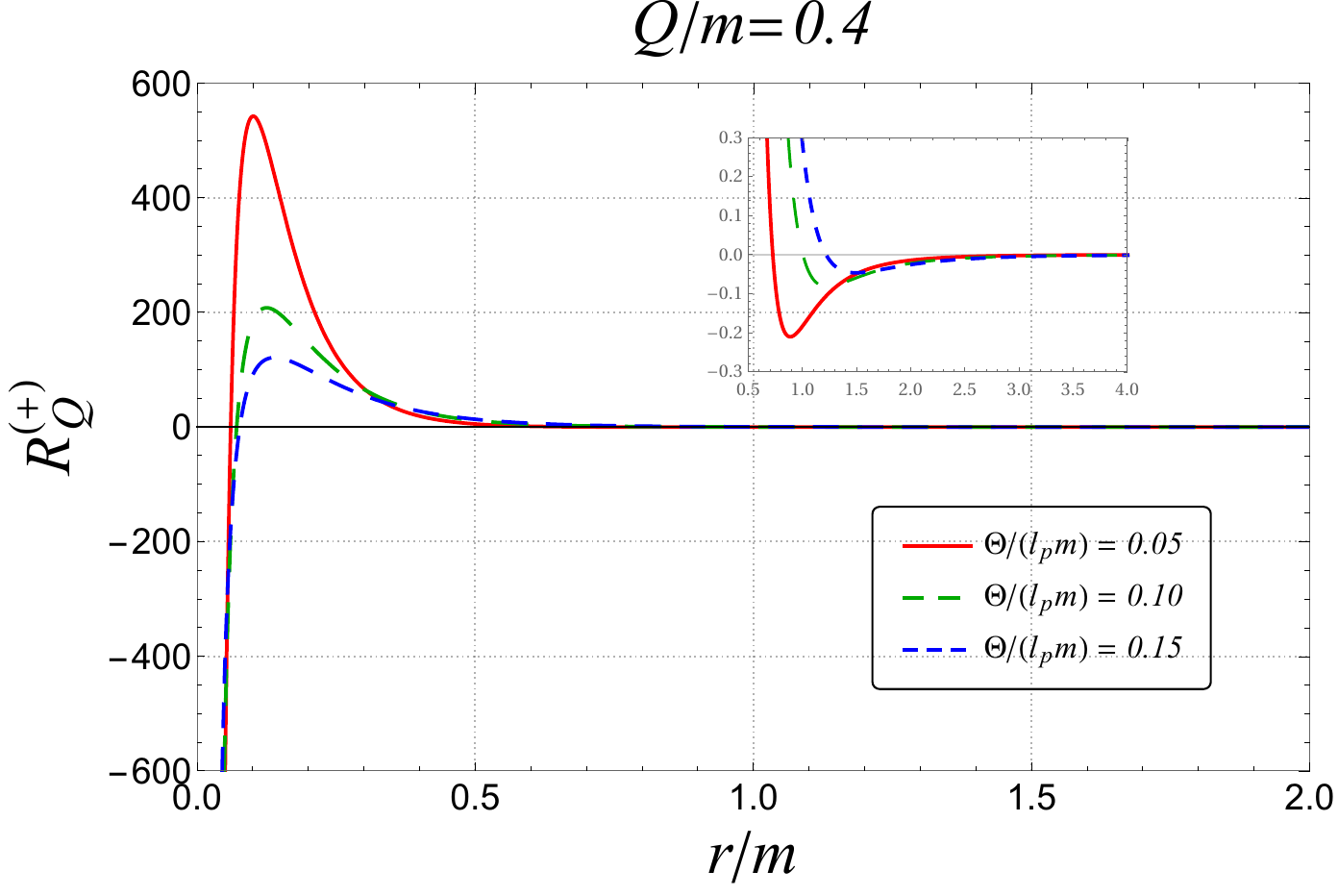}
		\includegraphics[width=0.45\textwidth]{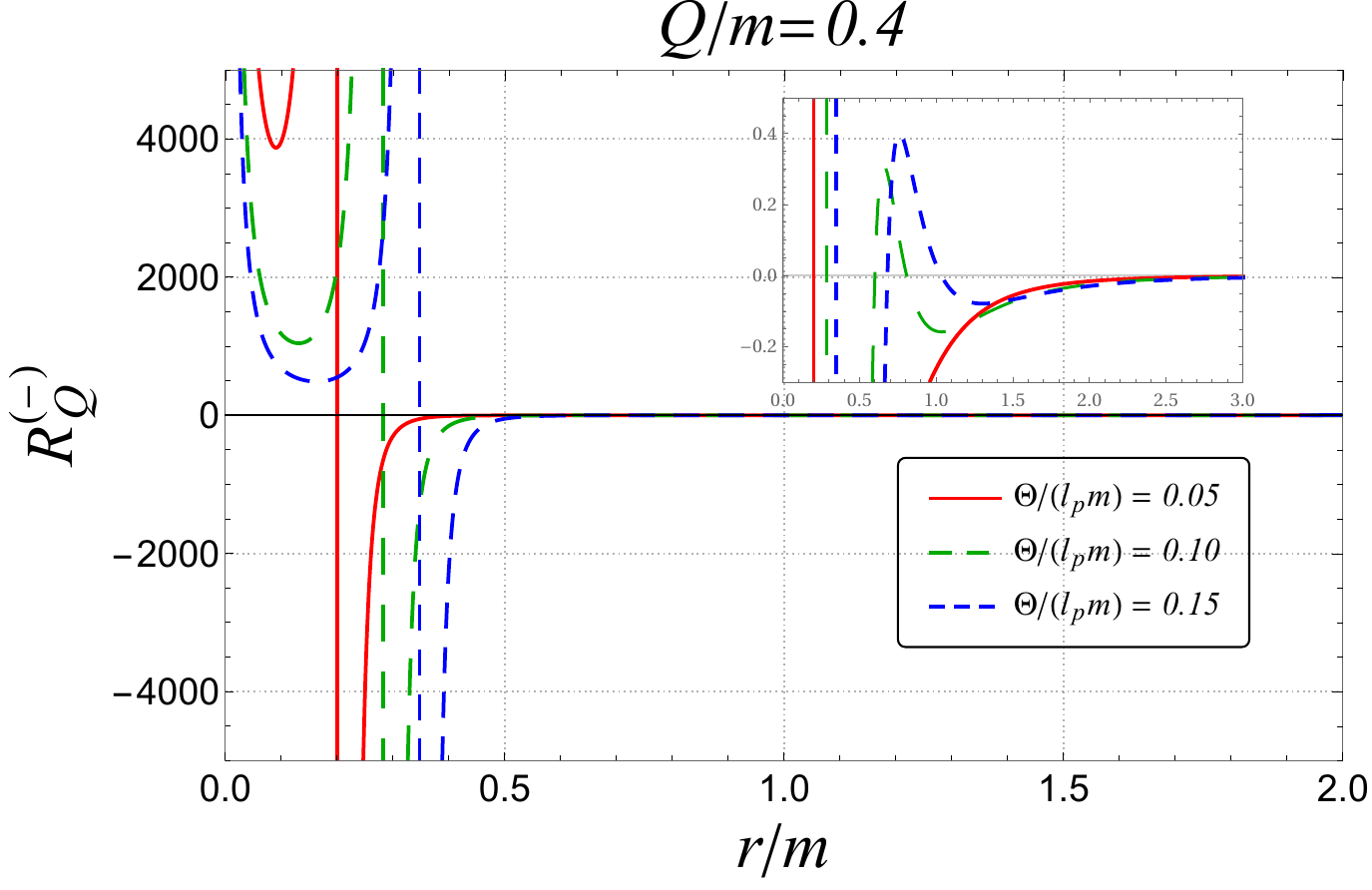}
		\caption{Behavior of the Ricci scalar function \(R^{(\pm)}_Q\) for the RN-like black hole as a function of the radius \(r\) for different NC parameters \(\tilde{\Theta}\). Left panel: attractive electric interaction. Right panel: repulsive electric interaction.}\label{fig:R2}
	\end{center} 
\end{figure}
In Fig.~\ref{fig:R2}, we present the variation of the Ricci scalar for both the attractive \(R^{(+)}_Q\) and repulsive \(R^{(-)}_Q\) NC RN-like black holes as a function of \(r\), and for different NC parameters, shown in the left and right panels, respectively. For the attractive electric interaction (left panel), the Ricci scalar shows two different extrema: the first corresponds to the minimum arising from the NC correction to the gravitational interaction, as discussed above in Fig.~\ref{fig:R-1}, and the second corresponds to the maximum associated with the attractive electric interaction, where the electric charge converts the positive divergence into a negative one near the singularity and leads to a new maximum, which decreases as the NC parameter increases. In the case of repulsive electric interaction, we observe a new divergence at \(r=\sqrt{2\tilde{\Theta}Q}\), where outside this radius we have the same behavior as in the attractive case, while inside this region we have a positive divergence of the Ricci scalar. In this type of interaction, the sign of the scalar curvature changes at the physical singularity \(r=0\), as shown by Eq.~\eqref{eq:NCRNRSM3}.

For the Kretschmann scalar, we use the same definition as in expression \eqref{eq:KS1}, together with the curvature function given by 
\begin{align}
	K^{(\mp)}_Q&=\frac{4 \left(\tilde{\Theta} m Q (3 Q\pm4  r)+r^2 \left(Q^2-2 m r\right)\right)^2}{r^4 \left(3 \tilde{\Theta} m+r^2\right)^2 \left(r^2\mp2\tilde{\Theta}Q\right)^2}+4 \left(\frac{2 m \left(r^2-3 \tilde{\Theta} m\right)}{r^2 \left(3 \tilde{\Theta} m+r^2\right)^2}-\frac{2 Q^2}{r \left(r^2\mp2\tilde{\Theta}Q\right)^2}\right)^2\notag\\
	&+\left(\frac{4 m \left(9 \tilde{\Theta} m r-r^3\right)}{\left(3 \tilde{\Theta} m+r^2\right)^3}-\frac{2 Q^2 \left(\mp2 \tilde{\Theta} Q-3 r^2\right)}{\left(r^2\mp2\tilde{\Theta}Q\right)^3}\right)^2,
\end{align}

In order to investigate the physical singularity at the origin, we consider the dominant term of the Kretschmann scalar for small radii. Then we find:
\begin{equation}
	\lim_{r\to0}K^{(\mp)}_Q=\frac{Q^2}{\tilde{\Theta}^2 r^4}\bigg|_{r=0},\label{eq:NCKSRNM1}
\end{equation}
It is clear that this geometry weakens the singularity at the origin of the RN-like solution compared with the commutative RN metric, which has a stronger singularity at the origin \(K\propto1/r^8\), whereas in this model it is reduced to \(K\propto1/r^4\) only. As observed above, even the leading-order NC correction in this model succeeds in weakening the central singularity of the RN-like black hole, and this motivates us to further investigate higher-order NC corrections, which may eventually enable us to remove the singularity from the Schwarzschild and RN black holes.

\subsection{Energy conditions}

In this subsection, we investigate the classical energy conditions of the present model. According to standard general relativity, there are four energy conditions that the energy-momentum tensor must satisfy. Before analyzing the effect of non-commutativity on the energy conditions, we first need to obtain the corresponding compact form of the energy-momentum tensor for the gravitational and electric interactions. For this purpose, we use our compact metric function $f^{(\pm)}_Q$ given by Eq. \eqref{eq:NCRNM1}, together with the Einstein equations \eqref{eq:Eeq1}. The compact form of the energy-momentum tensor components \eqref{eq:EMT1} for the gravitational sector is written as follows:
\begin{equation}
	\rho_m(r)=\frac{3c^2m^2\tilde{\Theta}}{2\pi G\,r\left(r^2+3m\tilde{\Theta}\right)^2},\quad p_r=-\rho_m,\quad p_t=-\rho_m-\frac{r}{2}\partial_r\rho_m,\label{eq:EDm}
\end{equation}
where, at first order in the NC parameter, this expression reduces to the one given in Eq. \eqref{eq:NCrhom1}, and in the limit $\tilde{\Theta}=0$ the vacuum energy-momentum tensor is recovered. For the electric sector, the usual Maxwell energy-momentum tensor is not valid in the presence of non-commutativity. As we have seen above, the Ricci scalar is non-zero for both the gravitational and electric sectors, which implies that the trace of the energy-momentum tensor in NC geometry should also be non-zero. This is satisfied by the matter term $T_m$, whereas the Maxwell energy-momentum tensor has conformal symmetry, which means that its trace vanishes. This is in contradiction with the Ricci scalar, and at this point the Einstein equations are not satisfied. The same problem was first observed for NC smeared energy with Gaussian distributions of electric charge and matter in Ref. \cite{alkac2025}, where it was noted that the angular Einstein equations are not satisfied when the Maxwell energy-momentum tensor is used, which is also the case in our approach. To fix this problem, we impose the covariant conservation of the energy-momentum tensor and the contracted Bianchi identities in the Einstein equations and obtain
\begin{equation}
	\nabla_\mu G^\mu_\nu=0,\qquad \text{and}\qquad \nabla_\mu T^\mu_\nu=0,
\end{equation}
The solution of the second equation is an anisotropic fluid \cite{alkac2025}, which allows us to write the energy-momentum tensor for the NC electric interaction in a form similar to the gravitational sector \eqref{eq:EMT1}:
\begin{equation}\label{eq:EMTE2}
	[\hat{T}^\mu{}_\nu]_Q=\operatorname{diag}(-\hat{\rho}|_Q,\hat{p}_r|_Q,\hat{p}_t|_Q,\hat{p}_t|_Q).
\end{equation}
For our model, we have:
\begin{equation}
	\rho_Q^{(\mp)}(r)=\frac{c^2}{8\pi G}\,
	\frac{Q^2\left(r^2\pm 2\tilde{\Theta}Q\right)}
	{r^2\left(r^2\mp 2\tilde{\Theta}Q\right)^2},\quad p_r^{(\mp)}|_Q=-\rho_Q^{(\mp)},\quad p_t^{(\mp)}|_Q=-\rho_Q^{(\mp)}-\frac{r}{2}\partial_r\rho_Q^{(\mp)},\label{eq:EDe}
\end{equation}
It is clear that, in the first-order approximation, the above expression reduces to the one given by Eq. \eqref{eq:TNC_corrected}. Also, the commutative expression is recovered when we set $\tilde{\Theta}=0$.

Next, we investigate the effect of this geometry on the standard energy conditions \cite{Book-enerycondition,enerycondition1,enerycondition2,enerycondition3}.

\subsubsection{Weak energy condition}

The weak energy condition (WEC) states that any observer in spacetime measures a non-negative energy density for any matter distribution, which excludes the presence of exotic matter. This condition implies:
\begin{equation}
	\rho \geq 0, \quad \text{and}\qquad\rho+p_i\geq 0,\quad i=1,2,3.
\end{equation}
In our system, the energy density is obtained from both gravitational and electric interactions, \(\rho=\rho_m+\rho_Q^{(\mp)}\). We now study the first condition to determine whether the NC geometry respects the non-negativity of the energy density.

\begin{figure}[h]
	\begin{center}
		\includegraphics[width=0.3\textwidth]{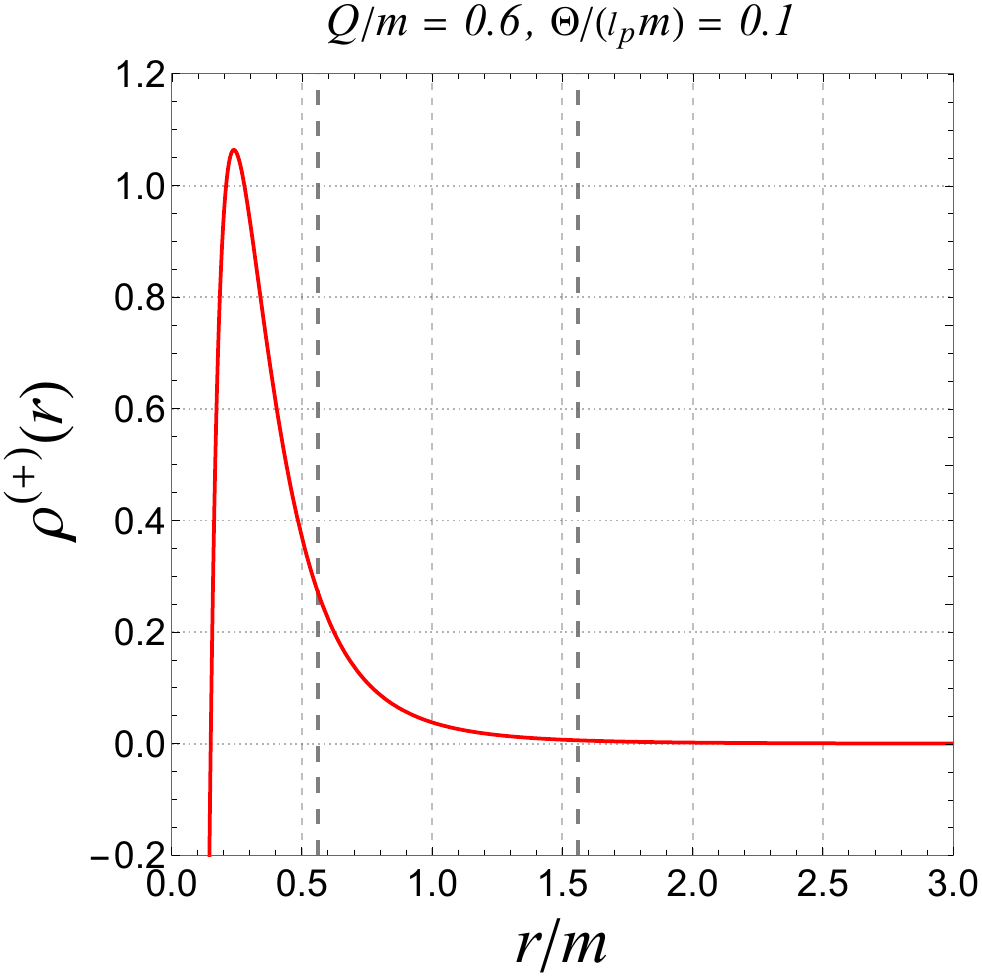}
		\includegraphics[width=0.295\textwidth]{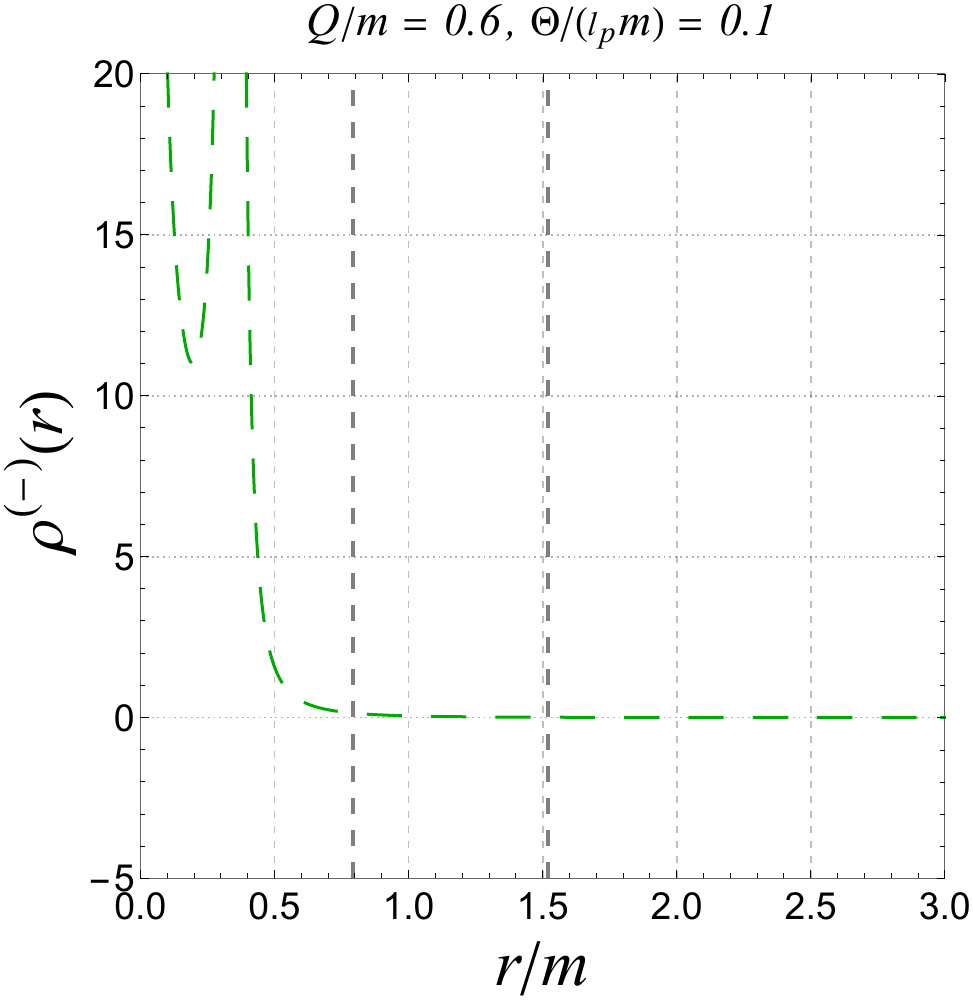}
		\includegraphics[width=0.29\textwidth]{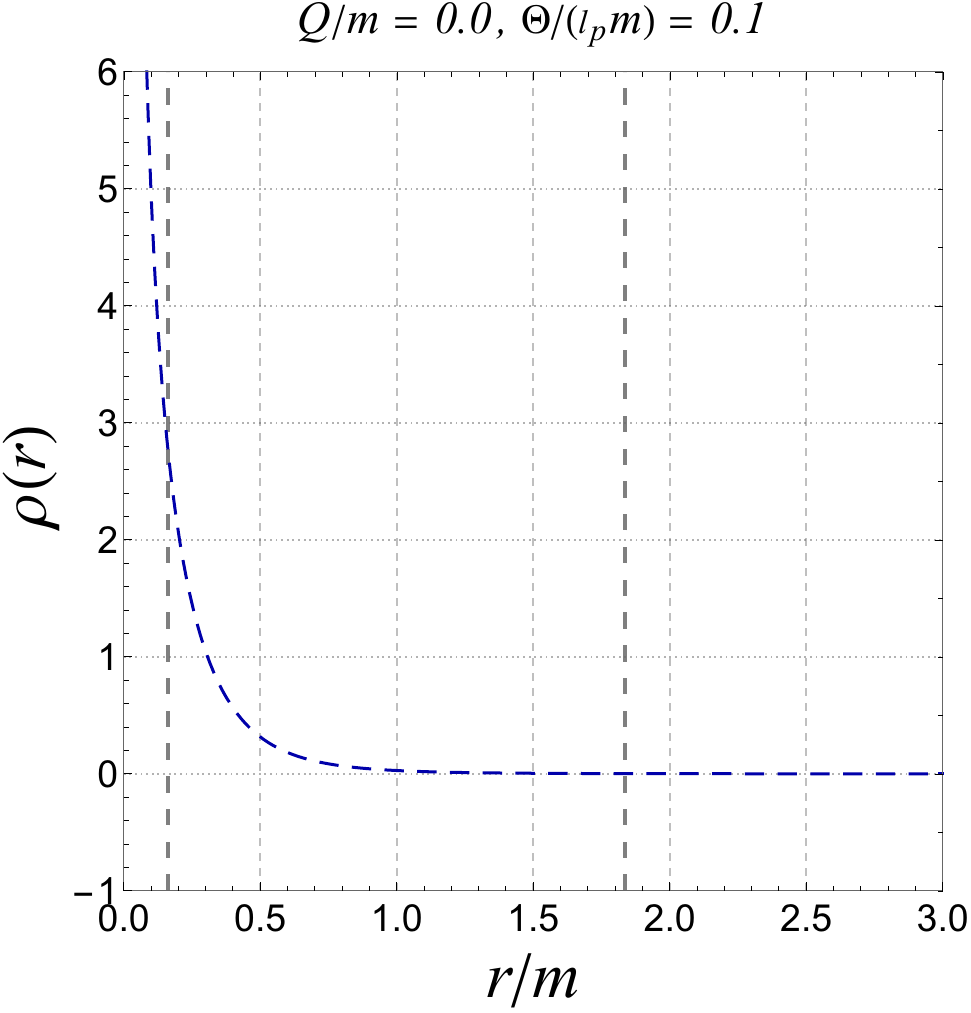}
		\caption{Variation of the energy density function \(\rho^{(\pm)}\) for the energy-momentum tensor in the present model as a function of the radius \(r\) for different types of interaction. Left panel: attractive electric interaction. Middle panel: repulsive electric interaction. Right panel: uncharged black hole $Q=0$. The dashed line represents the inner and outer event horizons for the selected parameters, from left to right respectively in each panel.}\label{fig:WEC1}
	\end{center} 
\end{figure}
In Fig.~\ref{fig:WEC1}, we illustrate the behavior of the energy density of the RN-like black hole for different types of interaction. It is clear that, in all panels, outside both event horizons (inner and outer), the left inequality is satisfied, where the energy density is positive, except for the attractive electric interaction in the left panel, which shows a negative energy density inside the inner horizon in the presence of non-commutativity. In this case, the combination of two attractive forces and the presence of NC geometry creates an effective energy density that behaves differently from the ordinary one, producing a repulsive-like behavior near the singularity; however, this region is not directly accessible to any observer. By contrast, in the middle panel, the repulsive electric interaction creates a new divergent behavior inside the inner horizon at \(r=\sqrt{2\tilde{\Theta}Q}\), while the energy density remains positive. This is due to the combination of repulsive and attractive interactions together with non-commutativity. In the absence of electric interaction, the NC deformation of the gravitational interaction keeps the energy density positive in all regions for all values of the NC parameter, which makes the first inequality valid everywhere in this spacetime. 

The second inequality must be verified for all pressure components \(p_i\). It is clear that, for the radial pressure, this inequality is satisfied since \(\rho+p_r=0\), leaving only the two identical inequalities for the tangential pressure, \(\rho+p_t=-\frac{r}{2}\partial_r\rho\).

\begin{figure}[h]
	\begin{center}
		\includegraphics[width=0.3\textwidth]{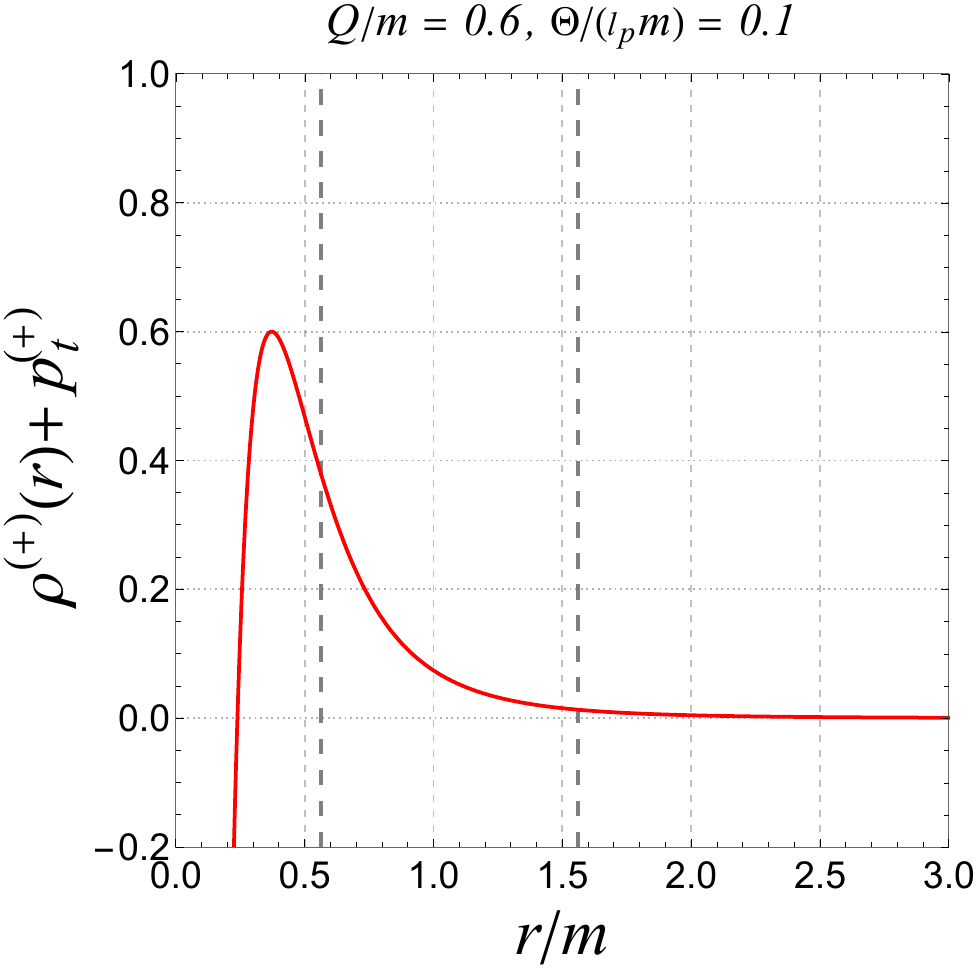}
		\includegraphics[width=0.295\textwidth]{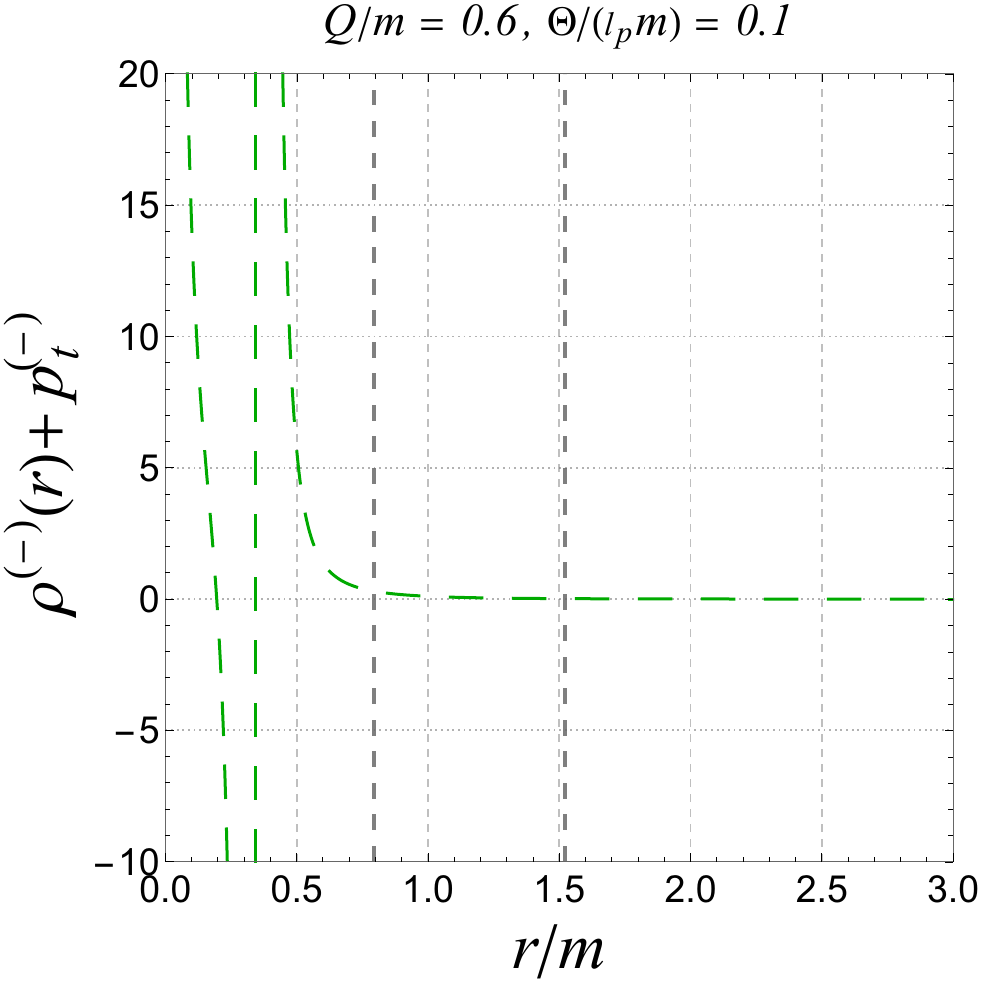}
		\includegraphics[width=0.285\textwidth]{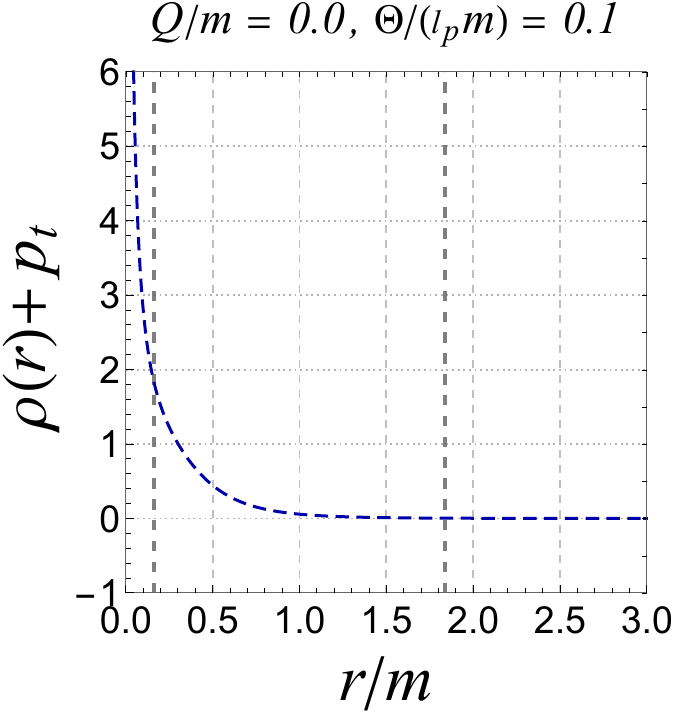}
		\caption{Variation of the second inequality of the WEC, \(\rho+p_t\), for the energy-momentum tensor in the present model as a function of the radius \(r\) for different types of interaction. Left panel: attractive electric interaction. Middle panel: repulsive electric interaction. Right panel: uncharged black hole $Q=0$. The dashed line represents the inner and outer event horizons for the selected parameters, from left to right respectively in each panel.}\label{fig:WEC2}
	\end{center} 
\end{figure}
The behavior of \(\rho+p_t\) for the energy-momentum tensor in the present model as a function of the radius \(r\) for different types of interaction is shown in Fig.~\ref{fig:WEC2}. Note that the condition \(\rho+p_t\geq 0\) is satisfied outside both event horizons for the first two cases, namely attractive and repulsive electric interactions shown in the left and middle panels, while in the right panel, corresponding to the pure gravitational case (\(Q=0\)), this condition is satisfied everywhere. From the left panel, we observe a violation of the second WEC, where the attractive electric interaction together with NC geometry causes a violation of the first and second WEC. The same violation in the repulsive case is observed inside the inner horizon (see the middle panel), in the region \(0<r<\sqrt{2\tilde{\Theta}Q}\), and this is caused by the contribution of the pressure, since in this region the pressure dominates, whereas near the singularity the situation is reversed and the energy density dominates again. In the absence of electric charge (\(Q=0\)), the WEC is satisfied everywhere in this model.

\subsubsection{Null energy condition}

The null energy condition (NEC) represents the minimal requirement for the focusing of null geodesics. The NEC implies:
\begin{equation}
	\rho+p_i\geq 0,\quad i=1,2,3.
\end{equation}
This condition follows directly from the WEC analyzed above. Therefore, the NEC is also satisfied in our model outside the inner horizon, while in the uncharged case it is valid everywhere in spacetime.

\subsubsection{Strong energy condition}

The strong energy condition (SEC) states that, in standard general relativity, ordinary matter produces attractive gravity that converges nearby time-like geodesics. 
\begin{equation}
	\rho+p_r+2p_t \geq 0, \quad \text{and}\quad\rho+p_i\geq 0,\quad i=1,2,3.
\end{equation}
Note that the right inequality was verified in the WEC subsection and is valid in the present model outside the horizon region, while its violation is hidden inside the inner horizon in the case of charged black holes, while for the uncharged case, it is satisfied everywhere in this spacetime. Furthermore, the left inequality tells us about the nature of the interaction: it is positive for an attractive interaction and negative for a repulsive one. 

\begin{figure}[h]
	\begin{center}
		\includegraphics[width=0.3\textwidth]{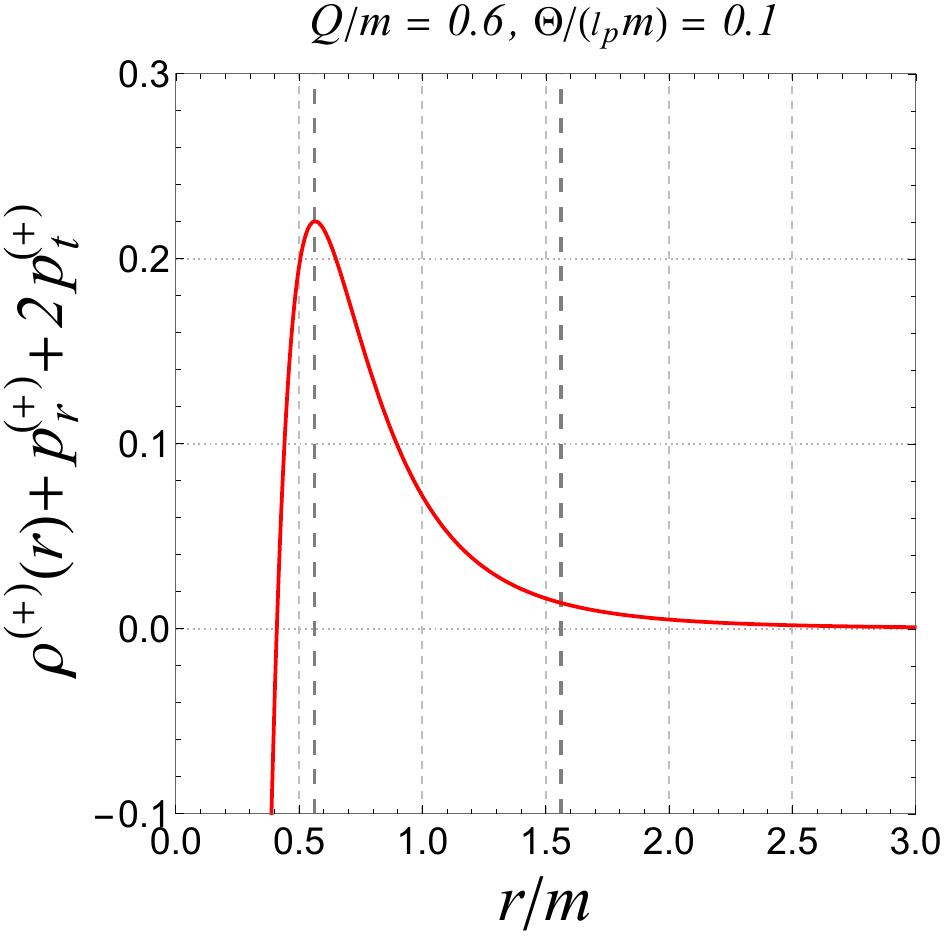}
		\includegraphics[width=0.3\textwidth]{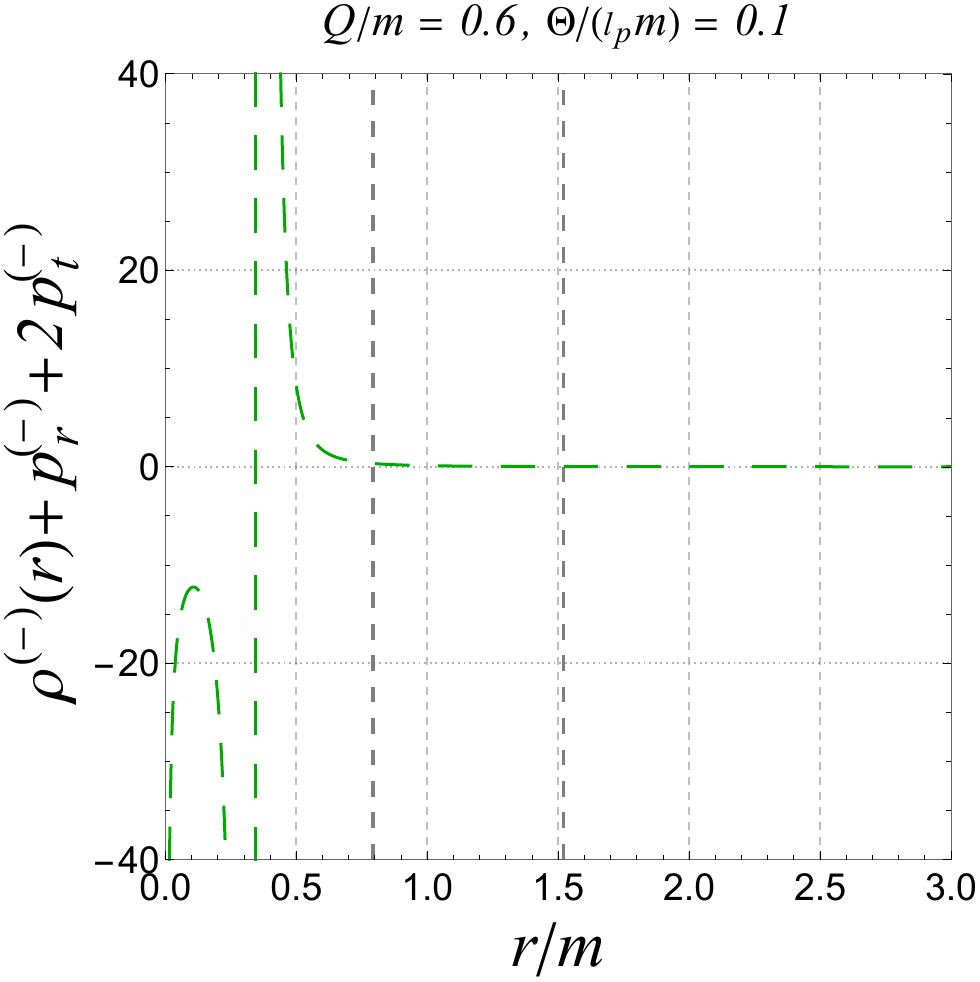}
		\includegraphics[width=0.3\textwidth]{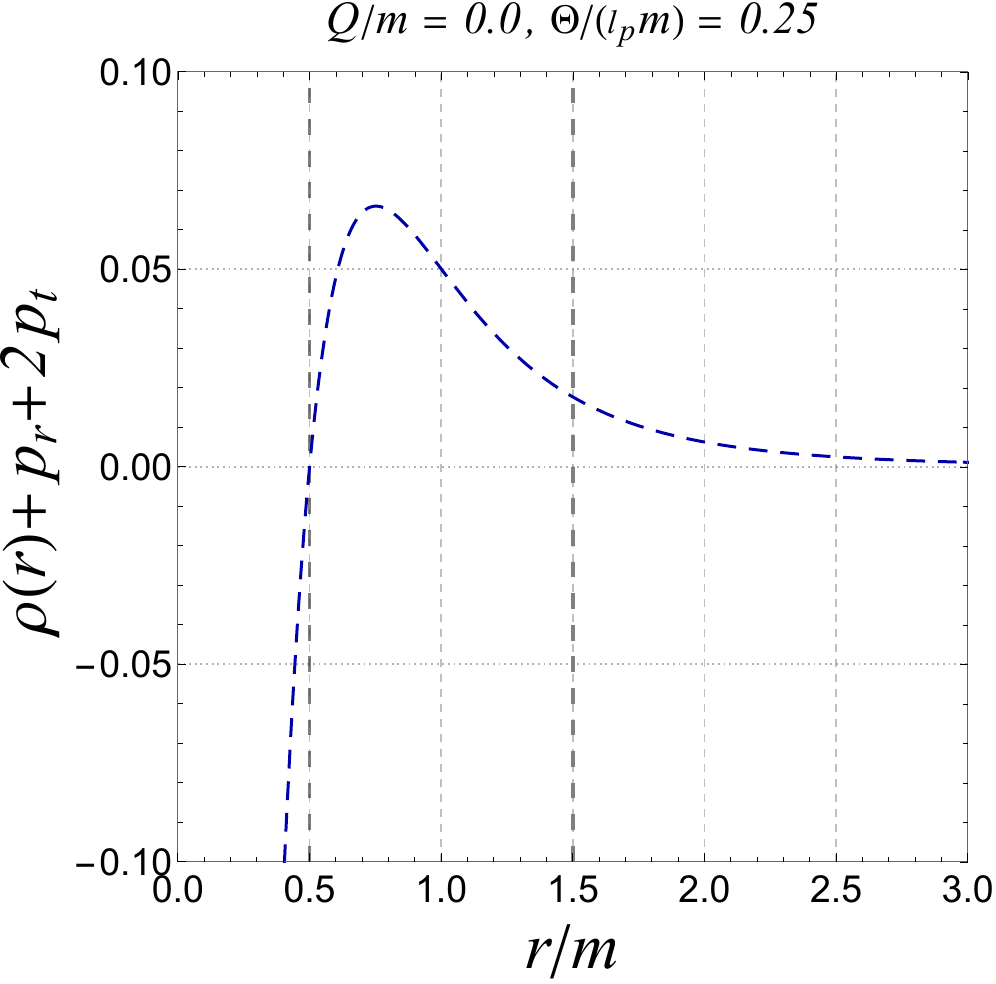}
		\caption{Variation of the SEC for the energy-momentum tensor in the present model as a function of the radius \(r\) for different types of interaction. Left panel: attractive electric interaction. Right panel: repulsive electric interaction. The dashed line represents the inner and outer event horizons for the selected parameters, from left to right respectively in each panel.}\label{fig:SEC}
	\end{center} 
\end{figure}
Fig.~\ref{fig:SEC} shows the behavior of the SEC for the energy-momentum tensor in the present model for different types of interaction. It is clear that, for the attractive electric interaction, the SEC is violated inside the inner horizon for the selected parameters, while for other parameter values this violation can extend to the region between the inner and outer horizons. As is known, this condition tells us about the nature of the interaction created by the matter distribution. In the commutative case \(\tilde{\Theta}=0\), this condition is satisfied, meaning that we have gravitational interaction; whereas in our NC model, this condition is violated and becomes negative, so the resulting interaction becomes repulsive. This means that non-commutativity creates a repulsive behavior that helps avoid a singularity at the origin \(r=0\). This behavior is similar to that of some regular black holes, but in this case the curvature is not finite at the origin, and the repulsive behavior increases as one approaches the singularity, leading to divergent behavior that prevents anything from reaching the singularity. In the middle panel, for the repulsive case, we also see a violation of the SEC inside the inner horizon for the selected parameters (for some other values of \(Q/m\) and \(\tilde{\Theta}/m\), this condition can be violated outside the inner horizon, similar to Ref. \cite{alkac2025}), where the SEC becomes negative when \(r<\sqrt{2\tilde{\Theta}Q}\). This corresponds to a repulsive behavior generated by the effective energy-momentum tensor in the presence of non-commutativity. Note that the selected parameters \(Q/m\) and \(\tilde{\Theta}/m\) satisfy the conditions for the existence of a black hole solution, as discussed in the geometrical properties subsections (see \ref{Sub:GP1} and \ref{Sub:GP2}). The SEC for uncharged matter is illustrated in the right panel, which shows the same behavior as the attractive electric interaction in the left panel; this similarity results from the same nature of the interaction considered. The violation of the SEC coincides with the inner event horizon, where the repulsive behavior starts from the inner horizon, meaning that this geometry begins to prevent anything from reaching the singularity from the inner horizon for \(\tilde{\Theta}/m=0.25\). Moreover, the violation of the SEC can be hidden inside the inner horizon only for the range \(m/4\leq\tilde{\Theta}\leq m/3\), and its location shifts outside the inner horizon for \(\tilde{\Theta}\leq m/4\) (this condition on the NC parameter was obtained from the event horizon equation \eqref{eq:eh1} and Eq. \eqref{eq:EDm}), while for \(\tilde{\Theta}>m/3\) there is no black hole solution. 

It is clear that, in this model, there exist suitable parameter values that can hide the location where the SEC violation occurs inside the inner horizon and make the exterior region physically acceptable, contrary to Ref. \cite{alkac2025}.

\subsubsection{Dominant energy condition}
The dominant energy condition (DEC) states that the energy measured by an observer should be non-negative and that its propagation should follow null or time-like world lines, which means that the propagation of matter/energy is never faster than light.

\begin{equation}
	\rho\geq0,\quad\text{and}	\quad\rho\geq|p_i|,\quad i=1,2,3.
\end{equation}
The first condition is verified by the WEC above, and for the second inequality, \(\rho\geq|p_r|\), the DEC is satisfied outside the inner horizon for attractive electric charge, while for the repulsive and uncharged scenarios it is valid everywhere because \(p_r=-\rho\). 

\begin{figure}[h]
	\begin{center}
		\includegraphics[width=0.3\textwidth]{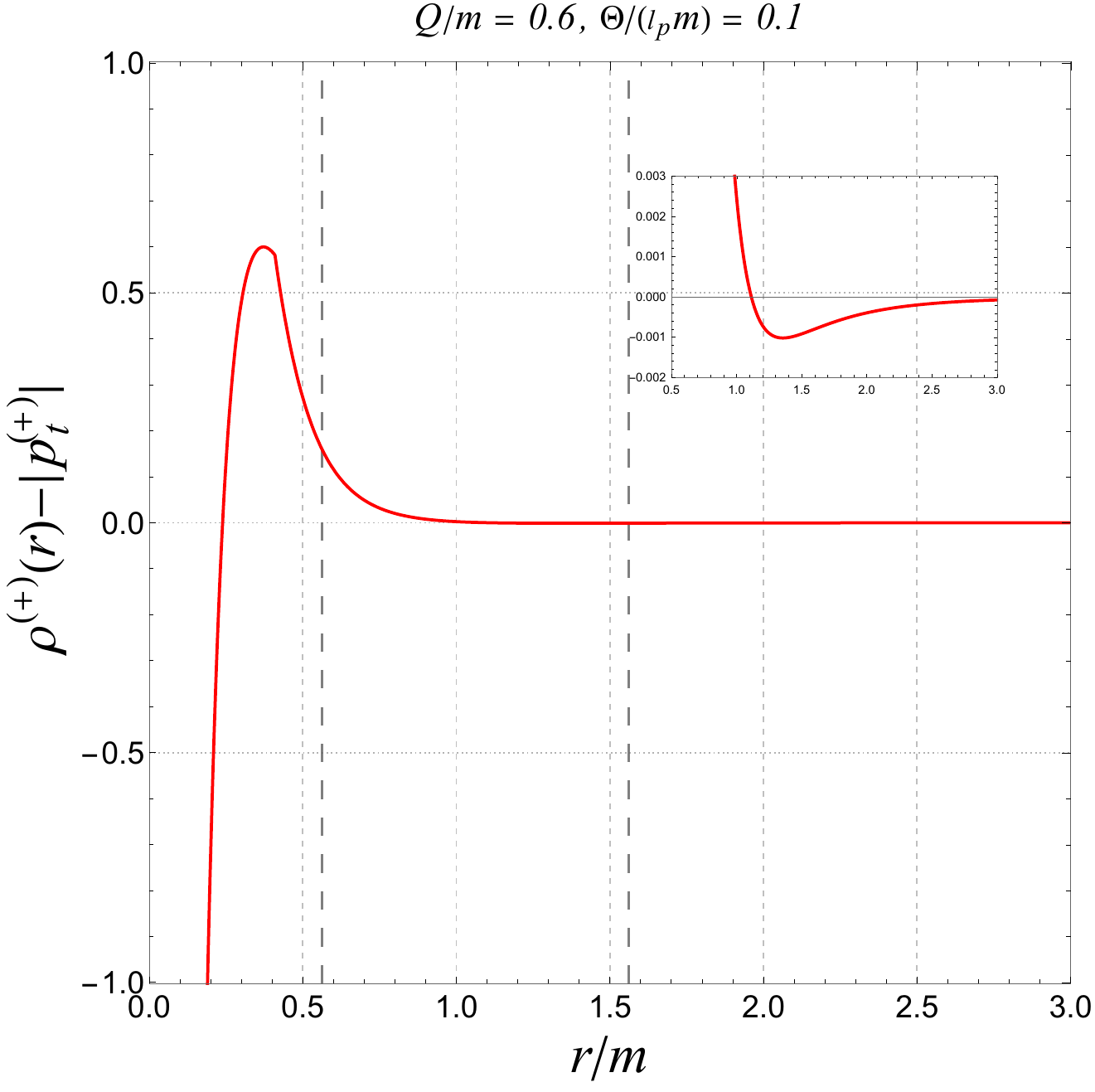}
		\includegraphics[width=0.3\textwidth]{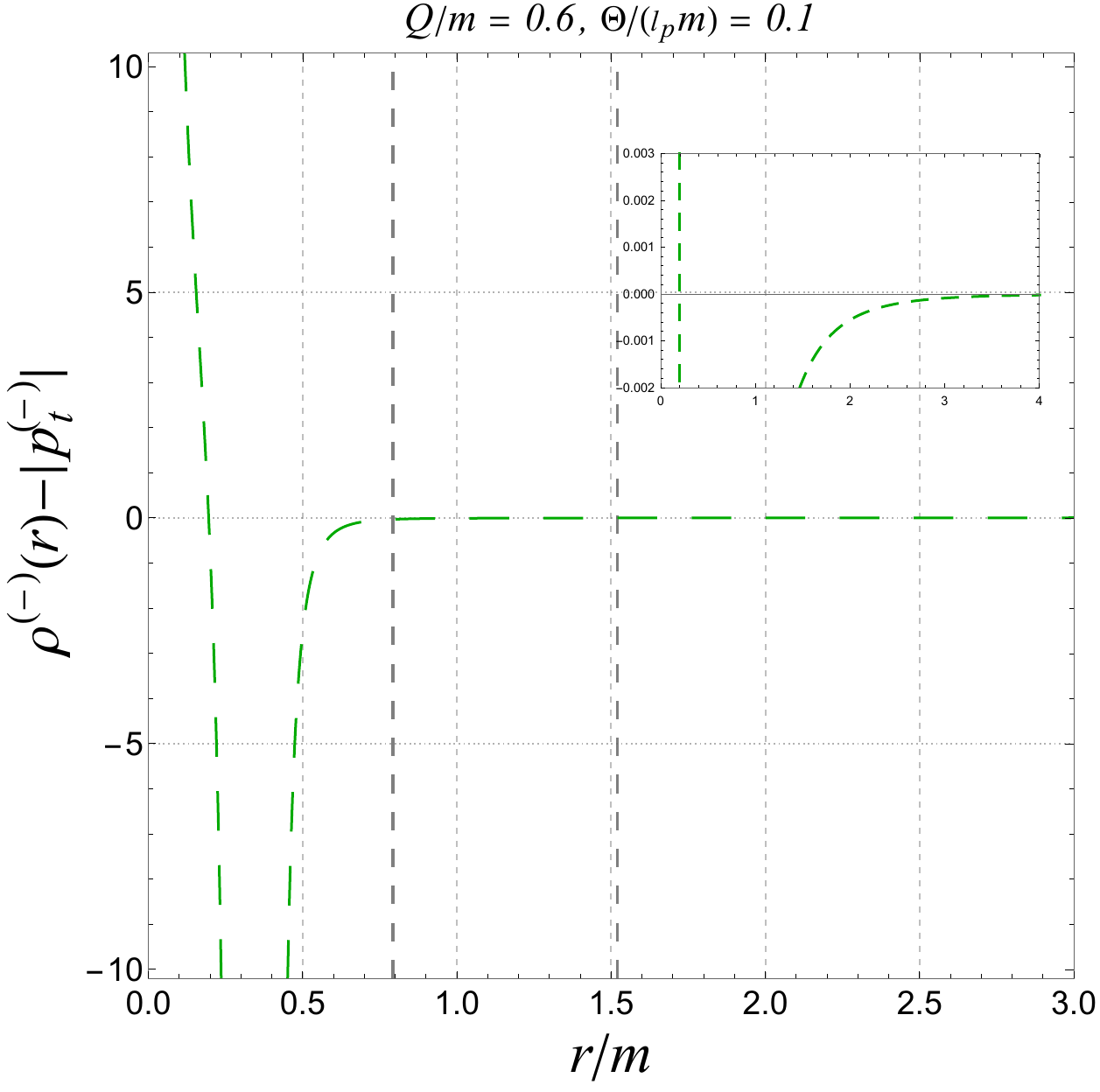}
		\includegraphics[width=0.31\textwidth]{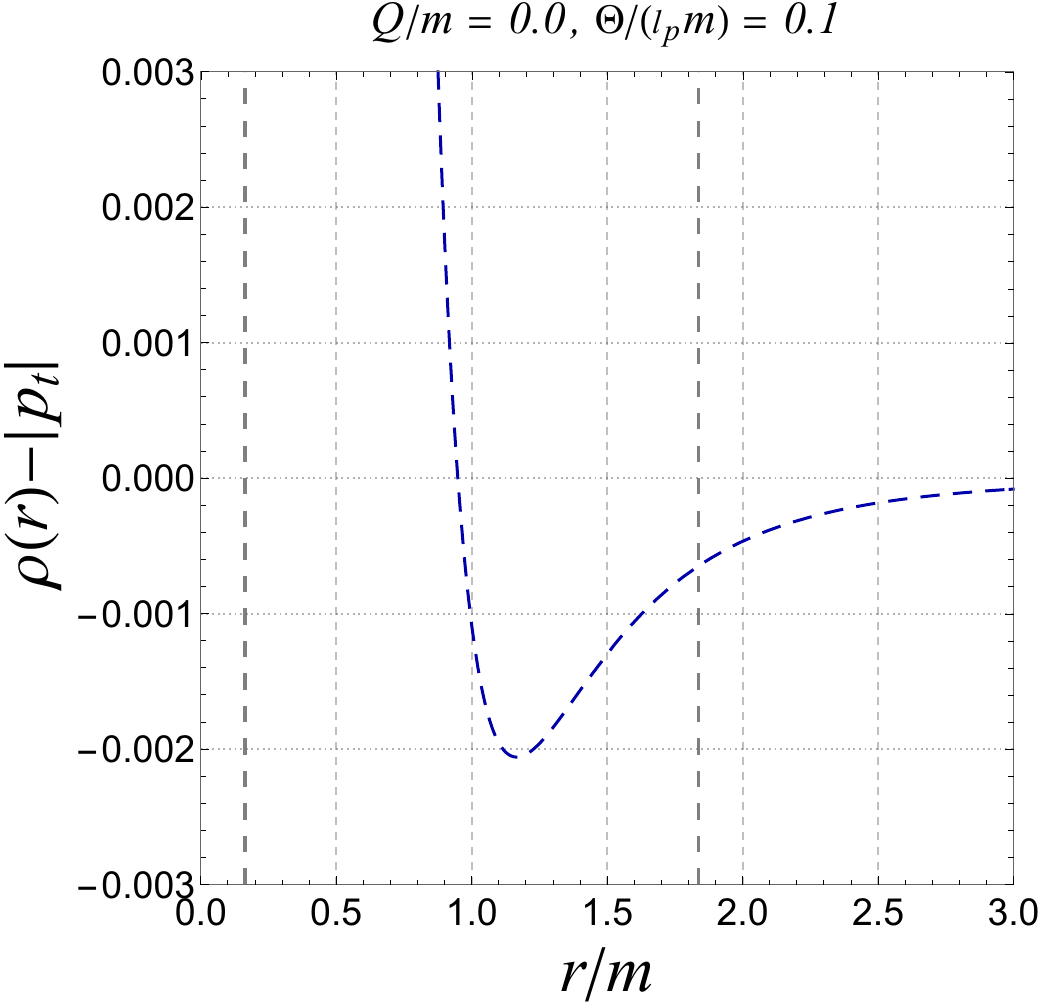}
		\caption{Variation of the DEC for the energy-momentum tensor in the present model as a function of the radius \(r\) for different types of interaction. Left panel: attractive electric interaction. Right panel: repulsive electric interaction. The dashed line represents the inner and outer event horizons for the selected parameters, from left to right respectively in each panel.}\label{fig:DEC}
	\end{center} 
\end{figure}
The second inequality in the right relation is strongly violated by the tangential pressure, for which \(|p_t|>\rho\), and this violation occurs outside the outer event horizon. This means that the energy-momentum tensor is strongly anisotropic, as shown in Fig.~\ref{fig:DEC}. This violation of the DEC is observed in all interaction scenarios, from the left to the right panels, which shows that the effective tangential pressure in this geometry is stronger than the energy density. Thus, this geometry affects the tangential pressure strongly, whereas in the commutative case this condition is well satisfied.

\section{Thermodynamic properties}\label{Sec:TP}

In this section, we investigate in detail the thermodynamic properties induced by the NC deformation via the SW map for the Schwarzschild black hole. Previous works have studied NC effects by deforming the metric itself \cite{abdellah2,abdellah5,Hassanabadi1,zaim1}. By contrast, here we introduce non-commutativity at the level of the interaction potential before solving the Einstein equations (see Sec.~\ref{Sec:NCSBH}), which yields new NC black hole solutions that are simpler to analyze. In this investigation, we use the standard geometric approach to black hole thermodynamics. The first law is written as
\begin{equation}
	dm=T\,dS+\Phi \,dQ,\label{eq:flbht}
\end{equation}
where \(m\), \(T\), \(S\), and \(Q\) denote the black hole mass, Hawking temperature, entropy, and electric charge, respectively.

\subsection{Black hole mass, Hawking temperature, Entropy}

As a first step, we compute the mass of the NC RN-like black hole by solving \(f_Q(r_+)=0\) for \(m\), where \(r_+\) is the outer NC event horizon (see Eq.~\eqref{eq:ehdmdq1} for the repulsive case and Eq.~\eqref{eq:ehdmdq1} for the attractive case). The black hole mass is
\begin{equation}\label{eq:NCQm1}
	m_Q^{(\pm)}=\frac{r_+^2 \left(\left(r_+^2\pm 2 \tilde{\Theta} Q\right)+Q^2\right)}{-3 \tilde{\Theta} Q^2-3 \tilde{\Theta} \left(r_+^2\pm 2 \tilde{\Theta} Q\right)+2 r_+ \left(r_+^2\pm 2 \tilde{\Theta} Q\right)},
\end{equation}
where \(m_Q^+\) and \(m_Q^-\) describe the black hole mass for the attractive and repulsive RN-like black holes, respectively. It is clear that this expression diverges when the denominator becomes zero; we denote this point by \(r_+=r_s\). In the uncharged case \(Q=0\), this expression reduces to that of the NC Schwarzschild black hole with event horizon given by \eqref{eq:eh1}:
\begin{equation}\label{eq:NCm1}
	m=\frac{r_+^2}{2r_+-3\tilde{\Theta}}.
\end{equation}

\begin{figure}[h]
	\begin{center}
		\includegraphics[width=0.45\textwidth]{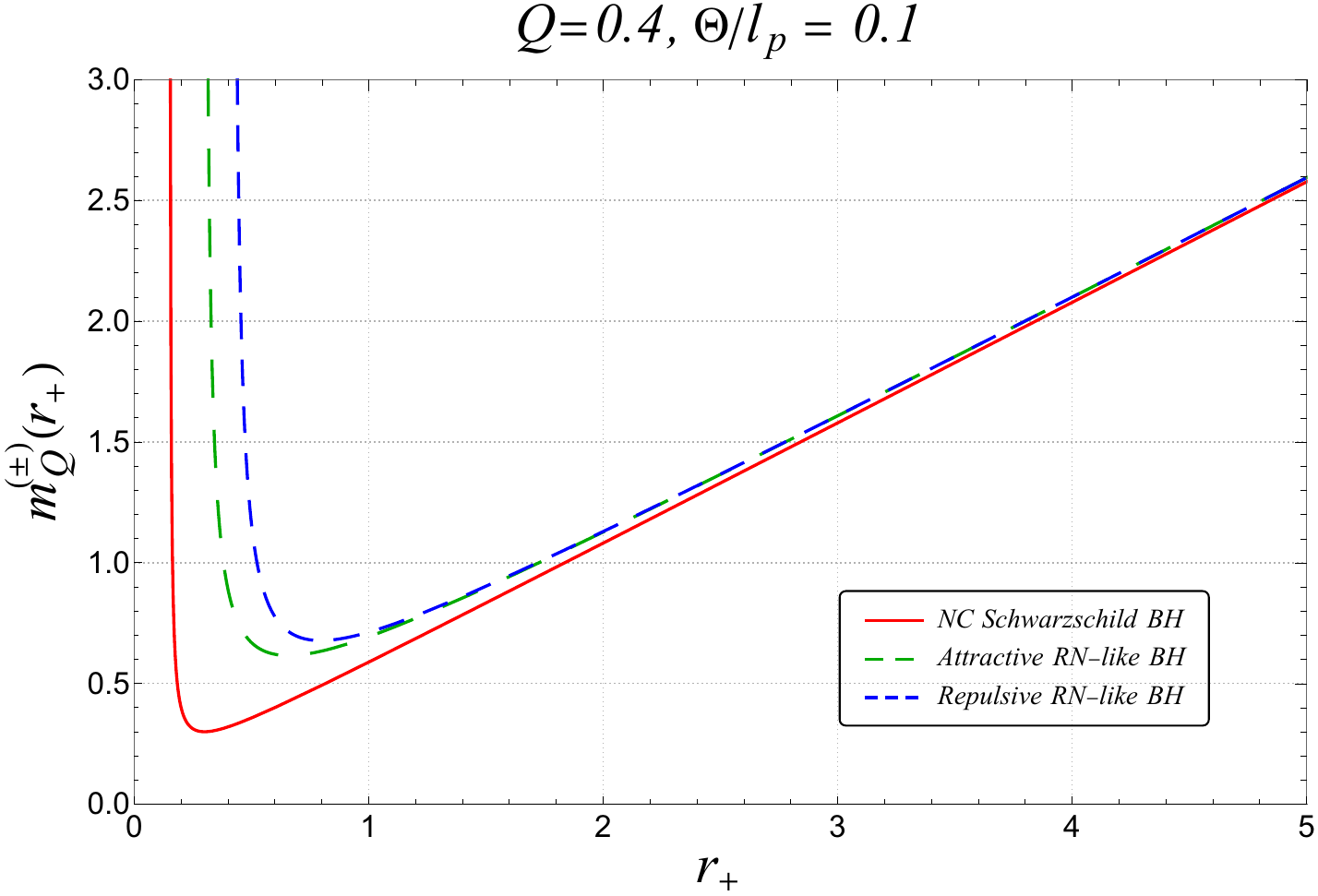}
		\includegraphics[width=0.45\textwidth]{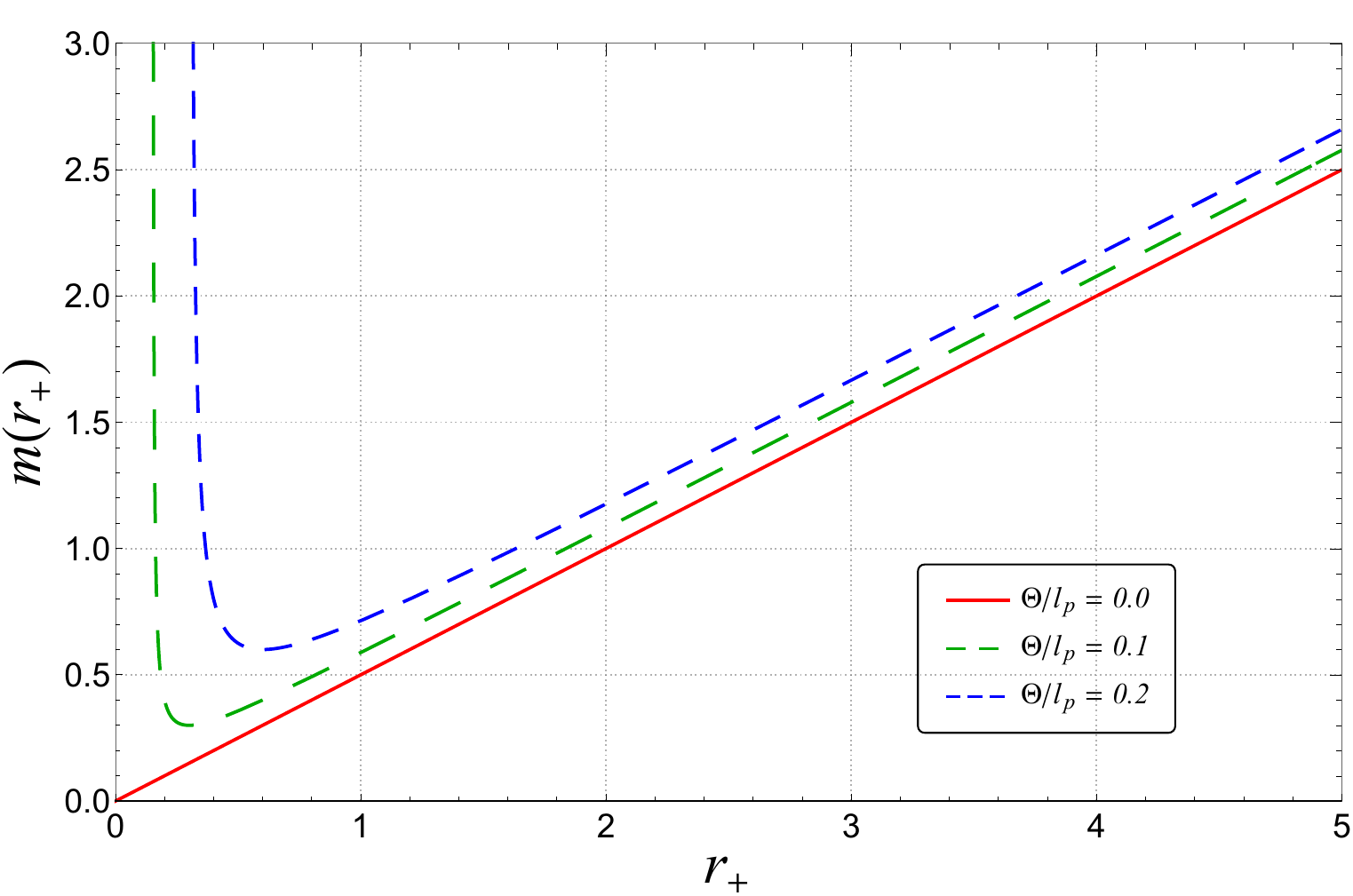}
		\caption{Behavior of the black hole mass \(m_Q^\pm(r_+)\) as a function of the NC outer horizon \(r_+\) for different types of interaction in the left panel and for different values of the NC parameter \(\tilde{\Theta}\) for the uncharged case \(Q=0\) in the right panel.}\label{fig:m}
	\end{center}
\end{figure}
Fig.~\ref{fig:m} shows the black hole mass profile \(m_Q^\pm(r_+)\) as a function of the NC horizon \(r_+\) for several types of interaction in the left panel and for different values of \(\tilde{\Theta}\) in the right panel (we plot only the positive branch for \(r_+>r_s\) in the left panel and for \(r_+>3\tilde{\Theta}/2\) in the right panel). From the left panel, we see that non-commutativity creates a minimum for the Schwarzschild black hole mass, similar to the commutative charged case, and this behavior is also present in the RN-like black hole, where the electric charge increases this minimum. When comparing the different types of interaction, attractive gravity (the NC Schwarzschild black hole) and both charged branches show similar mass behavior. However, the repulsive RN-like black hole exhibits the largest minimum mass, which is shifted to larger event horizons compared to the attractive case. For sufficiently large event horizons, their masses coincide. In the right panel, for the uncharged case, we compare with the commutative limit \(\tilde{\Theta}=0\), where we recover the linear relation \(m=r_+/2\), which vanishes as \(r_+\to0\). In the NC model, the mass profile develops a minimum \(m^{\min}=3\tilde{\Theta}\) attained at \(r_+^{\min}=3\tilde{\Theta}\), consistent with analogous results in NC gauge-theory models \cite{abdellah2,Hassanabadi1,zaim1}. We emphasize that thermodynamic quantities should be studied as functions of the NC event horizon \(r_+\) (the physical size of the NC black hole) rather than the commutative horizon, since using the commutative radius can lead to incorrect expressions for the NC mass and therefore misleading thermodynamic conclusions.

\subsubsection{Hawking temperature}\label{Subsub:NCT1}

The black hole temperature is obtained from the surface gravity via \(T_Q^{(\pm)}=\kappa/(2\pi)=\partial_r f^{(\pm)}_Q(r_+)/(4\pi)\). Evaluating the derivative at the outer horizon gives
\begin{align}
	T^{(\pm)}_Q&=\frac{1}{4\pi}\partial_r f^{(\pm)}_Q(r)\bigg|_{r=r_+}=\frac{-3 \tilde{\Theta} Q^4+Q^2 (r_+-6 \tilde{\Theta}) \left(r_+^2\pm 2 \tilde{\Theta} Q\right)+(r_+-3 \tilde{\Theta}) \left(r_+^2\pm 2 \tilde{\Theta} Q\right)^2-2 Q^2 r_+^3}{4 \pi  r_+^2 \left(r_+^2\pm 2 \tilde{\Theta} Q\right)^2}.\label{eq:NCtQ1}
\end{align}
For the uncharged case \(Q=0\), this reduces to the NC Schwarzschild black hole temperature:
\begin{equation}
	T=\frac{1}{4\pi r_+}-\frac{3\tilde{\Theta}}{4\pi r_+^2}.\label{eq:NCt1}
\end{equation}

\begin{figure}[h]
	\begin{center}
		\includegraphics[width=0.45\textwidth]{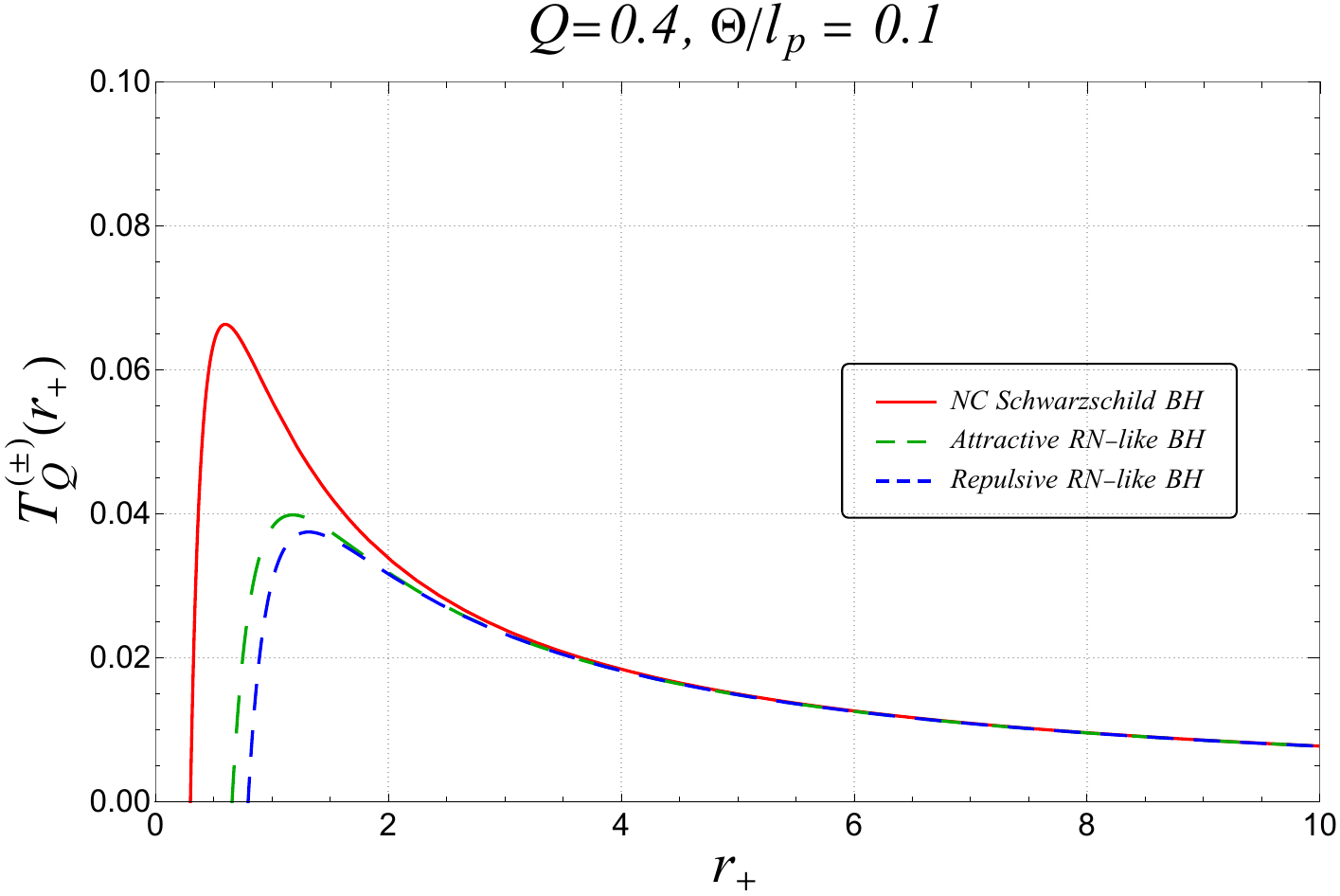}
		\includegraphics[width=0.45\textwidth]{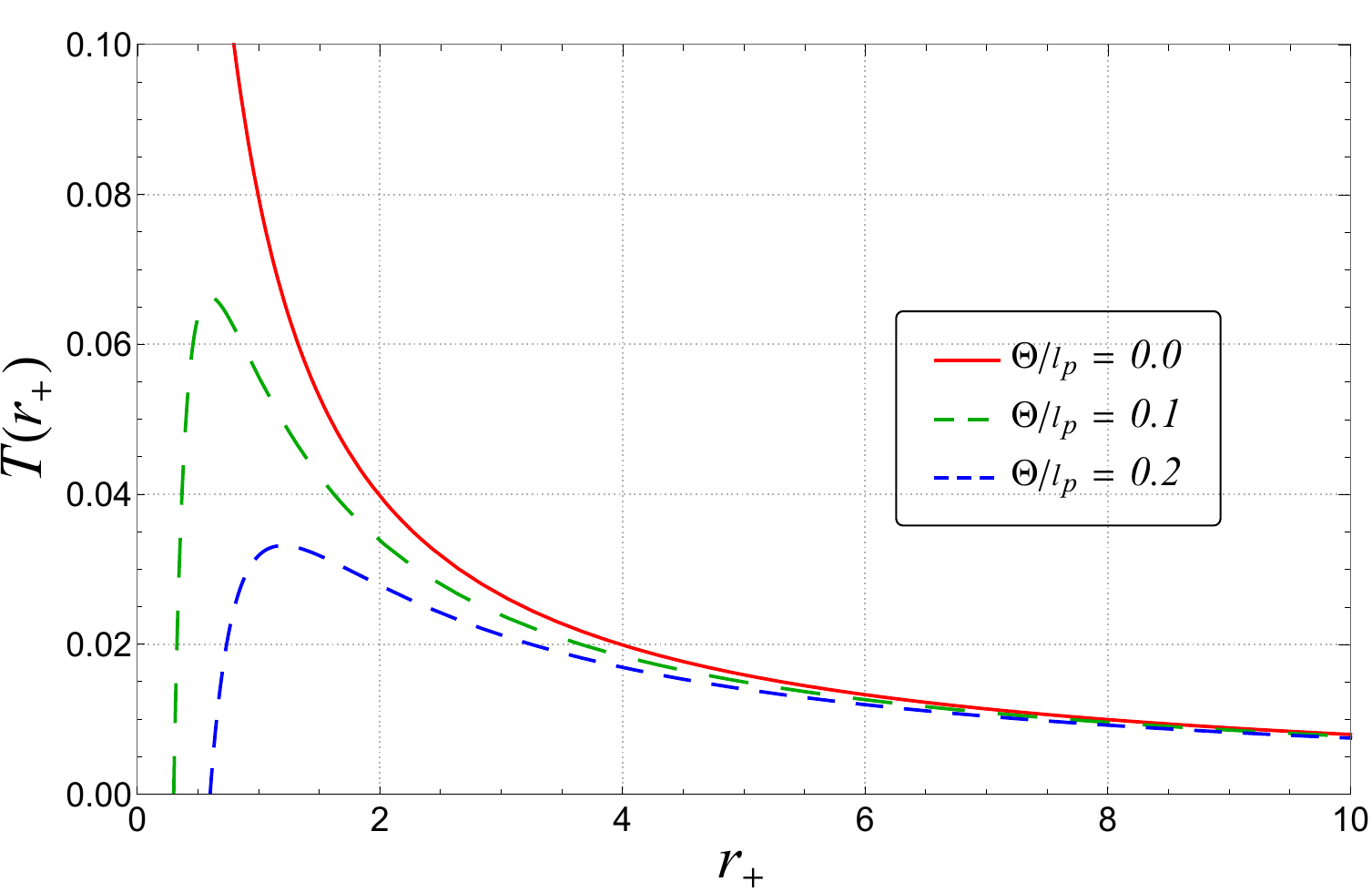}
		\caption{Behavior of the Hawking temperature \(T^{(\pm)}_Q(r_+)\) for the RN-like black hole as a function of the NC outer horizon \(r_+\) for different types of interaction in the left panel and for different values of the NC parameter \(\tilde{\Theta}\) for the uncharged case \(Q=0\) in the right panel.}\label{fig:NCHT1}
	\end{center}
\end{figure}
Fig.~\ref{fig:NCHT1} displays the NC Hawking temperature \(T^{(\pm)}_Q(r_+)\) for the RN-like black hole as a function of \(r_+\) for several types of interaction in the left panel and for various values of \(\tilde{\Theta}\) in the right panel. From the right panel, we see that the NC correction removes the divergence of the commutative temperature and produces a maximal temperature \(T_{\max}\) attained at the critical radius \(r_+^{\rm crit}=6\tilde{\Theta}\); numerically, one finds \(T_{\max}\simeq 0.01/\tilde{\Theta}\) for the parameter choices used in the plots. The temperature then rapidly falls to zero at the minimal horizon \(r_+^{\min}=3\tilde{\Theta}\), coinciding with the minimum mass \(m^{\min}\) shown in Fig.~\ref{fig:m}. Thus, evaporation halts and a remnant with radius \(r_+^{\min}\) and mass \(m^{\min}\) remains. These qualitative features agree with earlier NC gauge-theory-of-gravity results \cite{abdellah2,Hassanabadi1}, while the present method yields simpler analytic expressions for the thermodynamic quantities. The comparison between the NC Schwarzschild black hole and the attractive and repulsive RN-like black holes shows similar temperature behavior, where the presence of electric charge decreases the maximal temperature (\(T^{(\pm)\max}_Q<T_{\max}\)), and they coincide for large black holes, while the difference is observed for small ones. Note that the temperature emitted by the repulsive RN-like black hole is smaller than that emitted by the attractive one for small event horizons, due to pair-particle creation near the event horizon: the attractive RN-like black hole absorbs more particles, allowing more pairs to escape, whereas the repulsive one absorbs fewer particles and allows more pairs to annihilate, which results in a lower temperature compared with the attractive case. Furthermore, the repulsive RN-like black hole stops radiating before the attractive one, and black holes in the presence of electric charge stop radiating before the uncharged one.

Now, returning to the right panel for the NC Schwarzschild black hole, we estimate a bound on the NC parameter using the back-reaction scale. Taking the critical radius \(r_+^{\rm crit}=6\tilde{\Theta}\), the characteristic thermal energy is \(E\sim T_{\max}\simeq 0.01/\tilde{\Theta}\), and the black hole mass at that scale (in natural units \(\hbar=k_B=c=1\)) is \(M=\tfrac{1}{2G}r_+^{\rm crit}=3\tilde{\Theta}M_p^2\). Equating the energy scales gives an estimate\footnote{We keep two digits after the decimal point.} for \(\tilde{\Theta}=\Theta/l_p\):
\begin{equation}\label{eq:NCp1}
	\tilde{\Theta}=\frac{\Theta}{l_p}\simeq1.47\times10^{-35}\,\mathrm{m}\sim l_p,
\end{equation}
and hence
\begin{equation}\label{eq:NCp2}
	\sqrt{\Theta}=\sqrt{\tilde{\Theta}l_p}\simeq1.54\times10^{-35}\,\mathrm{m}\sim l_p.
\end{equation}
These estimates place the NC scale close to the Planck length, consistent in order of magnitude with other NC gauge-theory-of-gravity bounds obtained from black hole thermodynamics \cite{abdellah2,abdellah4,abdellah5}.

It is worth noting that in recent cosmological studies, the formation of stable, non-radiating cold black hole remnants from the evaporation of primordial black holes in the early universe serves as a viable candidate for cold dark matter (CDM) \cite{CDM,CDM1,CDM2,CDM3}. Within our non-commutative (NC) framework, the mass of this cold remnant is completely independent of the black hole's initial mass and is directly related to the NC parameter by $m^{\text{min}}=3\tilde{\Theta}$. Given that this parameter is fundamentally linked to the Planck length scale \eqref{eq:NCp1}, the predicted remnant mass naturally falls within the Planck scale, specifically yielding $M^{\text{min}}\simeq2.73 M_{P}$. According to recent studies \cite{CDM3}, remnants at this mass scale remain highly viable dark matter candidates. Crucially, these remnants are thermally stable, and their mass profile is uniquely dictated by the fundamental NC scale through $M^{\text{min}}=\frac{3c^2}{G}\tilde{\Theta}$, providing a purely gravitational dark matter candidate, thus remaining perfectly consistent with current cosmological density constraints.

\subsubsection{Entropy}

We first comment on entropy in NC geometries. In many implementations of non-commutativity that include the mass source in the energy-momentum tensor \cite{FLBHTherViolation1,FLBHTherViolation2}, or that generate higher-order coupling mass terms through geometry corrections \cite{abdellah2,Hassanabadi1}, the NC corrections modify the area law: the entropy receives corrections and, when obtained from the first law, typically exhibits logarithmic terms at second order in the NC parameter \cite{abdellah2}. 

In the present model, the solid angle is not modified by non-commutativity, so the horizon area retains the commutative form with the NC event horizon,
\begin{equation}
	A_+=\iint\sqrt{g_{\theta\theta}g_{\phi\phi}}\,d\theta d\phi=4\pi r_+^2.
\end{equation}
In the Bekenstein-Hawking picture, the entropy is proportional to the area, so at leading order one has
\begin{equation}
	S=\frac{A_+}{4}=\pi r_+^2,
\end{equation}
with \(r_+\) the NC outer event horizon. In the commutative case, this relation is sufficient to derive thermodynamic quantities and is consistent with the first law \(T=\partial m/\partial S\). However, when NC corrections to the mass are included via the energy-momentum tensor in Eq.~\eqref{eq:emtdmass}, the naive area law may no longer ensure the first law in its simple form. Thus, one finds in general that
\begin{equation}
	T\neq\Big(\frac{\partial m}{\partial S}\Big)
\end{equation}
unless the mass and entropy expressions are adjusted consistently. This violation of the naive area law has been observed in other NC models, including NC gauge theory of gravity \cite{abdellah2,Hassanabadi1} and Lorentzian and Gaussian matter distributions \cite{FLBHTherViolation1,FLBHTherViolation2}. Below, we show that the NC RN-like solution preserves the area law and the first law in the present construction when the mass has no correction term.

The correct entropy must be obtained from the first law \cite{flbht1,abdellah5},
\begin{equation}
	S=\int \frac{dm}{T},\label{eq:FBHE}
\end{equation}
and using the relations \eqref{eq:NCQm1} and \eqref{eq:NCtQ1} one obtains:
\begin{align}
	S_Q^{(\pm)}=\pi  r_+^2+6 \pi  \tilde{\Theta} r_++\frac{27}{2} \pi  \tilde{\Theta}^2 \log (r_+)-\frac{6 \pi  \tilde{\Theta}
		Q^2}{r_+}-\frac{27 \pi  \tilde{\Theta}^2 Q^2}{2 r_+^2}\pm\frac{4 \pi  \tilde{\Theta}^2 Q^3}{r_+^3}-\frac{27 \pi  \tilde{\Theta}^2 Q^4}{8 r_+^4},\label{eq:NCSQ1}
\end{align}
This expression was obtained up to second order in the NC parameter, due to the difficulty of the integral. 

Now, if one uses the uncharged mass and temperature given by Eqs.~\eqref{eq:NCm1} and \eqref{eq:NCt1}, respectively, one obtains
\begin{align}
	S&=\pi r_+^2+6\pi \tilde{\Theta} r_++\frac{27\pi}{2}\tilde{\Theta}^2\log(2r_+-3\tilde{\Theta}),\notag\\
	&=\pi r_+^2+6\pi \tilde{\Theta} r_++\mathcal{O}(\tilde{\Theta}^2).\label{eq:NCS}
\end{align}
This expression can be obtained from Eq.~\eqref{eq:NCSQ1} in the limit of an uncharged black hole \(Q=0\), with the only difference being the presence of \(\log(2)\), which comes from the compact expression that we integrated. It is also similar to that obtained in Ref. \cite{saleem2026thermal} for another NC model.
At second order, the uncharged entropy therefore contains a logarithmic correction, as found in other NC gauge-gravity studies \cite{abdellah4,abdellah5}. Because the present work focuses on first-order effects, we shall use the truncated expression (last line) in what follows. Note that the logarithmic term implies that the argument \(2r_+-3\tilde{\Theta}\) must be positive; this is compatible with the restriction \(r_+>3\tilde{\Theta}/2\) used elsewhere.

Importantly, the first-order correction implies that the entropy at the minimal horizon \(r_+^{\min}\) is nonzero, so the remnant has nonzero entropy. To force the entropy to vanish at the remnant, one must include higher-order (second-order) NC corrections \cite{abdellah4,abdellah5,abdellahPhD}, which will provide further insight into the thermodynamic fate of the remnant.

\subsubsection{Heat capacity and phase transition}

To test the thermal stability of the RN-like black hole, we compute the heat capacity, defined by
\begin{align}
	C_Q^{(\pm)}&=T\bigg(\frac{\partial S}{\partial T}\bigg)\notag\\
	&=\frac{2 \pi  \mathcal{B}_1(r_+) \left(r^2\pm 2 \tilde{\Theta} Q\right) \left(\tilde{\Theta}^2 \mathcal{B}_2(r_+)+\tilde{\Theta} \mathcal{B}_3(r_+)+r^3 \left(Q^2-r^2\right)\right)}{r \left(4 \tilde{\Theta}^2
		\mathcal{B}_4(r_+) Q^2\pm 8 \tilde{\Theta}^3 Q^3 (r-6 \tilde{\Theta})-6 \tilde{\Theta} r^2 \left(Q^2+r^2\right) \left(3 Q^2+r^2\right)\pm 6 \tilde{\Theta} \mathcal{B}_5(r_+) Q-3 Q^2
		r^5+r^7\right)},
\end{align}
with
\begin{align}
	&\mathcal{B}_1(r_+)=3 a \left(Q^2+r^2\right)+r^3,\quad \mathcal{B}_2(r_+)=12 a Q^2\pm 4 Q \left(3 \left(Q^2+r^2\right)\mp Q r\right),\notag\\
	&\mathcal{B}_3(r_+)=3 \left(Q^2+r^2\right)^2\mp 2 r \left(Q^3+2 Q r^2\right),\quad\mathcal{B}_4(r_+)=r \left(Q^2+3 r^2\right)-6 a \left(2 Q^2+3 r^2\right),\notag\\
	&\mathcal{B}_5(r_+)=r^5+2 Q^2 r^3-2 a \left(Q^4+6 Q^2 r^2+3 r^4\right).
\end{align}
Here we have used the entropy expressions to first order in Eqs.~\eqref{eq:NCSQ1} and \eqref{eq:NCtQ1}. In the uncharged limit \(Q=0\), the above expression reduces to the heat capacity of the NC Schwarzschild black hole:
\begin{align}
	C=-2 \pi  \frac{r_+^3}{r_+-6 \tilde{\Theta}},
\end{align}

\begin{figure}[h]
	\begin{center}
		\includegraphics[width=0.45\textwidth]{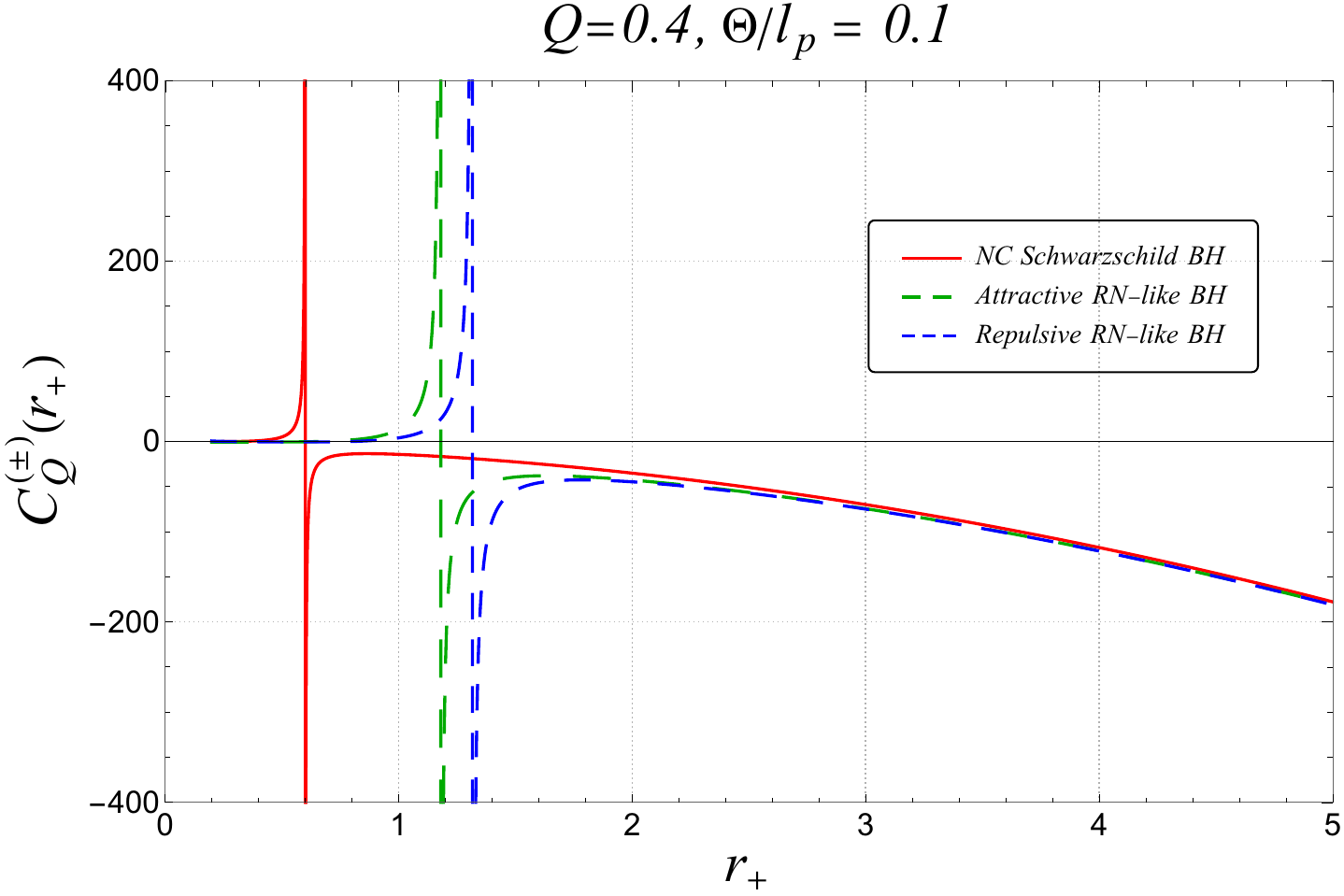}
		\includegraphics[width=0.45\textwidth]{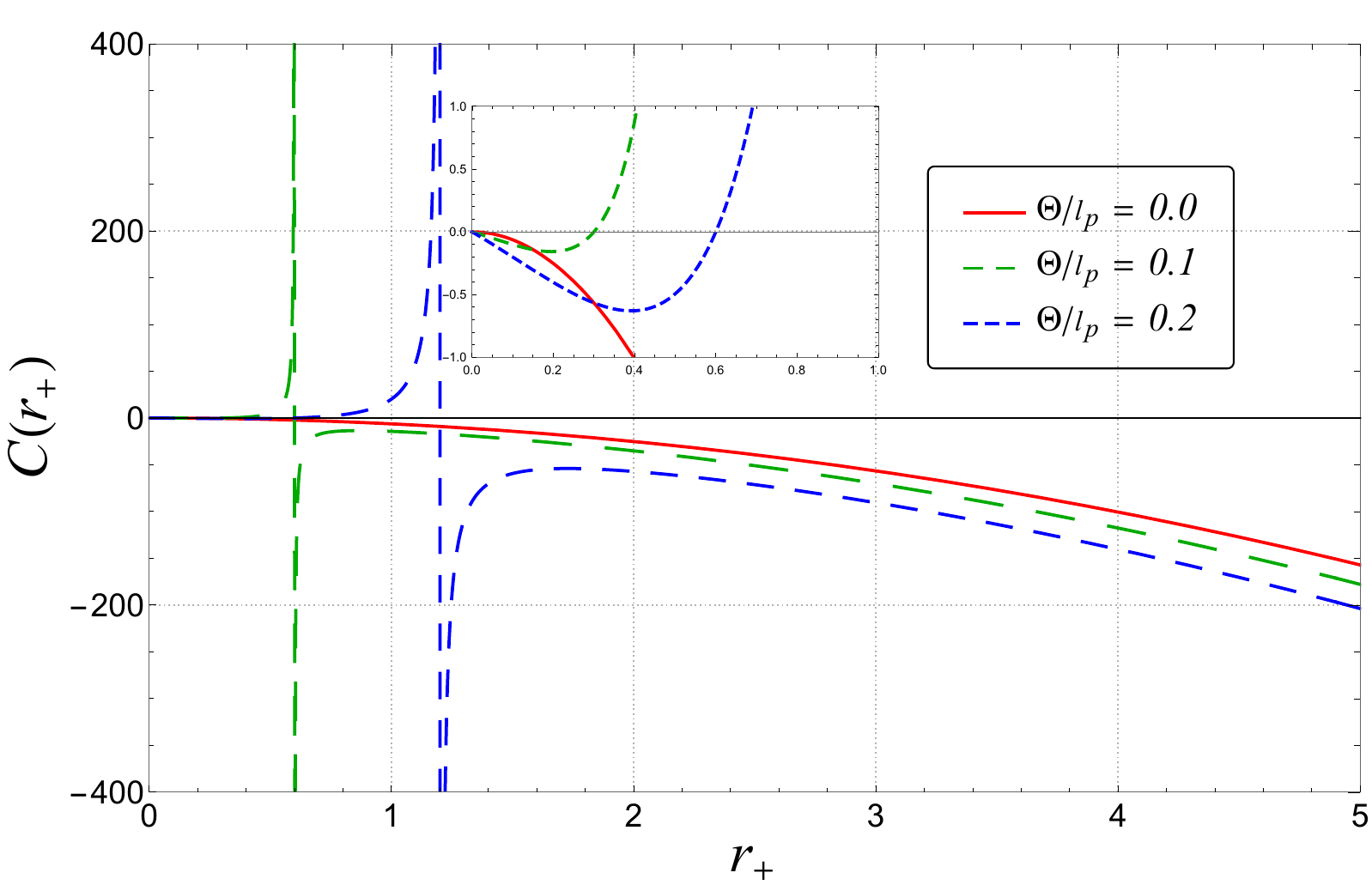}
		\caption{Behavior of the heat capacity \(C_Q^{(\pm)}(r_+)\) for the RN-like black hole as a function of the NC outer horizon \(r_+\) for different types of interaction in the left panel, and for different values of \(\tilde{\Theta}\) in the uncharged case \(Q=0\) in the right panel.}\label{fig:c}
	\end{center} 
\end{figure}
Fig.~\ref{fig:c} shows the heat capacity \(C_Q^{(\pm)}(r_+)\) of the RN-like black hole as a function of \(r_+\) for several types of interaction in the left panel, and for different values of \(\tilde{\Theta}\) in the uncharged case \(Q=0\) in the right panel. In the right panel, two characteristic points appear: (i) a divergence at \(r_+=r_+^{\rm crit}=6\tilde{\Theta}\), and (ii) the physical limiting point \(r_+=r_+^{\min}\). These points coincide with the back-reaction scale, where \(T=T_{\max}\), and the end of evaporation, where \(T=0\), as shown in Fig.~\ref{fig:NCHT1}. For \(r_+>r_+^{\rm crit}\), the heat capacity is negative and the black hole is thermodynamically unstable (non-equilibrium), whereas for \(r_+^{\min}<r_+<r_+^{\rm crit}\), the heat capacity is positive and the black hole is in a stable equilibrium. This behavior is consistent with results obtained in other NC gauge theory of gravity models \cite{abdellah2,abdellah5,Hassanabadi1}. The divergence at \(r_+^{\rm crit}\) signals a second-order phase transition from an unstable large black hole to a stable small black hole, and the critical radius increases with \(\tilde{\Theta}\). The stable region \(r_+^{\min}<r_+<r_+^{\rm crit}\) also grows with \(\tilde{\Theta}\), implying that larger values of the NC parameter prolong the final evaporation stage. In the left panel, we compare the uncharged and charged black holes in this model. It is clear that both the attractive and repulsive RN-like black holes stop radiating before the Schwarzschild black hole, and among the two RN-like solutions, the repulsive one stops radiating first. This result is consistent with the temperature profile in Fig.~\ref{fig:NCHT1}. The three types of black holes exhibit a second-order phase transition from an unstable to a stable phase.

\begin{figure}[h]
	\begin{center}
		\includegraphics[width=0.6\textwidth]{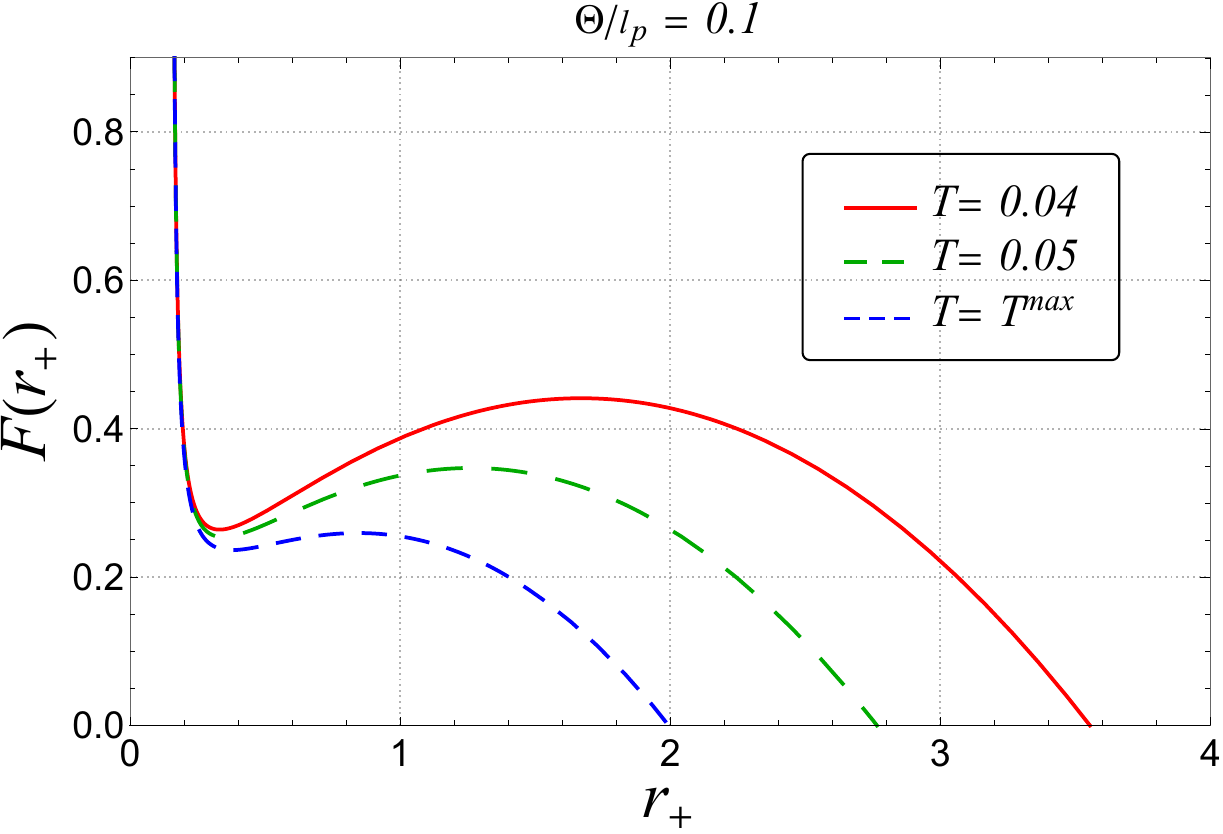}
		\caption{Behavior of the free energy \(F(r_+)\) as a function of \(r_+\) for different temperatures \(T\).}\label{fig:F1}
	\end{center} 
\end{figure}

To analyze the phase structure, we consider the Helmholtz free energy,
\begin{align}
	F=m-TS,
\end{align}
whose profile is shown in Fig.~\ref{fig:F1}.

Fig.~\ref{fig:F1} shows that non-commutativity modifies the free energy and generates a new minimum at the final evaporation stage \cite{abdellah2,Hassanabadi1}. The free energy typically exhibits two extrema corresponding to the stable (minimum) and unstable (maximum) branches, in agreement with the heat-capacity analysis. Furthermore, the Helmholtz free energy of the RN-like black hole shows a behavior similar to that of the uncharged case, and the repulsive branch has the same effect as that observed in the mass profile in Fig.~\ref{fig:m}.

\subsection{Gibbs free energy}

We now analyze the Gibbs free energy to study the effect of pressure on the phase structure of the NC black hole. We begin with the first law of black hole thermodynamics in the presence of pressure:
\begin{equation}\label{eqt3.18}
	dm=TdS+\Phi dQ-PdV,
\end{equation}
where \(V\) denotes the NC volume of the black hole. In our geometry, the volume is given by
\begin{align}\label{eq:NCV2}
	V=\frac{4\pi}{3}r_+^3,
\end{align}
and the pressure conjugate to the volume is defined as
\begin{align}\label{eq:NCP2}
	P_Q^{(\pm)}&=-\left(\frac{\partial m_Q^{(\pm)}}{\partial V}\right)
	=-\left(\frac{\partial m_Q^{(\pm)}}{\partial r_+}\right)\left(\frac{\partial V}{\partial r_+}\right)^{-1}
	\notag\\
	&=\frac{\left(12 \tilde{\Theta}^3 Q^2\pm 4 \tilde{\Theta}^2 Q \left(3 \left(Q^2+r_+^2\right)\mp Q r_+\right)\right)+\left(3 \tilde{\Theta} \left(Q^2+r_+^2\right)^2\mp 2 \tilde{\Theta} r_+ \left(Q^3+2
		Q r_+^2\right)\right)+r_+^3 \left(Q^2-r_+^2\right)}{2 \pi  r_+ \left(2 r_+^3-\left(\left(3 \tilde{\Theta} \left(Q^2+r_+^2\right)\mp 6 \tilde{\Theta}^2 Q\right)\pm 4 \tilde{\Theta} Q
		r_+\right)\right)^2}.
\end{align}
For the uncharged case \(Q=0\), we obtain
\begin{equation}
	P=-\frac{r_+-3 \tilde{\Theta}}{2 \pi  r_+ (2 r_+-3 \tilde{\Theta})^2}
	\simeq-\frac{1}{8\pi r_+^2}+\frac{9 \tilde{\Theta}^2}{32 \pi  r_+^4},
\end{equation}
where the leading term \(P=-\dfrac{1}{8\pi r_+^2}\) is the commutative contribution, while the NC correction appears only at second order in \(\tilde{\Theta}\). 

\begin{figure}[h]
	\begin{center}
		\includegraphics[width=0.6\textwidth]{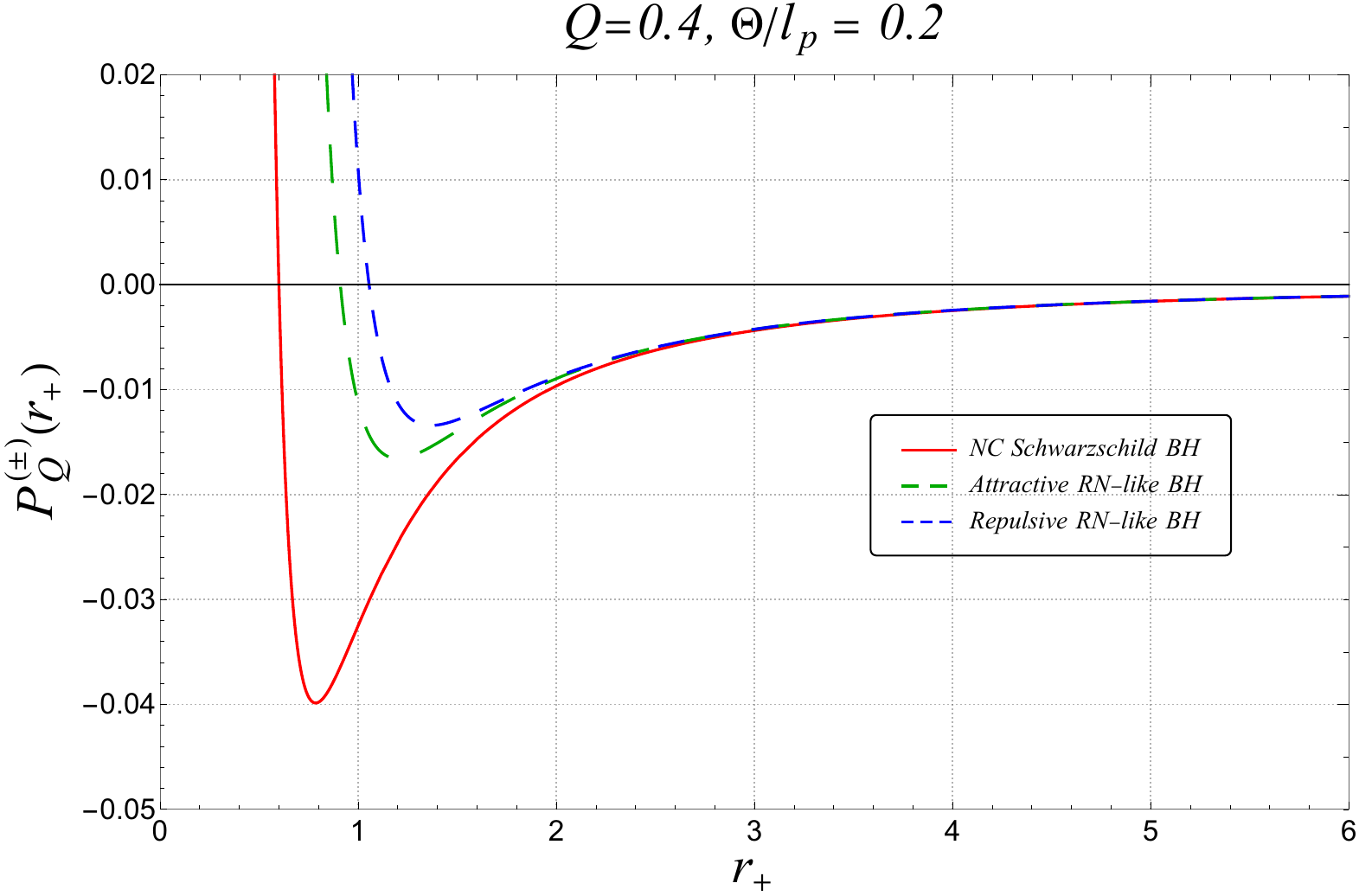}
		\caption{Behavior of the pressure \(P_Q^{(\pm)}\) of the RN-like black hole as a function of the NC event horizon \(r_+\) for different types of interaction.}\label{fig:P}
	\end{center}
\end{figure}
In Fig.~\ref{fig:P}, we illustrate the effect of the NC geometry and the different types of interactions on the black hole pressure. As shown, non-commutativity creates a new minimum in the black hole pressure for both uncharged and charged black holes. According to our previous work \cite{abdellah2}, a negative \(P\) represents the pressure exerted by the black hole on spacetime, and the minimum of \(P\) corresponds to the maximum pressure that the black hole can exert on spacetime. This pressure is induced by black hole radiation, and as the black hole begins to lose radiation, its pressure decreases until it reaches zero at the equilibrium radius, which corresponds to the point where the black hole stops radiating (see the temperature profile in Fig.~\ref{fig:NCHT1}). At that point, a remnant is formed. We also note that the repulsive RN-like black hole exerts a smaller pressure on spacetime, which is consistent with the behavior observed in the temperature profile in Fig.~\ref{fig:NCHT1}.

In this context, we use the Gibbs free energy to characterize the influence of pressure on phase transitions for the NC Schwarzschild black hole. The Gibbs free energy is defined as
\begin{equation}\label{eqt3.21}
	G=m-TS+VP.
\end{equation}

\begin{figure}[h]
	\begin{center}
		\includegraphics[width=0.32\textwidth]{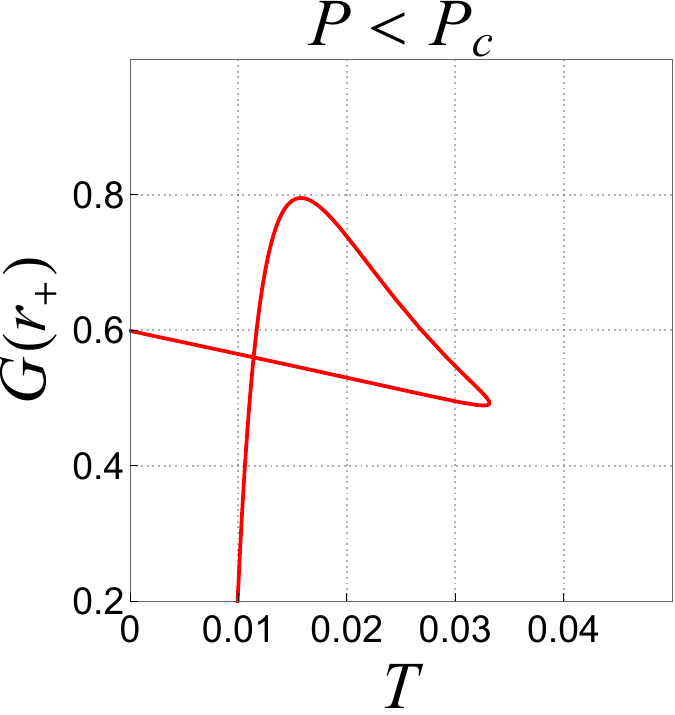}
		\includegraphics[width=0.32\textwidth]{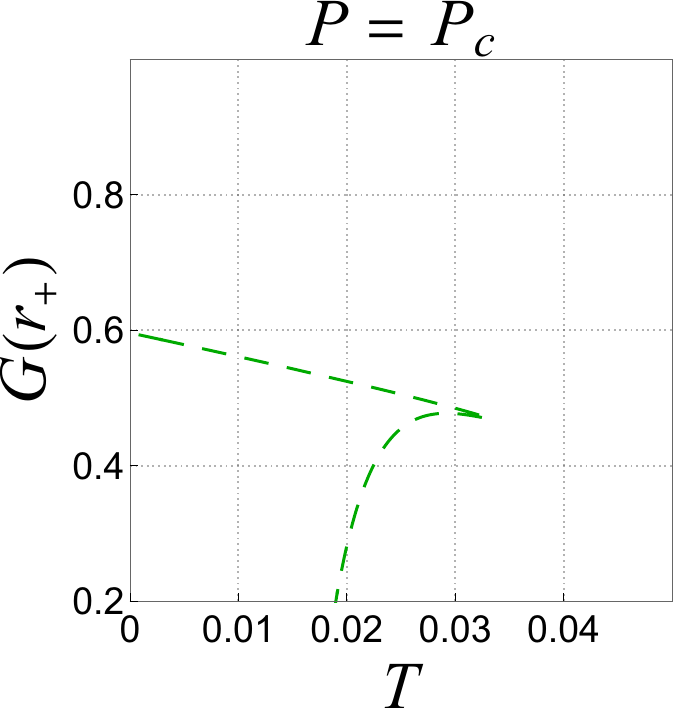}
		\includegraphics[width=0.32\textwidth]{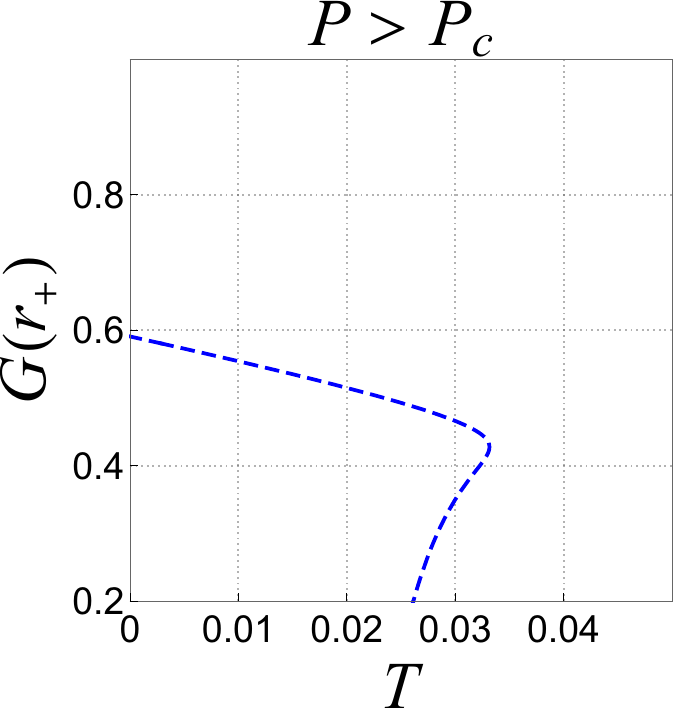}
		\caption{Behavior of the Gibbs free energy \(G(r_+)\) as a function of the temperature \(T\) for different pressures \(P\). The plots use \(\tilde{\Theta}=0.2\) and show a critical pressure \(P_c=-0.0042\).}\label{fig:G1}
	\end{center}
\end{figure}
Fig.~\ref{fig:G1} displays the Gibbs free energy for \(\tilde{\Theta}=0.2\) as a function of the temperature \(T\) at several values of the pressure \(P\). The overall behavior reproduces that observed in studies of NC gauge theory of gravity \cite{abdellah2}: the present approach retains the same qualitative thermodynamic features while providing a simpler framework for studying NC black holes. In particular, an inflection point signaling a Hawking--Page-like phase transition appears at a critical pressure \(P_c\approx-0.0042\). For pressures below \(P_c\), the curve shows a quasi-swallowtail structure with a smooth profile, whereas for pressures above \(P_c\), the inflection disappears and the Gibbs curve becomes smooth \cite{abdellah2}. At the critical point, the black hole undergoes a transition from a large unstable branch to a small stable branch under thermal radiation. Note that the Gibbs free energy can also be analyzed for the RN-like black hole in both the attractive and repulsive cases, where it shows the same behavior as in the uncharged case.

\subsection{Non-commutative susceptibility and Linear Response}\label{sebsec:slrNC}

In this subsection, we examine the linear response of the uncharged black hole to variations in the NC parameter \(\tilde\Theta\). In this case, the first law of black hole thermodynamics in the presence of the NC potential reads:
\begin{equation}
	dm=TdS+\mathcal{A}_{\tilde{\Theta}}\, d\tilde{\Theta},
\end{equation} 
where \(\mathcal{A}_{\tilde{\Theta}}\) represents the NC potential conjugate to the NC parameter \(\tilde{\Theta}\), and is defined by \cite{deformedpotential1,abdellahPhD}:
\begin{equation}
	\mathcal{A}_\Theta=\bigg(\frac{\partial m}{\partial \tilde{\Theta}}\bigg)=\frac{3r_+^2}{(2r_+-3\tilde{\Theta})^2},
\end{equation}
It is clear that this expression has the same divergence behavior as the black hole mass at the point \(r_+=\tfrac{3}{2}\tilde{\Theta}\). This quantity measures the geometric response to variations in the NC parameter.

The susceptibility associated with the NC parameter can be evaluated using the following definition \cite{susceptibility1,susceptibility2}:
\begin{align}
	\chi_{\tilde{\Theta}}&=\bigg(\frac{\partial \mathcal{A}_{\tilde{\Theta}}}{\partial \tilde{\Theta}}\bigg)
	=\bigg(\frac{\partial^2 m}{\partial\tilde{\Theta}^2}\bigg)\notag\\
	&=\frac{18r_+^2}{(2r_+-3\tilde{\Theta})^3}.
\end{align}

\begin{figure}[h]
	\begin{center}
		\includegraphics[width=0.5\textwidth]{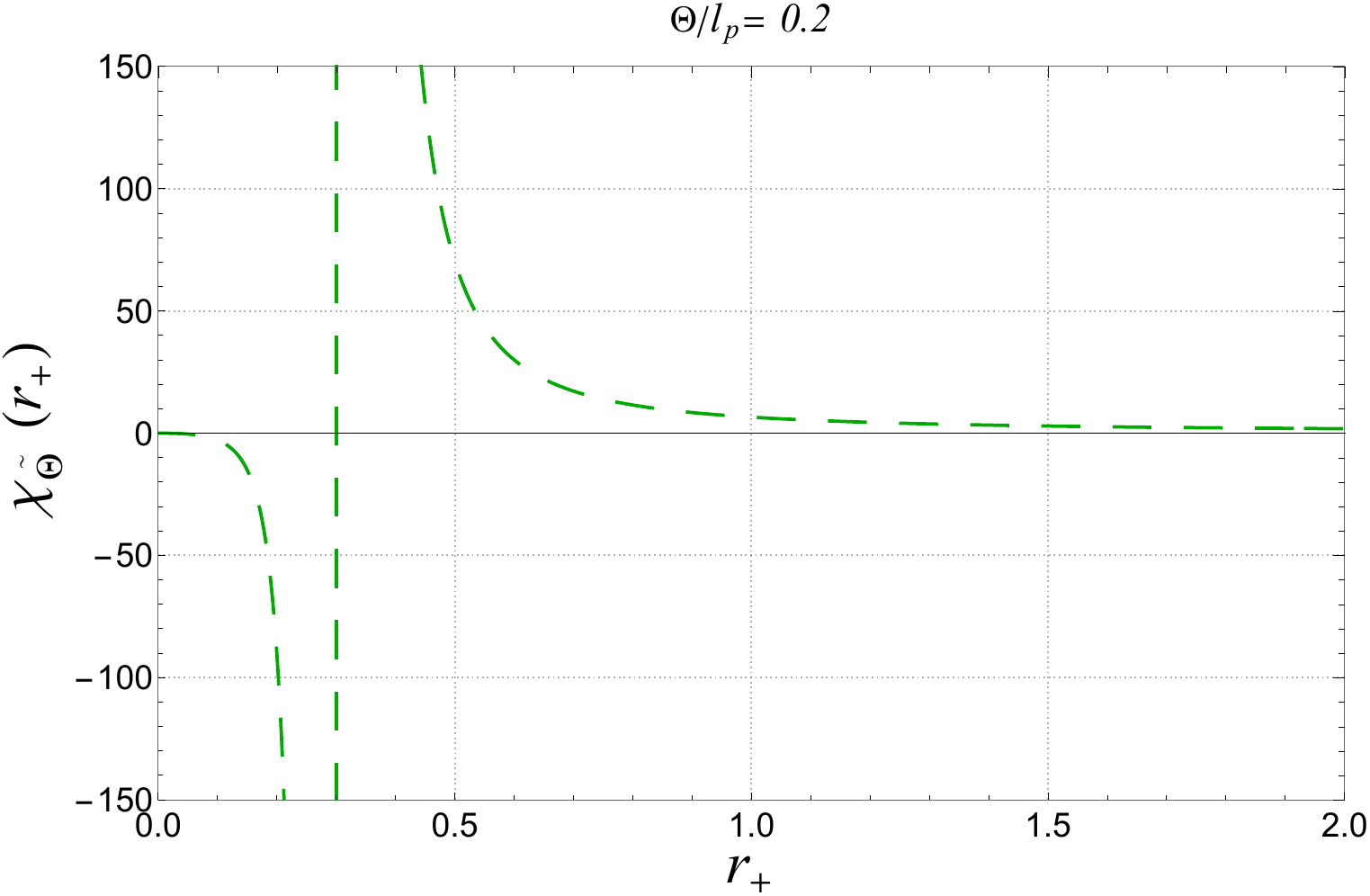}
		\caption{Behavior of the NC susceptibility \(\chi_{\tilde{\Theta}}\) as a function of the NC event horizon \(r_+\).}\label{fig:chi1}
	\end{center}
\end{figure}
Fig.~\ref{fig:chi1} shows the behavior of the susceptibility \(\chi_{\tilde{\Theta}}\) as a function of the NC event horizon \(r_+\). The function \(\chi_{\tilde{\Theta}}\) diverges at \(r_+=\tfrac{3}{2}\tilde{\Theta}\), which would indicate a continuous second-order critical point in the NC sector. However, this divergence lies inside the remnant region: the black hole stops radiating at \(r_+=3\tilde{\Theta}\), so the singular point remains in a thermally inaccessible region within the semi-classical evaporation process. It is possible that a purely quantum evaporation process, which forces the black hole to complete its evaporation ($m_{min} \rightarrow 0$), could reach this region. In this case, the divergence may be interpreted as a phase transition in the NC sector. Since our model predicts a remnant at \(r_+=3\tilde{\Theta}\), the physically accessible region is only the exterior of the remnant, namely \(r_+\geqslant 3\tilde{\Theta}\). In this region, the NC susceptibility is positive, \(\chi_{\tilde{\Theta}}>0\), indicating a normal and stable linear response of the geometry to the deformation induced by non-commutativity. For large black holes, the NC susceptibility is small, indicating that the system becomes less sensitive to the NC parameter. This behavior is consistent with the mass profile in Fig.~\ref{fig:m} and the temperature profile in Fig.~\ref{fig:NCHT1}, where relatively large changes in \(\tilde{\Theta}\) produce only small variations in the thermodynamic quantities. By contrast, for small black holes, i.e. in regions closer to the remnant, the NC susceptibility becomes large, indicating that the system is highly sensitive to changes in \(\tilde{\Theta}\): even a small variation in the NC parameter leads to significant changes in the thermodynamic properties.

\subsection{Thermodynamic properties of the NC charged black hole}\label{SubSec:TPNCRN}

In this subsection, we compute several thermodynamic quantities in order to verify the consistency of the first law of black hole thermodynamics. We use the metric obtained above, without a deformed mass in Eq.~\eqref{eq:NCRN1}, and work within the approximation regime.

\paragraph{ADM mass:}
We begin with the ADM mass, obtained by solving \(f(r_+)=0\),
\begin{equation}
	m_Q = \frac{r_+^4+Q^2 r_+^2+2 a Q^3}{2 r_+^3},
\end{equation} 
which reduces to the RN mass in the commutative limit \(\tilde{\Theta}=0\). For \(Q=0\), this expression further reduces to the Schwarzschild mass.

\paragraph{Temperature:}
The NC-corrected black hole temperature is obtained from the surface gravity:
\begin{align}
	T_Q\simeq\frac{1}{4 \pi  r_+}-\frac{Q^2}{4 \pi  r_+^3}-\frac{3 \tilde{\Theta} Q^3}{2 \pi  r_+^5},
\end{align}
which reproduces the RN temperature when \(\tilde{\Theta}=0\), and the Schwarzschild temperature for \(Q=0\).

\paragraph{Entropy:}
We now use the first law of black hole thermodynamics in Eq.~\eqref{eq:flbht} to derive the entropy. From \(S_Q=\int dm/T\), and after expressing \(dm/T\) in terms of \(r\), we find
\begin{align}
	S_Q&=\int_0^{r_+}2 \pi  r\frac{r^4-Q^2 \left(6 \tilde{\Theta} Q+r^2\right)}{r^4-Q^2 r^2-6 \tilde{\Theta} Q^3}\,dr
	=2\pi\int_{0}^{r_+}r\,dr,\notag\\
	&=\pi r_+^2.
\end{align}
Thus, the entropy satisfies the Bekenstein--Hawking area law, \(S_Q=\pi r_+^2\). This result shows that, in the present setup, non-commutativity preserves both the area law and the first law of black hole thermodynamics. This is because the NC correction enters through the \(U(1)\) interaction (the gauge field) rather than through a deformed mass contribution in the energy-momentum tensor, as in the previous uncharged case.

\section{Quantum tunneling process}\label{Sec:QTP}

As demonstrated in the previous section, the black hole in the presence of non-commutativity exhibits a macroscopic second-order phase transition at critical point $r^{\text{crit}}_+=6\tilde{\Theta}$. This finding motivates us to investigate the microscopic mechanism responsible for the evaporation process. For that, the semi-classical quantum tunneling process can provide a good microscopic description of the evaporation dynamics that drives the black hole to change its thermal phase toward this critical point.

In this section, we investigate quantum tunneling from the new NC black hole solution for both thermal and non-thermal radiation. To describe the tunneling process, we employ regular (stationary) coordinates \cite{tunn1,tunn2,tunn3} that do not exhibit a coordinate singularity at the horizon, in contrast to Schwarzschild-like coordinates. This choice ensures energy conservation during the derivation of the radiation spectrum. Concretely, we adopt the Painlevé--Gullstrand form \cite{tunn1,tunn2,tunn3,tunn4} to represent our NC black hole metric. For radial motion, the line element \eqref{eq:metric1} becomes
\begin{align}\label{eq:PG-line}
	ds^2=-f(r)cdt^2+2h(r)dtdr+dr^2+r^2d\theta^2+r^2\sin^2\theta\,d\phi^2,
\end{align}
where
\begin{equation}
	h(r)=\sqrt{1-f(r)}.
\end{equation}

In the semi-classical tunneling picture, the emission rate and the imaginary part of the action are related by \cite{tunn1,tunn2,tunn3}
\begin{equation}\label{eq:tunn1}
	\Gamma\sim e^{-2\operatorname{Im}\hat{S}}.
\end{equation}
Here, \(\hat{S}\) denotes the NC tunneling action of the particle, defined for a particle moving freely in curved spacetime by
\begin{equation}
	\hat{S}=\int p_{\mu}\,dx^{\mu},
\end{equation}
with conjugate momentum \(p_{\mu}=\hat{g}_{\mu\nu}\dfrac{dx^{\nu}}{d\lambda}\) and affine parameter \(\lambda\). We consider a massless particle moving along a radial trajectory in the equatorial plane \(\theta=\pi/2\). The imaginary part of the action then reduces to
\begin{align}\label{eq:ims}
	\operatorname{Im}\hat{S}=\operatorname{Im}\int\!(p_{t}\,dt+p_r\,dr)
	=\operatorname{Im}\int_{r_i}^{r_f}\!\int_{0}^{p_r}\!dp'_r\,dr .
\end{align}

The term involving \(p_t\) is real and therefore does not contribute to the imaginary part. Using Hamilton's equation \(\dot{r}=\dfrac{dH}{dp_r}\) (where \(\dot{r}=\dfrac{dr}{dt}\)) and taking the system Hamiltonian as \(H=m-\omega'\)\footnote{Where \(\omega'\) represents the emitted particle energy.}, Eq.~\eqref{eq:ims} becomes
\begin{align}\label{eq:ims2}
	\operatorname{Im}\hat{S}=\operatorname{Im}\int_{m}^{m-\omega}\!\int_{r_i}^{r_f}\frac{d(m-\omega)}{\dot{r}}\,dr .
\end{align}

We assume that the tunneling particle follows the radial null geodesic, obtained from \(\hat{g}_{\mu\nu}U^\mu U^\nu=0\), whose radial velocity is
\begin{align}\label{eq:radial}
	\frac{dr}{dt}=1-\sqrt{1-f(r)}.
\end{align}
Expanding near the outer horizon \(r_+\) gives
\begin{align}\label{eq:radial2}
	\frac{dr}{dt}\simeq \frac{1}{2}f'(r_+)\,(r-r_+)+\mathcal{O}((r-r_+)^2),
\end{align}
where the surface gravity is \(\kappa(r_+)=\tfrac{1}{2}f'(r_+)\). Substituting this into Eq.~\eqref{eq:ims2} leads to
\begin{align}\label{eq:ims3}
	\operatorname{Im}\hat{S}=\operatorname{Im}\int_{m}^{m-\omega}\!\int \frac{d(m-\omega)}{\kappa(r_+)(r-r_+)}\,dr .
\end{align}

\subsection{Pure thermal radiation}

In the pure thermal radiation case, we have \(\hat{g}_{01}=\hat{h}\) and \(dH=d(m-\omega)=-d\omega\). After rearranging Eq.~\eqref{eq:ims3}, we obtain
\begin{align}\label{eq:ims4}
	\operatorname{Im}\hat{S}=\operatorname{Im}\int_{0}^{\omega}(-d\omega)\int \frac{dr}{\kappa(r_+)(r-r_+)}.
\end{align}
The radial integral exhibits a simple pole at the event horizon \(r=r_+\). We evaluate it using a contour deformation around the pole, namely the residue theorem. Working in the low-frequency regime \(\omega\ll m\), so that the Boltzmann approximation \(\Gamma\sim e^{-2\operatorname{Im}\hat S}\sim e^{-\beta\omega}\) holds, and following the standard steps in Refs.~\cite{tunn9,tunn7}, we find
\begin{align}\label{eq:ims5}
	\operatorname{Im}\hat{S}&=\pi \int_{0}^{\omega}\frac{d\omega}{\kappa(r_+)}=\frac{\pi\omega}{\kappa(r_+)}.
\end{align}

\subsubsection{Tunneling rate in thermal radiation}

Substituting Eq.~\eqref{eq:ims5} into the tunneling-rate expression \eqref{eq:tunn1} yields
\begin{align}\label{eq:tunneling1}
	\Gamma&\sim\exp\bigg[-4\pi\omega \frac{\big(\sqrt{m (m-3 \tilde{\Theta})}+m\big)^2}{\sqrt{m (m-3 \tilde{\Theta})}-3 \tilde{\Theta}+m}\bigg],
\end{align}
and, to leading order in the NC parameter, one obtains
\begin{align}\label{eq:pttunneling2}
	\Gamma\sim \exp\left[-2\pi\omega\left(4m+3\tilde{\Theta}\right)\right].
\end{align}
In the commutative limit \(\Theta \rightarrow 0\), this reproduces the usual weak-energy result \((\omega\ll m)\) \cite{tunn1,tunn2}. The above expression can be generalized to the charged NC black hole by using the corresponding surface gravity.

\begin{figure}[h]
	\begin{center}
		\includegraphics[width=0.32\textwidth]{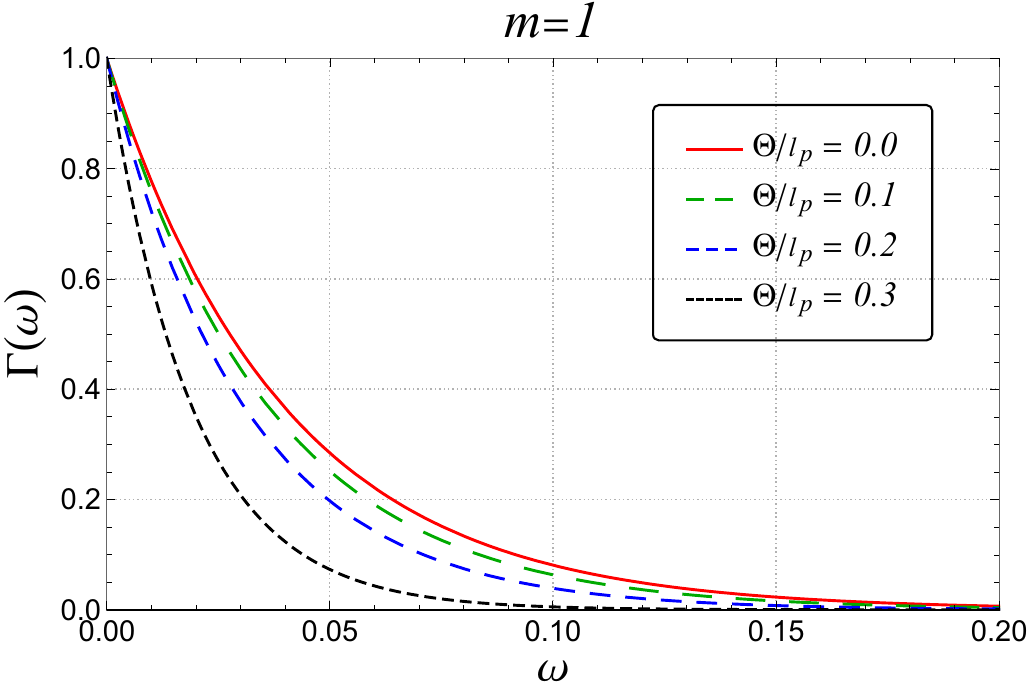}
		\includegraphics[width=0.32\textwidth]{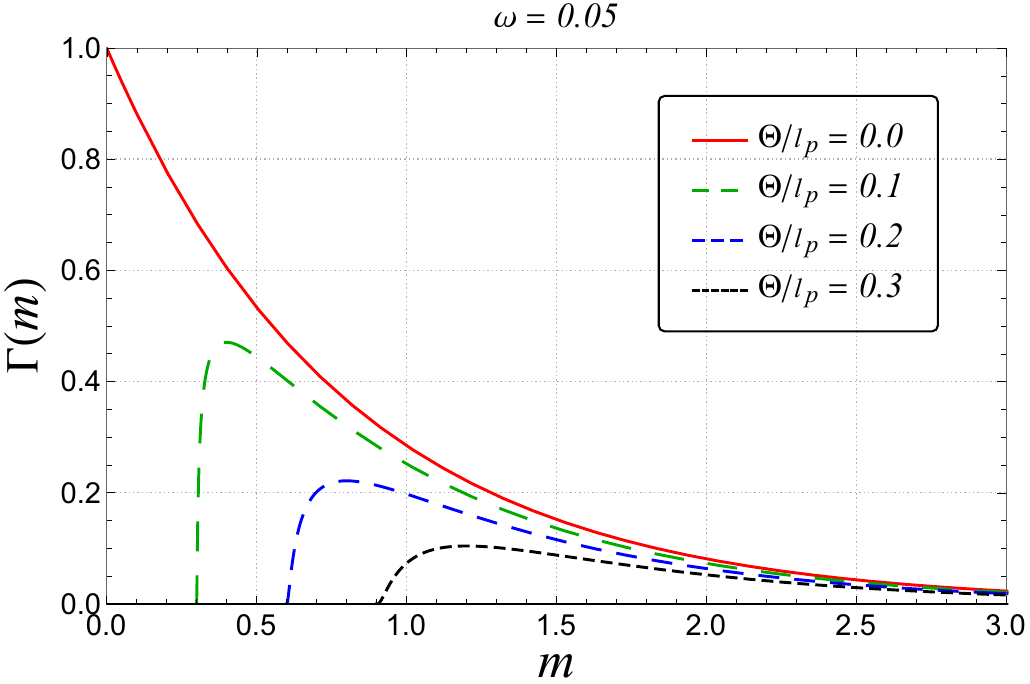}
		\includegraphics[width=0.32\textwidth]{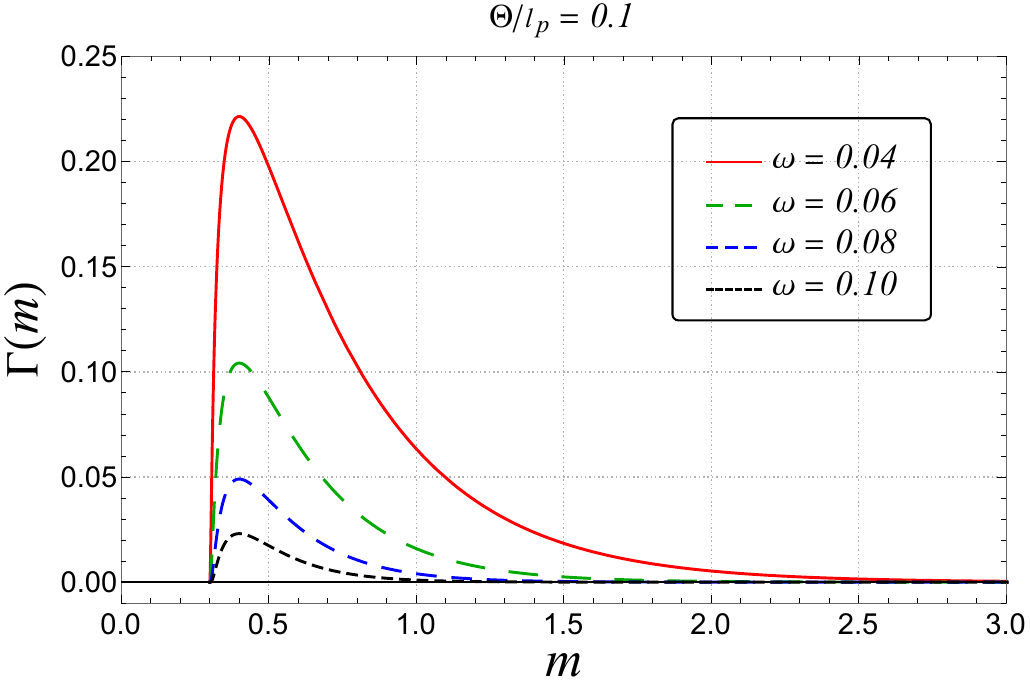}
		\caption{Behavior of the tunneling rate \(\Gamma\) as a function of the NC outer event horizon \(r_+\) for different NC parameters \(\tilde{\Theta}\).}\label{fig:Gamma}
	\end{center} 
\end{figure}
Fig.~\ref{fig:Gamma} shows the tunneling rate $\Gamma$ across the NC event horizon. The left panel displays $\Gamma$ as a function of the particle frequency $\omega$ for different values of $\tilde{\Theta}$; the center panel shows $\Gamma$ as a function of the black hole mass $m$ for fixed $\omega$, and the right panel shows $\Gamma$ as a function of $m$ for various values of $\omega$. For the parameter ranges shown, the NC correction reduces the tunneling rate, acting effectively as a barrier against particle escape. The qualitative behavior of the middle and right panels mirrors the temperature profile $T(r_+)$ in Fig.~\ref{fig:NCHT1}: during evaporation, the tunneling rate increases to a maximum, corresponding to the maximum temperature, and then decreases to zero when the black hole ceases to radiate. Moreover, in this model, the deformed potential generates an effective quantum-like barrier that prevents particles from escaping, and this barrier becomes stronger during the evaporation process. At the critical mass, this potential prevents the tunneling rate from exceeding its maximum value, which reflects the maximum radiation captured by the temperature profile in Fig.~\ref{fig:NCHT1} and the critical point in the heat-capacity profile in Fig.~\ref{fig:NCC1}, which signals a second-order phase transition. As the black hole mass approaches its minimum value $m^{\text{min}}$, the NC quantum-like barrier becomes dominant and entirely suppresses particle emission, causing the tunneling rate to vanish. This behavior explicitly reflects the heightened sensitivity of small black holes to the NC parameter, as detailed in SubSec.~\ref{sebsec:slrNC}. At this point, the formation of a cold finite remnant appears in the macroscopic thermodynamic description. This shows that the microscopic tunneling process and the macroscopic thermodynamic evolution are governed by the same NC deformation.

\paragraph{NC temperature}
Assuming a purely thermal spectrum (Boltzmann factor), \(\Gamma\sim e^{-\beta\omega}\) with \(\beta=1/T\), we obtain the compact temperature expression
\begin{align}\label{eq:newtemp1}
	T= \frac{\sqrt{m (m-3 \tilde{\Theta})}-3 \tilde{\Theta}+m}{4\pi\big(\sqrt{m (m-3 \tilde{\Theta})}+m\big)^2},
\end{align}
which, to leading order, reduces to the previously derived expression \eqref{eq:NCt1}:
\begin{align}\label{eq:newtemp2}
	T= \frac{1}{8 \pi m}-\frac{3 \tilde{\Theta}}{32 \pi m^2}=\frac{1}{4\pi r_+}-\frac{3\tilde{\Theta}}{4\pi r_+^2}.
\end{align}

\subsubsection{Particle-number density}

According to Refs.~\cite{number1, number2, number3,ntunn1, tunn4}, the emitted particle-number density can be computed from the tunneling rate. For our NC result \eqref{eq:tunneling1}, the emitted particle-number density is
\begin{align}\label{eq:dn1}
	n=\frac{\Gamma}{1-\Gamma}=\frac{1}{e^{4\pi\omega \frac{\big(\sqrt{m (m-3 \tilde{\Theta})}+m\big)^2}{\sqrt{m (m-3 \tilde{\Theta})}-3 \tilde{\Theta}+m}}-1},
\end{align}
which, to leading order in the NC parameter, becomes
\begin{align}\label{eq:dn2}
	n=\frac{1}{e^{8\pi\omega\frac{4 m^2}{4m-3\tilde{\Theta}} }-1}.
\end{align}
It is worth noting that this expression has the same form as the Planck distribution for black-body radiation and depends explicitly on the NC parameter that characterizes the deformed geometry. In the commutative limit \(\Theta\to 0\), this particle-number density reduces to the standard expression \cite{hawking1}:
\begin{equation}\label{eq:dn3}
	n=\frac{1}{e^{8\pi m\omega}-1}.
\end{equation}

\begin{figure*}[htb]
	\centering
	\includegraphics[clip=true,width=0.335\textwidth]{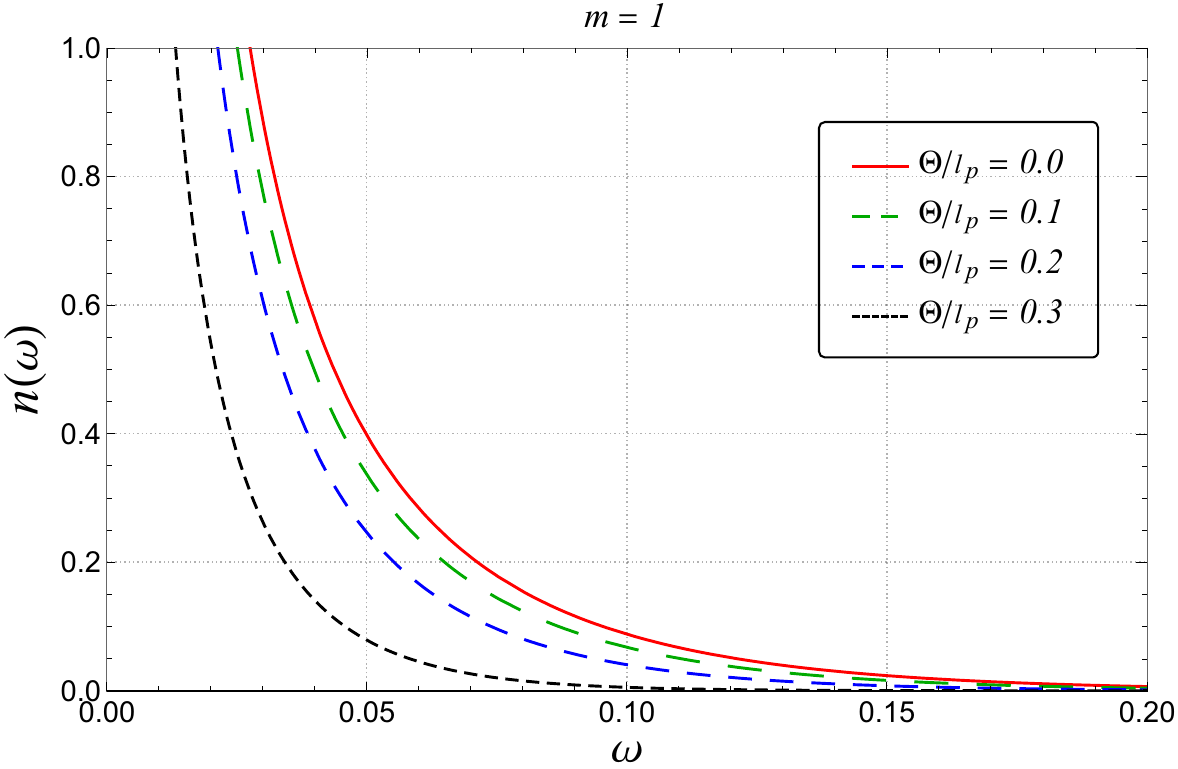}
	\includegraphics[clip=true,width=0.32\textwidth]{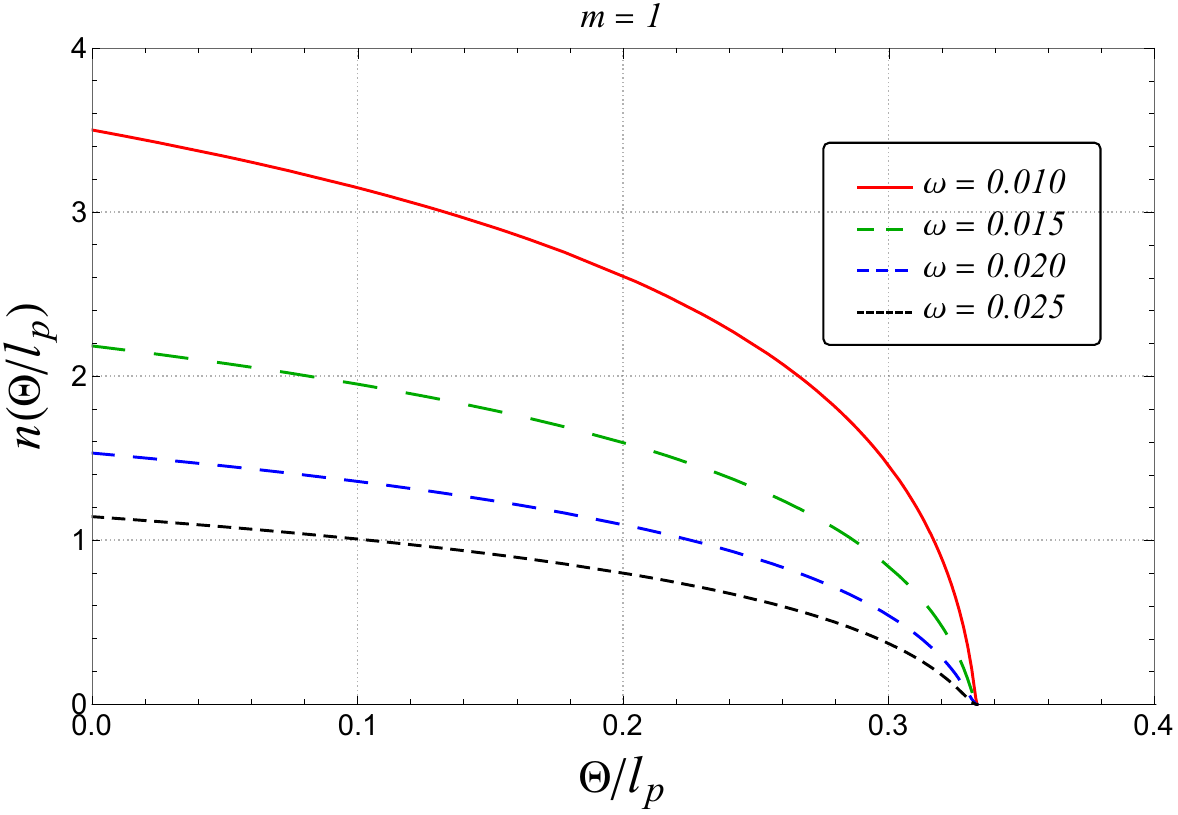}
	\includegraphics[clip=true,width=0.33\textwidth]{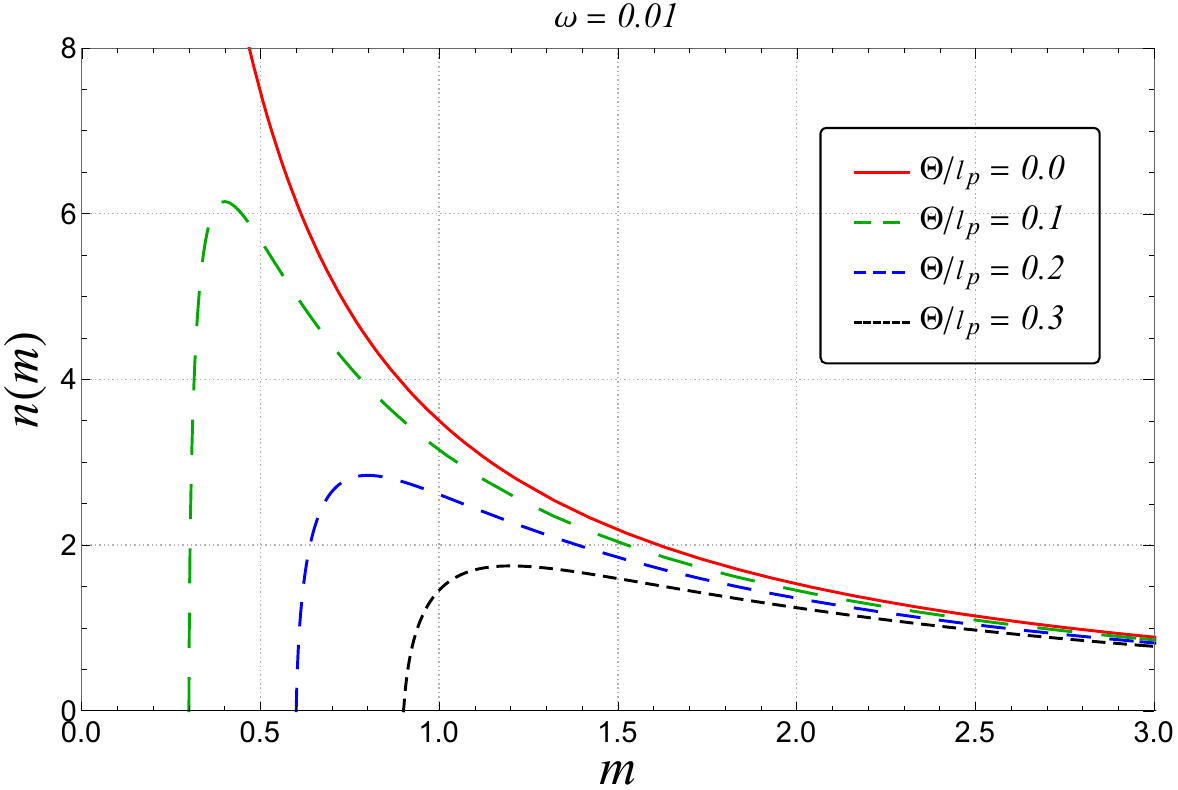}
	\caption{Variation of the particle-number density \(n\) emitted from an NC black hole. Left: \(n(\omega)\) for different \(\tilde{\Theta}\) (fixed \(m=1\)). Middle: \(n(\tilde{\Theta})\) for different \(\omega\) (fixed \(m=1\)). Right: \(n(m)\) for different \(\tilde{\Theta}\) (fixed \(\omega=0.01\)).}\label{fig3}
\end{figure*}
Fig.~\ref{fig3} shows the particle-number density \(n\) emitted by the NC Schwarzschild black hole as a function of the particle frequency \(\omega\) in the left panel. A key observation is the effect of non-commutativity: increasing the NC parameter \(\Theta\) reduces the emitted particle-number density. This suppression resembles the effect of an effective potential barrier in quantum mechanics. The middle panel illustrates how non-commutativity lowers the emission density as \(\tilde{\Theta}\) increases for several fixed frequencies. The right panel displays behavior similar to the temperature profile \(T(r_+)\) (see Fig.~\ref{fig:NCHT1}): as the black hole evaporates, the emission density reaches a maximum, corresponding to the maximum temperature, and then decreases to zero when the black hole stops radiating. Thus, non-commutativity effectively enhances the barrier against particle escape and suppresses the emission density, in agreement with the behavior of the tunneling rate. This suppression is also consistent with the thermodynamic picture, since both the tunneling rate and the heat-capacity analysis point to the same final evaporation stage and remnant formation.

\subsection{Non-thermal radiation}

In this subsection, we consider the large-frequency regime \(\omega\) and incorporate energy conservation. To proceed, we use Eq.~\eqref{eq:ims3} and integrate the relation \eqref{eq:ims5} over the quantity \(d(\hat{m}-\omega)\), which gives
\begin{align}\label{eq:ims7}
	\text{Im}\hat{S}&=\pi \int_{m}^{m-\omega}\frac{d(m-\omega)}{\kappa(m-\omega)}
\end{align}

According to Refs.~\cite{tunn1,tunn2}, the black hole mass in \eqref{eq:ims5} is replaced by \(m-\omega\). This substitution leads to the following expression:
\begin{align}\label{eq:ims6}
	\text{Im}\hat{S}=&-\pi \int_{m}^{m-\omega}\left(\frac{\big(\sqrt{(m-\omega) ((m-\omega)-3 \tilde{\Theta})}+(m-\omega)\big)^2}{\sqrt{(m-\omega) ((m-\omega)-3 \tilde{\Theta})}-3 \tilde{\Theta}+(m-\omega)}\right)d(m-\omega),\\
	=&-\pi\Bigg((m-\omega)^2+\frac{1}{2}  \sqrt{(m-\omega) ((m-\omega)-3 \tilde{\Theta})} (9 \tilde{\Theta}+2 (m-\omega))-m^2- \frac{1}{2}  \sqrt{m (m-3 \tilde{\Theta})} (9 \tilde{\Theta}+2 m) \notag \\
	&+\frac{27}{4}   \tilde{\Theta}^2 \log \left(2 \left(\sqrt{(m-\omega) ((m-\omega)-3 \tilde{\Theta})}+(m-\omega)\right)-3 \tilde{\Theta}\right)-\frac{27}{4}   \tilde{\Theta}^2 \log \left(2 \left(\sqrt{m (m-3 \tilde{\Theta})}+m\right)-3 \tilde{\Theta}\right)\Bigg),
\end{align}
This equation represents the usual relation between the tunneling rate and entropy \cite{tunn1,tunn2,tunn3}:
\begin{align}\label{eq:tunneling2}
	\hat{\Gamma}\sim e^{-2Im\hat{S}}=e^{\Delta\hat{S}_{BH}}.
\end{align}
Here, \(\Delta\hat{S}_{BH}= \hat{S}_{BH}(m-\omega)-\hat{S}_{BH}(m)\). From this relation, we obtain the following entropy expression:
\begin{align}\label{eq:entropy1}
	\hat{S}_{BH}=2\pi m^2 +\pi \sqrt{m (m-3 \tilde{\Theta})} (9 \tilde{\Theta}+2 m)\simeq \pi r_+^2+6\tilde{\Theta} r_+.
\end{align}
This entropy coincides with the expression obtained in Eq.~\eqref{eq:NCS}; in the commutative limit, we recover the Schwarzschild entropy. Up to second order, we also observe the logarithmic correction mentioned above and found in gauge theory of gravity models \cite{abdellah4,abdellah5,abdellahPhD}.

\subsubsection{Correlations}

In what follows, we investigate the effect of this NC approach on the statistical correlations between particles tunneling across the event horizon of the NC Schwarzschild black hole. To capture correlations, we must expand the entropy to second order in the NC parameter, since there is no modification of the correlation at first order. Using Eq.~\eqref{eq:tunneling2} together with Eq.~\eqref{eq:ims6}, the change in entropy is
\begin{align}\label{eq:deltaentropy}
	\Delta S=&-4\pi m\omega\left(1-\frac{\omega}{2m}\right)+\frac{1}{2}  \sqrt{(m-\omega) ((m-\omega)-3 \tilde{\Theta})} (9 \tilde{\Theta}+2 (m-\omega))- \frac{1}{2}  \sqrt{m (m-3 \tilde{\Theta})} (9 \tilde{\Theta}+2 m)\notag\\
	&+\frac{27}{2}\tilde{\Theta}^2 \log \left(\frac{2 \left(\sqrt{(m-\omega) ((m-\omega)-3 \tilde{\Theta})}+(m-\omega)\right)-3 \tilde{\Theta}}{2 \left(\sqrt{m (m-3 \tilde{\Theta})}+m\right)-3 \tilde{\Theta}}\right),
\end{align}
and substituting this into Eq.~\eqref{eq:tunneling2} gives
\begin{align}\label{eq:tunuling2'}
	\hat{\Gamma}
	&=\left(\frac{2 \left(\sqrt{(m-\omega) ((m-\omega)-3 \tilde{\Theta})}+(m-\omega)\right)-3 \tilde{\Theta}}{2 \left(\sqrt{m (m-3 \tilde{\Theta})}+m\right)-3 \tilde{\Theta}}\right)^{\frac{27\pi}{2}\tilde{\Theta}^2}\exp\bigg[-4\pi m\omega\left(1-\frac{\omega}{2m}\right)+\frac{1}{2}  \sqrt{(m-\omega) ((m-\omega)-3 \tilde{\Theta})}\notag\\
	&\times (9 \tilde{\Theta}+2 (m-\omega))- \frac{1}{2}  \sqrt{m (m-3 \tilde{\Theta})} (9 \tilde{\Theta}+2 m)\bigg].
\end{align}
Using the estimate \(\tilde{\Theta}\sim l_{\mathrm{Planck}}\) from Eq.~\eqref{eq:NCp1} and expanding the entropy difference to second order in the NC parameter, the tunneling rate can be written as
\begin{align}\label{eq:tunneling3}
	\hat{\Gamma}=e^{\Delta\hat{S}_{BH}}=\bigg(\left(1-\frac{\omega}{m}\right)^{\frac{27\pi}{2}l_{Planck}}e^{-6\pi\omega}\bigg)^{l_{Planck}}e^{-8\pi m\omega\left(1-\frac{\omega}{2m}\right)}.
\end{align}
This expression contains an exponential factor that reproduces the standard commutative non-thermal result \cite{tunn5}, together with a multiplicative correction depending on the Planck length. The latter represents a quantum modification similar to that found in Ref.~\cite{abdellah4}. The additional factor \(e^{-6\pi\omega}\) originates from the first-order NC correction. In the semi-classical limit \(l_{\mathrm{Planck}}\to 0\), the correction vanishes and the commutative result is recovered \cite{tunn5}.

We now investigate the statistical correlations between successive emissions of energies \(\omega_1\) and \(\omega_2\). Denote the single-emission rates by \(\ln\Gamma_{\omega_1}\) and \(\ln\Gamma_{\omega_2}\), and the combined emission by \(\ln\Gamma_{\omega_1+\omega_2}\). Following Refs.~\cite{correlation1, correlation2, correlation3}, the statistical correlation is
\begin{align}\label{eq:correlation1}
	\chi(\omega_1+\omega_2;\omega_1,\omega_2)&=\ln\left(\frac{\ln\hat{\Gamma}_{\omega_1+\omega_2}}{\ln\hat{\Gamma}_{\omega_1}+\ln\hat{\Gamma}_{\omega_2}}\right)\notag \\
	&=4\pi\omega_1\omega_2+\pi\sqrt{m (m-3 \tilde{\Theta})} (9\tilde{\Theta}
	+2m)-\pi\sum_{i=1}^{2}\sqrt{(m-\omega_i) (-3 a+m-\omega_i)} (9 \tilde{\Theta}+2   m-2  \omega_i)\notag\\
	&+\pi\sqrt{(m-\omega_1-\omega_2) (-3 a+m-(\omega_1+\omega_2))} (9\tilde{\Theta}+2 m-2 (\omega_1+\omega_2))\notag\\
	&+\frac{27}{2} \pi  \tilde{\Theta}^2 \ln \Bigg[\frac{A(m) A(m-(\omega_1+\omega_2))}{A(m-\omega_1)A(m-\omega_2)}\Bigg],
\end{align}
where
\begin{equation}
	A(m)=2 \left(\sqrt{m (m-3 \tilde{\Theta})}+m\right)-3 \tilde{\Theta}.
\end{equation}
The statistical correlation function remains nonzero in this NC gauge theory approach, as in the commutative case \cite{correlation1, correlation2, tunn8, correlation3}. Thus, different radiation frequencies during black hole evaporation remain correlated (\(\chi\neq0\)). Non-commutativity reduces the statistical correlations between successive emissions: the modified geometry acts like an effective potential well that suppresses particle tunneling across the horizon and weakens their mutual correlations. The existence of correlations among emitted quanta indicates that information can be carried by Hawking radiation, which addresses the information-loss problem. In this NC framework, information is preserved in a cold stable remnant (see Sec.~\ref{Subsub:NCT1}); the geometric suppression of correlations is consistent with mechanisms by which information may escape the horizon, akin to “hidden messengers in Hawking radiation” \cite{correlation1}.

\begin{figure*}[htb]
	\begin{center}
		\includegraphics[width=0.5\textwidth]{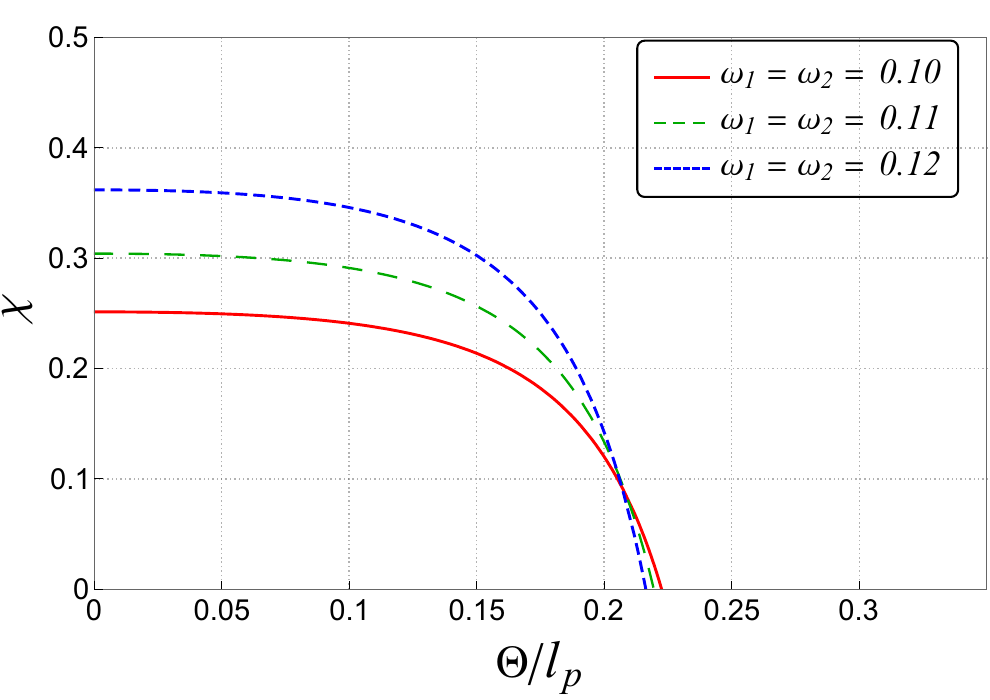}
		\caption{Variation of the statistical correlation function \(\chi\) as a function of the NC parameter \(\tilde{\Theta}\) for different frequencies \(\omega_i\).}\label{fig:NCC1}
	\end{center} 
\end{figure*}
Fig.~\ref{fig:NCC1} shows the behavior of the statistical correlation function \(\chi\) as a function of the NC parameter \(\tilde{\Theta}\) for different particle frequencies \(\omega\). Non-commutativity reduces the correlation between two successive emitted particles, while increasing the particle energies strengthens the correlation: higher-energy emissions carry more information, whereas lower-energy emissions carry less. Moreover, the leading-order expression for the correlation reads
\begin{align}\label{eq:correlation2}
	\chi(\omega_1+\omega_2;\omega_1,\omega_2) \simeq 8\pi\omega_1\omega_2+\frac{27\pi}{2}\tilde{\Theta}^2\ln\left(\frac{m(m-(\omega_1+\omega_2))}{(m-\omega_1)(m-\omega_2)}\right),
\end{align}
which matches the result obtained in NC gauge theory of gravity studies \cite{abdellah4,abdellahPhD}. In the limit \(\tilde{\Theta}=0\), we recover the commutative correlation function \cite{correlation1, correlation2, tunn8, correlation3}.

\section{Conclusion} \label{Sec:Conc}

In this work, we have presented a novel approach to constructing black hole solutions within the framework of NC gauge theory by applying the SW map to the interaction potential and then solving Einstein's field equations for the resulting deformed gauge potential. We analyzed the geometric properties of both NC Schwarzschild and Reissner-Nordstr\"om (RN) black holes, the latter exhibiting a genuine branch dependence between attractive and repulsive electric interactions, which is absent in the commutative limit. We also presented the metrics in both compact form and as expansions in terms of the NC parameter $\Theta$. Our results show that the compact NC metric shares similarities with certain regular black hole solutions and NC matter-distribution models. In the approximate limit, the curvature function exhibits behavior similar to the RN black hole, characterized by two horizons, an inner and an outer one. Moreover, this geometry leads to a reduction of the event horizon radius, such that $r_+<2m$. The analysis of the scalar invariants shows that non-commutativity softens the central singularity but does not remove it completely at leading order, which motivates the study of higher-order corrections that may be promising for eliminating the central singularity of spacetime. In addition, the energy-condition analysis shows that the physically relevant exterior region remains acceptable for the parameter ranges considered, whereas any violations are confined to the inner-horizon or near-singularity region, depending on the branch. This violation induces a repulsive behavior that helps to avoid the central singularity. The attractive/repulsive asymmetry in the charged sector also affects the energy conditions, a feature that is absent in the commutative limit.

Subsequently, we investigated the thermodynamic properties of both the NC Schwarzschild and RN-like black holes, as well as the influence of this geometry on thermal stability. We computed the ADM mass, Hawking temperature, and entropy for both NC Schwarzschild and RN-like black holes. Our findings indicate that non-commutativity produces a new minimum in the final stage of evaporation in the ADM mass profile, in agreement with other NC geometric models \cite{abdellah2}. This framework removes the temperature divergence and leads to a cold remnant as the black hole ceases to radiate. A comparison between the uncharged case and the charged case with attractive and repulsive electric interactions shows that the repulsive branch stops radiating before the attractive branch, while the uncharged black hole is the last to stop radiating. Moreover, the entropy displays a second-order logarithmic correction, deviating from the area law due to the inclusion of NC geometry through modified matter and electric distributions in the energy-momentum tensor. However, when this analysis is extended to the NC RN-like case without mass-term corrections, the area law and the first law of black hole thermodynamics remain valid (see SubSec.~\ref{SubSec:TPNCRN}). An estimate of the NC parameter from the temperature profile yields a bound at the Planck scale, $\Theta \sim l_p$, consistent with NC gauge-gravity models \cite{abdellah2,abdellah4,abdellah5}. This approach therefore provides a convenient framework for studying black hole physics in a way that is closely analogous to the commutative RN case.

Furthermore, we examined the heat capacity of the NC Schwarzschild black hole and both branches of the RN-like black hole, identifying a divergence at the critical point $r_+^{\rm crit}$, which corresponds to the state of maximum radiation, i.e. the maximum temperature $T^{\rm max}$. At this point, a second-order phase transition occurs from an unstable large black hole to a stable small black hole, a behavior absent in commutative solutions. Thermal stability was further confirmed through the Helmholtz free energy, which exhibits two extrema: a maximum representing the unstable large black hole and a minimum representing the stable small black hole, consistent with the heat-capacity profile in Fig.~\ref{fig:c}. These results indicate that large black holes are thermodynamically unstable, whereas small ones are stable, eventually leading to a cold remnant. Upon introducing pressure, we observed three distinct regimes around the critical pressure $P_C$: for $P<P_C$, a quasi-swallowtail structure appears with smooth curves; at $P=P_C$, an inflection point emerges, corresponding to a Hawking--Page-like phase transition \cite{abdellah2,abdellahPhD}; and for $P>P_C$, this structure disappears. This behavior signals a second-order phase transition consistent with standard thermodynamic theory. We also investigated the susceptibility and linear response of the black hole to variations in $\tilde{\Theta}$, finding that small black holes are highly sensitive to small changes in $\tilde{\Theta}$, whereas the sensitivity becomes weaker for larger black holes, in good agreement with the mass and temperature profiles.

Finally, we investigated the effect of non-commutativity on quantum tunneling from the event horizon of NC Schwarzschild black hole. We employed a semi-classical tunneling approach in two scenarios. First, we examined pure thermal radiation characterized by the emission of massless particles at low frequency. This process reproduces the thermal properties obtained through the geometric approach, confirming that non-commutativity preserves this equivalence \cite{abdellah4}. We found that the geometry reduces the tunneling rate, similarly to an effective potential barrier in quantum mechanics, and leads to a vanishing tunneling rate in the final stages of evaporation, further confirming the existence of a cold remnant (see Fig.~\ref{fig:NCHT1}). A similar effect was observed in the particle-number density, where the geometry enhances the barrier against tunneling particles, which leads to the emergence of a second-order phase transition in the macroscopic classical thermodynamic description. This establishes a direct connection between the microscopic emission process and the macroscopic phase structure of the NC black hole, in which the microscopic tunneling process and the macroscopic thermodynamic evolution are governed by the same NC deformation. In the second scenario, we examined non-thermal radiation by taking energy conservation into account \cite{tunn2}, especially for high-frequency emissions. We found that the non-thermal tunneling rate is consistent with a unitary quantum description and is related to the entropy difference. This entropy contains a second-order logarithmic correction in the NC parameter, with a coefficient linked to the Planck length $l_p$, representing a quantum scale effect. These results are in agreement with other QG candidates, including LQG, ST, the Generalized Uncertainty Principle (GUP), and Modified Dispersion Relations (MDR) \cite{QLG1, QLG2, ST3, nozari3, nozari4}, and are also similar to those obtained from thermal fluctuation corrections \cite{thermalfluc1,thermalfluc2,thermalfluc3}. Finally, we analyzed the statistical correlation between two successive emitted particles with energies $\omega_1$ and $\omega_2$. Our results show that non-commutativity reduces this correlation: as the NC parameter $\Theta$ increases, the correlation weakens, reaching zero for specific values of $\tilde{\Theta}$ and the frequencies, and eventually becoming negative (\(\chi<0\)) for sufficiently large values of $\tilde{\Theta}$.

Most notably, the present model predicts the emergence of a stable, cold remnant with profound cosmological implications. Within our framework, the mass scale of these remnants is governed directly by the non-commutative parameter via $M^{\text{min}}=\frac{3c^2}{G}\,\tilde{\Theta}$, which naturally leads to a mass at the Planck scale, specifically yielding $M^{\text{min}}\simeq2.73 M_{P}$. If such remnants were produced during the final stages of primordial black hole evaporation in the early universe, these stable, non-radiating objects could serve as an ideal, purely gravitational cold dark matter (CDM) candidate. This establishes a concrete bridge between non-commutative gauge theory and dark universe phenomenology, while naturally accounting for the missing mass of the universe. Furthermore, the preservation of quantum information within these stable remnants offers a compelling resolution to the black hole information loss paradox. Consequently, our model opens up several promising pathways for future investigations. Specifically, modeling the complete evaporation dynamics of primordial black holes, calculating the expected remnants abundance based on this specific mass scale, and analyzing their clustering behavior in the early universe represent vital directions for subsequent research.

Overall, these results provide a more complete description of the NC black hole geometry and offer a simpler, physically transparent framework that captures geometric, thermodynamic, and quantum features in a unified way, thereby complementing the analyses presented in this work.

\acknowledgments
The author expresses his sincere gratitude to colleagues for their valuable assistance in proofreading the manuscript and improving the English phrasing. Additionally, the author would like to thank the anonymous reviewers for their constructive comments and insightful suggestions, which greatly helped to improve the quality and physical motivation of this paper.

\appendix

\section{Non-commutative correction to the Schwarzschild-De-Sitter black hole} \label{appA}

In this appendix, we present the NC correction to the Schwarzschild-de Sitter black hole. The commutative solution is given by
\begin{align}
	f(r)= 1-\frac{2m}{r}-\frac{\Lambda}{3}r^2,
\end{align}
and the corresponding Einstein equation reads
\begin{equation}\label{eq:EEC}
	G^\mu_\nu-\Lambda\delta^\mu_\nu=T^\mu_\nu.
\end{equation}
Since the cosmological term is constant, it does not receive a gauge correction. Therefore, we use the NC matter correction given by Eq.~\eqref{eq:emtdmass}, which is obtained from the NC Newton potential \(\hat{\Phi}_N\). The resulting NC curvature function is then written as
\begin{align}\label{eq:NClambda1}
	f_\Lambda(r)&=1-\frac{2mr}{\big(r^2+3m\tilde{\Theta}\big)}-\frac{\Lambda}{3}r^2\notag\\
	&\simeq 1-\frac{2m}{r}+\frac{6m^2}{r^3}\tilde{\Theta}-\frac{\Lambda}{3}r^2,
\end{align}
which clearly reduces to the Schwarzschild-de Sitter black hole in the commutative limit \(\tilde{\Theta}=0\) \cite{Chandr1}.

If one wishes to compute the NC correction to the cosmological term, it is necessary to introduce the cosmological constant directly into the Newtonian potential, rather than through the Einstein equation \cite{hobson2006}:
\begin{equation}
	\Phi_{N,\Lambda}=-\frac{GM}{r}-\frac{\Lambda c^2 r^2}{6}.
\end{equation}
By following the same procedure as in Sec.~\ref{Sec:NCSBH}, we obtain the NC gauge potential for the Newtonian plus cosmological term:
\begin{equation}
	\hat{\Phi}_{N}=-\frac{GM}{r}-\frac{\Lambda c^2 r^2}{6}+\frac{\Theta}{l_p c^2}\left(\frac{G^2 M^2}{r^3}+\frac{\Lambda c^2 G M}{2}-\frac{\Lambda^2 c^4 r^3}{18}\right).
\end{equation}
The total NC density for the region \(r>0\) becomes
\begin{equation}
	\hat{\rho}_{N}+\hat{\rho}_{\Lambda}=\tilde{\Theta}\frac{3GM^2}{2\pi c^2 r^5}
	+\frac{\Lambda c^2}{8\pi G}-\tilde{\Theta}\frac{\Lambda^2 c^2\,r}{6\pi G},\label{eq:NCDMCC}
\end{equation}
from which we see that there is no correction to first order in the cosmological constant. The corresponding NC curvature function for the Schwarzschild-de Sitter black hole is therefore
\begin{align}
	f_\Lambda(r)&=1-\frac{2mr}{\big(r^2+3m\tilde{\Theta}\big)}-\frac{\Lambda r^2}{3(1+\Lambda\tilde{\Theta})},\notag\\
	&\simeq 1-\frac{2m}{r}+\frac{6m^2}{r^3}\tilde{\Theta}-\frac{\Lambda}{3}r^2+\frac{\Lambda^2 r^3 \tilde{\Theta}}{3}.
\end{align}
As is well known, the cosmological constant is extremely small, \(\Lambda\propto10^{-51}m^{-2}\), and when it is multiplied by the NC parameter, which is expected to be of the order of the Planck length, the second-order correction to the cosmological constant becomes negligible compared with the other terms at small distances. In this case, we recover the solution in Eq.~\eqref{eq:NClambda1}. On the other hand, at very large distances, \(r\to \infty\), this term may become dominant.

\section{Non-commutative correction to the Reissner-Nordström-De-Sitter black hole} \label{appB}

The NC metric of the RN-like de Sitter black hole can be derived in a similar way. In this case, we consider two different scenarios. In the first one, we compute the NC correction only to the Coulomb gauge potential, while using the NC Maxwell energy-momentum tensor in Eq.~\eqref{eq:TNC_corrected}. This yields the following NC RN-dS black hole solution:
\begin{align}
	f_{Q,\Lambda}(r)&=1-\frac{2m}{r}+\frac{Q^2}{(r^2-2\tilde{\Theta}Q)}-\frac{\Lambda}{3}r^2,\notag\\
	&\simeq 1-\frac{2m}{r}+\frac{Q^2}{r^2}+\frac{2\tilde{\Theta}Q^3}{r^4}-\frac{\Lambda}{3}r^2.\label{eq:NCRNL2}
\end{align}
In the limit \(\tilde{\Theta}\to0\), we recover the commutative solution \cite{li}.

The second scenario includes both the NC matter correction, with the cosmological constant inserted into the Newton potential through Eqs.~\eqref{eq:EMT1} and \eqref{eq:NCDMCC}, together with Eq.~\eqref{eq:TNC_corrected}. In that case, we obtain
\begin{align}
	f_{Q,\Lambda}(r)&=1-\frac{mr}{\big(r^2+3m\tilde{\Theta}\big)}+\frac{Q^2}{(r^2-2\tilde{\Theta}Q)}-\frac{\Lambda r^2}{3(1+\Lambda\tilde{\Theta})},\notag\\
	&\simeq 1-\frac{2m}{r}+\frac{Q^2}{r^2}+\bigg(\frac{6m^2}{r^3}+\frac{2Q^3}{r^4}\bigg)\tilde{\Theta}-\frac{\Lambda}{3}r^2+\frac{\Lambda^2 r^3 \tilde{\Theta}}{3}.\label{eq:NCRNML2}
\end{align}
For small distances, the second-order correction in the cosmological constant becomes negligible, and this leads to
\begin{align}
	f_{Q,\Lambda}(r)\simeq 1-\frac{2m}{r}+\frac{Q^2}{r^2}+\bigg(\frac{6m^2}{r^3}+\frac{2Q^3}{r^4}\bigg)\tilde{\Theta}-\frac{\Lambda}{3}r^2.\label{eq:NCRNM3}
\end{align}

\bibliography{ref}

\end{document}